\documentclass[12pt,a4paper]{article}
\usepackage[utf8]{inputenc}

\usepackage[margin=3cm,includefoot,footskip=30pt]
{geometry}
\usepackage[french,english]{babel} 
\usepackage[backend=biber,style=numeric]{biblatex}
\usepackage{amsmath}
\usepackage{amsfonts}
\usepackage{amssymb}

\usepackage{stmaryrd}
\usepackage{enumitem}

\usepackage{tikz}
\usetikzlibrary{arrows}
\usetikzlibrary{math} 
\newcommand{\mbullet[1]}[.6]{\mathbin{\vcenter{\hbox{\scalebox{#1}{$\bullet$}}}}}
\newcommand{\sbullet[1]}[.2]{\mathbin{\vcenter{\hbox{\scalebox{#1}{$\bullet$}}}}}

\usepackage[bookmarks=true,colorlinks,unicode,pdfpagelabels]{hyperref}
\usepackage{array,longtable}
\newcolumntype{C}{>{$}c<{$}} 
\newcolumntype{L}{>{$}l<{$}}
\newcolumntype{R}{>{$}r<{$}}
\usepackage[small,compact]{titlesec}
\titleformat{\subsection}[runin]{\normalfont\itshape}{\thesubsection}{.5em}{}[]
\usepackage{aliascnt}
\addto\extrasenglish{

}

\newcommand{\rv}{random variable}
\newcommand{\lb}{\left\{}
\newcommand{\rb}{\right\}}
\newcommand{\nun}{\lb\rb}
\newcommand{\eun}{\emptyset}
\newcommand{\setmin}{\! \setminus \!}
\newcommand{\ind}[1]{\mathbf{1}_{#1}}
\newcommand{\expec}[1]{{\mathbb E}\left(#1\right)}
\newcommand{\expb}[2]{{\mathbb E}_{#1}\left(#2\right)}
\newcommand{\var}[1]{{\mathbb V}\text{ar}\left(#1\right)}
\newcommand{\varb}[2]{{\mathbb V}\text{ar}_{#1}\left(#2\right)}
\newcommand{\cov}[1]{{\mathbb C}\text{ov}\left(#1\right)}
\newcommand{\covb}[2]{{\mathbb C}\text{ov}_{#1}\left(#2\right)}
\newcommand{\cvc}[1]{{\mathbb C}\text{vc}\left(#1\right)}

\newcommand{\jN}{j\in \mathbb{N}}
\newcommand{\kN}{k\in \mathbb{N}}
\newcommand{\lN}{l\in \mathbb{N}}
\newcommand{\mN}{m\in \mathbb{N}}
\newcommand{\nN}{n\in \mathbb{N}}
\newcommand{\intt}[1]{\llbracket #1 \rrbracket}
\newcommand{\floor}[1]{\left \lfloor #1 \right \rfloor}
\newcommand{\bec}{\raisebox{-.45em}{\begin{tikzpicture} [scale=.19] \draw (0,0) node {$\mbullet$} (-.5,.87) node {$\mbullet$} (.5,.87)  node {$\mbullet$}; \end{tikzpicture}}}
\DeclareMathOperator{\sgn}{sgn}

\makeatletter
\newcommand\Autoref[1]{\@first@ref#1,@}
\def\@throw@dot#1.#2@{#1}
\def\@set@refname#1{
    \edef\@tmp{\getrefbykeydefault{#1}{anchor}{}}%
    \xdef\@tmp{\expandafter\@throw@dot\@tmp.@}%
    \ltx@IfUndefined{\@tmp autorefnameplural}%
         {\def\@refname{\@nameuse{\@tmp autorefname}s}}%
         {\def\@refname{\@nameuse{\@tmp autorefnameplural}}}%
}
\def\@first@ref#1,#2{%
  \ifx#2@\autoref{#1}\let\@nextref\@gobble
  \else%
    \@set@refname{#1}
    \@refname~\ref{#1}
    \let\@nextref\@next@ref
  \fi%
  \@nextref#2%
}
\def\@next@ref#1,#2{%
   \ifx#2@ and~\ref{#1}\let\@nextref\@gobble
   \else, \ref{#1}
   \fi%
   \@nextref#2%
}
\makeatother

\newcommand{\ndeux}{\lb\nun\rb}
\newcommand{\edeux}{\lb\eun\rb}
\newcommand{\ntrois}{\lb\ndeux\rb}
\newcommand{\etrois}{\lb\edeux\rb}
   \newcommand{\bquatre}{\lb\ndeux\nun\rb}
\newcommand{\nquatre}{\lb\ndeux,\nun\rb}
\newcommand{\equatre}{\lb\edeux,\eun\rb}
\newcommand{\ncinq}{\lb\ntrois\rb}
\newcommand{\ecinq}{\lb\etrois\rb}
   \newcommand{\bsix}{\lb\bquatre\rb}
\newcommand{\nsix}{\lb\nquatre\rb}
\newcommand{\esix}{\lb\equatre\rb}
   \newcommand{\bsept}{\lb\ntrois\nun\rb}
\newcommand{\nsept}{\lb\ntrois,\nun\rb}
\newcommand{\esept}{\lb\etrois,\eun\rb}
   \newcommand{\bhuit}{\lb\bquatre\nun\rb}
\newcommand{\nhuit}{\lb\nquatre,\nun\rb}
\newcommand{\ehuit}{\lb\equatre,\eun\rb}
   \newcommand{\bneuf}{\lb\ntrois\ndeux\rb}
\newcommand{\nneuf}{\lb\ntrois,\ndeux\rb}
\newcommand{\eneuf}{\lb\etrois,\edeux\rb}
   \newcommand{\bdix}{\lb\bquatre\ndeux\rb}
\newcommand{\ndix}{\lb\nquatre,\ndeux\rb}
\newcommand{\edix}{\lb\equatre,\edeux\rb}
   \newcommand{\bonze}{\lb\bquatre\ntrois\rb}
\newcommand{\nonze}{\lb\nquatre,\ntrois\rb}
\newcommand{\eonze}{\lb\equatre,\etrois\rb}
   \newcommand{\bdouze}{\lb\ntrois\ndeux\nun\rb}
\newcommand{\ndouze}{\lb\ntrois,\ndeux,\nun\rb}
\newcommand{\edouze}{\lb\etrois,\edeux,\eun\rb}
   \newcommand{\btreize}{\lb\bquatre\ndeux\nun\rb}
\newcommand{\ntreize}{\lb\nquatre,\ndeux,\nun\rb}
\newcommand{\etreize}{\lb\equatre,\edeux,\eun\rb}
   \newcommand{\bquatorze}{\lb\bquatre\ntrois\nun\rb}
\newcommand{\nquatorze}{\lb\nquatre,\ntrois,\nun\rb}
\newcommand{\equatorze}{\lb\equatre,\etrois,\eun\rb}
   \newcommand{\bquinze}{\lb\bquatre\ntrois\ndeux\rb}
\newcommand{\nquinze}{\lb\nquatre,\ntrois,\ndeux\rb}
\newcommand{\equinze}{\lb\equatre,\etrois,\edeux\rb}
   \newcommand{\bseize}{\lb\bquatre\ntrois\ndeux\nun\rb}
\newcommand{\nseize}{\lb\nquatre,\ntrois,\ndeux,\nun\rb}
\newcommand{\eseize}{\lb\equatre,\etrois,\edeux,\eun\rb}

\newcommand{\monemph}[1]{\noindent \textbf{#1: }}
\newcommand{\myitem}[2]{\vspace{.3cm} \subsection{#1}\hspace{-5pt}: #2\hfill $\clubsuit$}
\newcommand{\myproof}{\vspace{.2cm} \hspace{-1.5em} \bec \hspace{.1em} }

\newcounter{myenumi}

\title{Random pure sets}
\author{Michel Bauer\,$^1$}

\begin{document}

\maketitle

 \vspace{-.5cm}

\begin{center}
\footnotesize $^1$ Université Paris-Saclay, CNRS, CEA, \\ Institut de Physique Théorique, \\ 91191 Gif-sur-Yvette, France, \\ email:michel.bauer@ipht.fr
 \end{center}

\begin{abstract}
We study from a statistical mechanics viewpoint some of the simplest mathematical objects, finite pure sets. Starting from the empty set, new generations are produced step by step, sets of the next generation being those whose elements are the sets of the current generation. We compute in particular correlations and limiting laws for chains of various lengths for the membership relation when the number of generations becomes large. 
\end{abstract}

\section{Motivations}

What are the salient features of a large random object? A great part of statistical mechanics is devoted to this question. The object is selected in a set of configurations, be they for instance positions and speeds of molecules in a box, possible shapes of polymers, orientations of spins in a magnetic material. Statistical methods are relevant because the set of configurations is large and full access to the details is well beyond our capacities. This work is devoted to a rather peculiar instance of this phenomenon: the simplest objects of set theory.  

Set theory is not part of the standard training of physicists and is unfamiliar to most theoretical physicists, including in particular the author. However, over the recent years, quantum computing and the theory of computability in general have aroused a wide interest in our community. It turns out that many issues in this domain are closely connected in one way or another to classical set theory.

The axioms of set theory\footnote{In fact, there are many, possibly conflicting, versions, but this is irrelevant for our discussion of finite sets. However, the vocabulary we use is tainted by the most commonly accepted framework.} were carefully build to be able to deal consistently (well, hopefully consistently) with infinite sets, but the author realized while struggling with the basics of quantum computability that even the simplest finite sets retained some mystery for him, and they still do after this work. At this point, apologies are in order for the naivety of the considerations that follow. They should at least convince readers not familiar with axiomatic set theory that no previous knowledge of it is required to understand the ensuing discussion. Let us also note that all the counting problems we deal with in the following can be rephrased as counting nodes and egdes on a particular family of rooted trees, the so-called identity trees. 

The only basic relation among sets, $\in$, expresses the fact that a set is an element of, or belongs to, or is a member of another. Set theory begins with the assumption that a set exists, the empty set $\nun$, that has no elements. Then comes a list of rules that allow to construct more complicated sets, and in the end most if not all mathematics can be rephrased in this framework. An axiom states that sets are fully determined by their elements. This allows to talk about inclusion, a set being included in another if each of its elements is an element of the other. The only further axiom that plays a crucial role in this work is the existence of the power set, which states that the collection of sets included in a given set builds in fact another set. Thus, starting from $\nun$, that we call the zeroth generation one builds a new set, the first generation, whose sole element is the empty set $\nun$, a set denoted by $\ndeux$. So in the first generation there are $2$ sets: $\ndeux$ and $\nun$. Then comes the second generation, whose elements are sets whose elements are chosen among $\ndeux$ and $\nun$, that is, whose elements are subsets of $\bquatre$.\footnote{Maybe more readable as $\nquatre$, but the notation without ``,'' is unambigous as the reader may check: it is all about the balance of opening braces  $\lb \right.$ and closing braces $\left. \rb$.} There are $4$ of them: $\bquatre$ itself, $\ntrois$, then $\ndeux$ and finally $\nun$ (the order is arbitrary). Those $4$ sets are the elements of $\bseize$,\footnote{Maybe more readable as $\nseize$.} whose subsets make the third generation which has $16$ members. And so it continues. The elements of the set of the $k$-th generation are called pure sets\footnote{It is unclear to the author whether the terminology ``Pure set'' is well-established. The author borrowed it from \cite{moore-1991}. The French ``Ensemble pur'' is also used in the recent \cite{leroy-2024}.} of depth less than or equal to $k$. The general procedure should be clear. Pure sets of depth $\leq k$ are automatically finite, and \textit{in the sequel the expression ``pure set'' stands for ``finite pure set'' unless otherwise stated}.  

The number of finite pure sets grows rapidly with the generation number and this motivates a statistical study. We shall assign the uniform probability measure (the counting measure up to normalization) to the family of pure sets of the $k$-th generation. We aim to study the distribution of some relevant observables in the large $k$ limit.  

As pure sets have elements (if any) that are again pure sets, one may consider chains for the membership relation $\in$ within a pure set: a chain in $x$ is a sequence $x\ni x'\ni x'' \ni...$.\footnote{The notation $x\ni x'$ means the same as $x'\in x$.} The number of chains of different types in a pure set are among the interesting observables, and chain counting will be one of our main goals in this work. A commonly (but not universally) accepted axiom of set theory is the foundation axiom, which implies that chains for $\in$ are always finite, and in particular cannot form cycles. Pure sets satisfy this axiom by construction, and pure sets in the $k$-th generation can be characterized as having no chain of length $>k$ ($x$ alone being a chain of length $0$ in $x$). 

The computation of the joint law of the number of chains of different types under the uniform probability measure for a large number of generations reveals a remarkable simplicity. The averages are easy to compute analytically for every $k$, the (co)variances are just slightly more challenging, and it turns out that in the large $k$ limit all the joint fluctuations are governed by just a single Bernoulli \rv\ and two independent Gaussian \rv s. This remarkable simplicity is our central result.

\vspace{.3cm}
This work does not claim compliance with mathematical standards. However, the main results are obtained via a number of small steps, sometimes involving lengthy computations. So we felt the need to organize matters in a more formal way than is usual in physics. We hope that our compromise enhances clarity and readability, and does not amount to take the worse of both worlds. We delimit within each section relevant fragments which we refer to as Items. The size of those Items is variable. They can cover a definition or the statement of a result and its justification for instance. Items start with a number and a short title, and end with a $\clubsuit$ sign. The beginning of a justification is indicated by a \bec, the \textit{because} symbol sometimes used in logic.

\vspace{.3cm}
Our main results are summarized in \autoref{i:lbcf}. Making this Item, which is of technical nature,  self-contained comes at the price of absorbing a handful of dry preliminary definitions which are detailed later.

\Autoref{sec:psbasics,sec:rpsc,sec:ffdb} are essentially descriptive. They provide the background. \autoref{sec:psbasics} gives some details on pure sets basics. \autoref{sec:rpsc} lists three alternative ways to represent pure sets. The most useful are binary strings and identity trees. \autoref{sec:ffdb} gives pictures illustrating the different representations for the first three generations of pure sets, and ends with considerations on the next generations, leading to some (hopefully entertaining) numerology.

The following sections contain the more technical discussions. \autoref{sec:spcf} gives some properties of the functions counting chains of various lengths for the $\in$ relation in a pure set. \autoref{sec:probmod} introduces the probabilistic model, which interprets chain counting functions as random variables. We give a number of illustrations, ranging from very simple to not so simple, meeting a few transcendental numbers on the way. This covers the first points of our main result,  \autoref{i:lbcf}. \autoref{sec:wgf} introduces weights and their generating functions, the goal being to manipulate easily the chain counting functions and their recursive properties. \autoref{sec:ie} is devoted to an application of these generating functions and their probabilistic use. \autoref{sec:mcvcf} contains the crucial computation of the correlations between chain counting functions for arbitrary lengths. \autoref{sec:lka} deals with the asymptotics of the correlations when the number of generation gets large and gives the final steps establishing the remaining statements of our main result, \autoref{i:lbcf}. \autoref{sec:ld} gives some results that go beyond the simplest correlations, mostly large deviation estimates and subleading fluctuations. We make a few general conjectures but we are able to justify only some special cases.

\section{Main results} \label{sec:mr}

\vspace{-.3cm}

This section collects the main results of this work, preceded by the basic definitions in dry form. Detailed definitions with examples can be found in later sections, together with the derivations.

\myitem{Pure sets}{The sequence of finite sets $\left({\mathcal S}_k\right)_{k\geq -1}$ is defined recursively by ${\mathcal S}_{-1}:=\nun$ and ${\mathcal S}_k={\mathcal P}({\mathcal S}_{k-1})$ (the power set) for $\kN$.

For $\kN$, a pure set of depth $\leq k$ is by definition any element of ${\mathcal S}_k$.

We set $Z_k:= \# {\mathcal S}_k$ (the cardinal of ${\mathcal S}_k$): $Z_{-1}=0$ and $Z_k=2^{Z_{k-1}}$ for $\kN$.
}\\
For details, see \autoref{sec:psbasics}. 
\myitem{Chains of pure sets\label{i:cps}}{A finite sequence $x_0,\cdots,x_n$ of pure sets is called a chain if $x_0\in x_1\in \cdots \in x_n$. Chains never close to make a cycle. We call $n$ the length of the chain.

The set $x_0$ (resp. $x_n$) is called the beginning (resp. the end) end of the chain, and the chain is called maximal if $x_0=\nun$ and $n>0$.\footnote{Excluding $n=0$ is just a convenient convention.}

If $x$ is a pure set, a chain in $x$ is, by definition, a chain $x_0,\cdots,x_n$ that ends in $x$, i.e. such that $x_n=x$.}\\
For details, see \autoref{sec:spcf}.

\myitem{Direct chain counting functions}{Direct chain counting functions are  denoted by  $G^{(n)}_k$ and $H^{(n)}_k$, $\kN$, $n\in {\mathbb Z}$. They map ${\mathcal S}_k$ to  ${\mathbb N}$. For $x\in {\mathcal S}_k$, $G^{(n)}_k(x)$ is the number of maximal chains of length $n$ in $x$, and $H^{(n)}_k(x)$ the  number of chains, maximal or not, of length $n$ in $x$.}\\
Later we shall also define inverse chain counting functions. For details, see \autoref{sec:spcf}.

\myitem{The probabilistic model\label{i:probmod}}{We endow ${\mathcal S}_k$, the set of pure sets of depth $\leq k$, with the uniform probability measure. So each element of  ${\mathcal S}_k$ has probability $1/Z_k$.

}\\
For details, see \autoref{sec:probmod}.

\myitem{Normalization of a \rv \label{i:nrv}}{The normalization $\widehat{X}$ of a real valued \rv\ $X$ (defined on some probability space) is obtained by subtracting the mean and dividing by the variance, under the assumption that they exist. Thus $\widehat{X}:=(X-\expec{X})/\sqrt{\var{X}}$, taken to be $0$ by convention if the variance vanishes.

}\\
To avoid some cluttering, if $X$ depends on some parameters that we need to make explicit, say $X(\lambda)$ or $X_{\lambda}$, we write $\widehat{X}(\lambda)$ or $\widehat{X}_{\lambda}$ instead of the more consistent $\widehat{X(\lambda)}$ or $\widehat{X_{\lambda}}$.

\vspace{.3cm}

Equipped with these basic definitions, we are now in position to state our main results. 

\myitem{\textbf{Large $k$ behavior of chain functions and their linear combinations}\label{i:lbcf}}{The chain counting functions have the following properties
\begin{enumerate}[label=(\roman*)]
\item The chain function  $H^{(0)}_k$ is the constant $1$ for $k\geq 0$. \label{enumitem:hzero}
\item Let $k\geq 2$. The chain functions $G^{(1)}_k$, $G^{(2)}_k$,  $H^{(1)}_k-G^{(1)}_k-G^{(2)}_k$ are independent \rv s. The law of $G^{(1)}_k$ is that of a symmetric Bernoulli \rv .  The \rv s  $G^{(2)}_k$, $H^{(1)}_k-G^{(2)}_k-G^{(1)}_k$ are binomial with parameters $(1/2,Z_{k-1}/2)$ and $(1/2,Z_{k-1}/2-1)$ respectively. The couple $(\widehat{G}^{(2)}_k,\widehat{H}^{(1)}_k)$ of normalized \rv s converges in law at large $k$ towards a Gaussian vector $(U,V)$ of $N(0,1)$ (i.e. normalized Gaussian) \rv s with $\cov{U,V}=1/\sqrt{2}$ and the normalized version of $H^{(1)}_k-G^{(2)}_k-G^{(1)}_k$ converges in law at large $k$ towards $\sqrt{2}V-U$. \label{enumitem:spec}
\item The chain function  $G^{(1)}_k$ is independent of all other chain counting functions (with the same $k$). \label{enumitem:gun}
\setcounter{myenumi}{\value{enumi}}
\end{enumerate}
For $k\geq 2$ let $X_k$ be a linear combination with real coefficients of the normalized chain functions $\widehat{G}^{(n)}_k$ with $n\geq 3$ and $\widehat{H}^{(n)}_k$ with $n\geq 1$. We denote by $\sigma_k$ the sum of the coefficients and by $|\sigma|_k$ the sum of the absolute values of the coefficients.
\begin{enumerate}[label=(\roman*)]
\setcounter{enumi}{\value{myenumi}}
\item Suppose that $\lim_{k\to +\infty}\sigma_k=0$ and $\lim_{k\to +\infty} |\sigma|_k/Z_{k-2}=0$. Then the random process $(X_k)_{k\geq 2}$ converges to $0$ in quadratic mean at large $k$.
\label{enumitem:gen}
\item Supposing now that $\lim_{k\to +\infty}\sigma_k=\sigma$ and $\lim_{k\to +\infty} |\sigma|_k/Z_{k-2}=0$, the random couple $(\widehat{G}^{(2)}_k,X_k)_{k\geq 2}$ converges in law towards the Gaussian vector $(U,\sigma V)$. 
\label{enumitem:undgauss}
\setcounter{myenumi}{\value{enumi}}
\end{enumerate}

\vspace{-.2cm}

\noindent The previous two statements can be restated informally as ``The space of real linear combinations of $\widehat{G}^{(n)}_k$s with $n\geq 3$ and $\widehat{H}^{(n)}_k$s with $n\geq 1$ becomes effectively 1-dimensional in the large $k$ limit''.
\begin{enumerate}[label=(\roman*)]
\setcounter{enumi}{\value{myenumi}}
\item Suppose that $\sum_k(|\sigma|_k) ^2/Z_{k-2}$ and $\sum_k(\sigma_k)^2$ are both finite. Then the random process $(X)_{k\geq 2}$ converges to $0$ almost surely.\label{enumitem:asconv}
\setcounter{myenumi}{\value{enumi}}
\end{enumerate} 
\begin{enumerate}[label=(\roman*)]
\setcounter{enumi}{\value{myenumi}}
\item Suppose that the coefficients in the definition of $X_k$ are all positive and that $X_k$ has a $k$-independent standard deviation $\sigma$.\footnote{It is enough that these conditions hold from some $k$ on.} Then $\lim_{k\to +\infty}\sigma_k=\sigma$ and $\lim_{k\to +\infty} |\sigma|_k/Z_{k-2}=0$. Moreover, the rate of convergence ensures that the process $(X_k-\sigma \widehat{H}^{(1)}_k)_{k\geq 2}$ converges to $0$ in mean square and almost surely.
  
In particular the differences  $\widehat{G}^{(n)}_k-\widehat{H}^{(1)}_k$ with $n\geq 3$ or $\widehat{H}^{(n)}_k-\widehat{H}^{(1)}_k$ with $n\geq 1$ go to $0$ in mean square and almost surely at large $k$. \label{enumitem:part}
\end{enumerate}
\vspace{-.6cm}
}

\vspace{.3cm}
\autoref{enumitem:hzero} does not deserve a proof. The derivation of \Autoref{enumitem:spec,enumitem:gun} is given in \Autoref{i:hungdeux,i:gunindep}. The (lengthy) derivation of \autoref{enumitem:gen} is given in \autoref{sec:mcvcf}. It is based on the explicit formul\ae\ for expectations and (co)variances of chain counting functions summarized in \autoref{i:ecvccf} and established in the rest of \autoref{sec:mcvcf}. \autoref{enumitem:asconv} has in itself little to do with random pure sets.  It is a very classical consequence of the Markov inequality and the Borel lemma. We recall a short proof in \autoref{i:ascp0} for completeness, but more important are the special cases in \autoref{enumitem:part}. See \Autoref{i:rfmaineqcor,i:ascpcor} for the final steps of yielding \autoref{enumitem:undgauss}, \autoref{enumitem:part} and more. 

\section{Pure sets basics} \label{sec:psbasics}

As usual, we denote a set whose elements are $a,b,c$ as $\lb a,b,c \rb$. The ordering is immaterial, and repetitions do not count so $\lb c,b,b,a,b,c \rb=\lb c,b,a\rb =\lb a,b,c \rb$ and if for instance $a=b$ then $\lb a,b,c \rb=\lb a,c \rb$.

We have not been explicit about where the elements $a,b,c$ are to be found. We are used to talk of the set of pupils in a class: the class is seen as a set and the pupils as its elements, the pupils themselves are not necessarily seen as sets.

However, in the framework of standard set theory, the elements of a set are themselves sets. There is only one basic relation, membership, denoted by $\in$: if $a$ and $x$ are sets, $a\in x$ is read as ``$a$ is an element of $x$'', or ``$a$ is a member of $x$'' or ``$a$ belongs to $x$'' and means that $x=\lb \cdots,a,\cdots\rb$; $x\ni a$ is a synonym for $a\in x$. Listing informally the most intuitive axioms for sets, we acknowledge:\footnote{The list is not exhaustive -- in fact, we leave aside infinitely many axioms -- but fits with our purposes.}\\
-- The existence of a set with no elements, the empty set $\lb\rb$, also known under the notation $\emptyset$,\\
-- The fact that sets are characterized by their elements: two sets with the same elements are equal.\footnote{Here, there is a mismatch with the informal notion of the set of pupils in a class: though we are able to distinguish the pupils, naively pupils have no elements, so they are not characterized by their elements. There are versions of set theory with so-called atoms, sets that differ from $\lb\rb$ but have no elements.} As a side product, if $x$ and $y$ are sets, we may define a concept of inclusion: $y$ is a subset of $x$ if every element of $y$ belongs to $x$ as well. Thus $y \subset x$ translates as $\forall a,  a\in y \Rightarrow a \in x$, or less formally $\forall a \in y, a \in x$. In particular, $\lb\rb$ is a subset of every set, \\
-- The existence of the power set: if $x$ is a set, there is another set usually denoted by ${\mathcal P}(x)$ (also denoted  $2^x$, see below) whose elements are precisely the subsets of x. Thus $y \in {\mathcal P}(x)$ translates as $y \subset x$.\\
-- The existence of unions: if $x$ is a set, there is a set denoted $\cup x$ and called the union over $x$, whose elements are exactly the sets that are elements of an element of $x$: $z\in \cup x$ translates as $\exists y, y \in x \text{ and } z\in y$ or less formally $\exists y\in x,\, z\in y$. In particular if $y,z$ are sets and $x:=\lb y,z\rb$ then $\cup x$ is also denoted as $y\cup z$ and called the union of $y$ and $z$. Thus $a\in x\cup y$ translates as $a \in x \text{ or } a\in y$.

The fact that sets are characterized by their elements leads to the uniqueness of the empty set, of the power set of any given set, of unions, and so on. 

\myitem{A word of caution\label{i:awoc}}{
  Let us observe that there are already some traps here. For instance, to talk about the inclusion $y \subset x$, we have to know in advance that $x$ and $y$ are sets. It takes further axioms to make sure that our intuitive idea, namely that we can pick up somewhat arbitrarily elements of a set $x$ to build a subset of $x$, is valid. Henceforth, the notation  ${\mathcal P}(x)$ is preferable to $2^x$ until one has a notion of function and one can prove that a subset $y$ of $x$ is essentially the same thing as a function from $x$ to $\nquatre$ (the set with two elements $\ndeux$ and $\nun$), the elements of $y$ being those whose image is $\ndeux$. One infers that if $x$ has $n$ elements then ${\mathcal P}(x)$ has $2^n$ elements.\footnote{See \autoref{i:vni} for the consistency of notation: set theory defines the integers as $\nun=:0$, $\ndeux=\lb 0 \rb=:1$ and $\nquatre=\lb 0,1\rb=:2$. Hence $2^x$ is indeed the set of functions from $x$ to $\lb 0,1\rb$.} The same kind of considerations apply to other notions we used in passing. For instance it requires a further axiom to state that if $y,z$ are sets there is a set $x$ whose elements are precisely $y$ and $z$. This set is denoted $\lb y,z\rb$.} 

\myitem{Pure sets\label{i:ps}}{We call pure sets precisely the sets that appear as elements of the successive powers sets of the empty set (one is allowed to make successive uses of the power set rule finitely many times).\footnote{The formal framework behind this definition is called ZF$_{finite}$ (were ZF stands for Zermolo and Frenkel) without assuming the existence of an infinite set, see e.g. \cite{leroy-2024}.}}

\vspace{.3cm}

We expand a little bit on this definition and take some notations. For $\kN$, we denote by ${\mathcal S}_k$ the family of sets obtained after $k$ iterations. Explicitly, the family ${\mathcal S}_0$ has only $1$ member, the empty set $\nun$ and for $k\geq 1$ the family ${\mathcal S}_k$ is made of all sets whose elements belong to ${\mathcal S}_{k-1}$. It follows immediately from the definition that ${\mathcal S}_k$ is itself a pure set: for instance ${\mathcal S}_0=\ndeux$, ${\mathcal S}_1=\nquatre$ (that can also be written without ambiguity, but less legibly,  $\bquatre$) and for  $k\geq 1$, ${\mathcal S}_k:=2^{{\mathcal S}_{k-1}}$ is the power set of ${\mathcal S}_{k-1}$. This makes ${\mathcal S}_{k-1}$ an element of ${\mathcal S}_k$ i.e. a pure set. It is convenient and consistent to set ${\mathcal S}_{-1}:=\nun$.

\vspace{.3cm}
If $x,y$ are pure sets, then so are $\lb x \rb$ and $\lb x,y \rb$ (that is simply $\lb x \rb$ if $x=y$). That they are sets is guaranteed by the general axioms, and purity is a consequence of the existence of the power set: if $x,y \in {\mathcal S}_{k-1}$ then $\lb x \rb, \lb x,y \rb\in {\mathcal S}_k$. Consequently, if $x,y$ are pure sets so is $\lb \lb x \rb,\lb x,y\rb\rb$. This set is a common definition of the ordered pair $(x,y)$, and belongs to ${\mathcal S}_{k+1}$ if $x,y \in {\mathcal S}_{k-1}$.

Though we use only the power set construction to define pure sets, pure sets are stable by (finite) unions, intersections, complement. For example, if $x \in {\mathcal S}_k$ and $x\ni y \ni z$, then $y\in {\mathcal S}_{k-1}$ and $z\in {\mathcal S}_{k-2}$ so that $\cup x \in {\mathcal S}_{k-1}$ because, by definition of the power set, ${\mathcal S}_{k-1}$ contains every set whose elements are in ${\mathcal S}_{k-2}$.\footnote{Despite appearances however, the existence of unions is not a consequence of the existence of power sets. This is related to the ``traps''  in \autoref{i:awoc} and connects loosely to the following remark. Our naive approach is to define a set by listing unambiguously its elements, the definition of sets ``in extension''. Another approach is to define a set ``in comprehension'', by giving a common property that characterizes its elements. We can see the definitions of the power set and the union as such definitions. But the fact that they lead indeed to legitimate sets are (separate) axioms.}

\vspace{.3cm}
The following result is important. We start with a definition.
\myitem{Transitive sets\label{i:ts}}{A set $x$ is called transitive if every element of $x$ is also a subset of $x$. Formally, if $z \in y \in x$ then $z\in x$. 
}

It is plain that the intersection or union of a family of transitive subsets of a given set $x$ is again transitive. In particular, if $x$ is itself transitive then any subset $y$ of $x$ is contained in a minimal transitive set, the intersection of all transitive subsets of $x$ that contain $y$. We shall not make use of this observation, but we shall return to transitive sets in \autoref{i:cts}. As usual, we are interested mainly in the case when $x$ is a pure set. Now for the important observation. 

\myitem{For $\kN$, ${\mathcal S}_{k}$ is transitive\label{i:eskssk}}{

\myproof This property can be shown by recursion. It holds obviously for $k=0$ because ${\mathcal S}_0=\ndeux$ has only one element, $\nun$ which has no element, so from the definition of inclusion $\nun$ is a subset of any set. Suppose it holds for $k-1$ for some $k\geq 1$ and let $x\in {\mathcal S}_{k}$. The set $x$ is a subset of ${\mathcal S}_{k-1}$ so if $y\in x$ then $y\in {\mathcal S}_{k-1}$ and by the recursion hypothesis $y\subset {\mathcal S}_{k-1}$, which means that $y\in {\mathcal S}_{k}$. Thus every element of $x$ belongs to ${\mathcal S}_{k}$, that is, $x \subset {\mathcal S}_{k}$ closing the recursion step. As any set is by definition an element of its power set, we have ${\mathcal S}_{k-1}\in {\mathcal S}_k$ and we conclude that ${\mathcal S}_{k-1}\subset {\mathcal S}_k$ for $k\geq 1$.}

\myitem{The collection of pure sets}{We set ${\mathcal S}_{\infty}:=\cup_{\kN} {\mathcal S}_k$.}

\vspace{.3cm}
As this work does not pretend to the rigor of mathematical logic (far from it), we employ freely an ambiguous notation: $\cup_{\kN} {\mathcal S}_k$ uses the union symbol in a generalized sense. From inside ZF$_{finite}$ the collection ${\mathcal S}_{\infty}$ is the universe, which cannot be a set. This collection would be $\cup x$ if ``$x=\lb {\mathcal S}_0,{\mathcal S}_1,{\mathcal S}_2,\cdots \rb $'' was a valid definition of a set. This is not granted by ZF$_{finite}$, nor is the existence of any infinite set like ${\mathbb N}$. This require a new axiom, the axiom of infinity. Alternatively, we can just interpret ${\mathcal S}_{\infty}$ from the naive point of view. In any case, we view ${\mathcal S}_{\infty}$  (definitely not a finite pure set) as an infinite transitive set. Then taking the power set and unions over already available sets again and again, many more sets become available, call them generalized pure sets. The simplest ones after ${\mathcal S}_{\infty}$ itself are the elements of $2^{{\mathcal S}_{\infty}}=:{\mathcal S}_{\infty+1}$.\footnote{In the standard set theory literature, ${\mathcal S}_k$ would be denoted as $V_k$, then ${\mathcal S}_{\infty}$ as $V_{\omega}$ , ${\mathcal S}_{\infty+1}$ by $V_{\omega+1}$, then comes $V_{\omega+2}:=2^{V_{\omega+1}}$ and so on.} Note in passing that most finite generalized pure sets are not finite pure sets! For instance $\lb {\mathcal S}_{\infty} \rb$ (an element of ${\mathcal S}_{\infty+1}$) is a finite generalized pure set, in fact a singleton, but not a pure set. We take this opportunity to stress that the confusion between $\in$ and $\subset$ is one of the common pitfalls. This confusion is encouraged by the fact both can occur at the same time as in transitive sets, or as in the following simple example: if $y$ is any set, $x:=\lb y,\lb y\rb \rb$ is a set with $\lb y \rb \in x$ but also $\lb y \rb \subset x$. The reader is invited to check that if $y$ is a pure set, then so is $x$.

\myitem{Depth\label{i:d}}{The depth of a pure set $x$ is by definition the smallest $k$ for which $x\in {\mathcal S}_k$. Then by the previous remark the elements of ${\mathcal S}_k$ are precisely the pure sets of depth $\leq k$. If $x$ has depth $k$ its elements have depth at most $k-1$ by construction. In particular there can be  no cycle in the $\in$ relation for pure sets, i.e. no finite sequence $x_0,\cdots,x_n$ such that $x_0\in x_1\in \cdots \in x_n$ with $x_n=x_0$.}

\vspace{.3cm}
We denote by $Z_k$ the number of pure sets of depth $\leq k$, so that $Z_0=1$ and $Z_k=2^{Z_{k-1}}$ for $k\geq 1$. In accordance with our convention ${\mathcal S}_{-1}:=\nun$, we set $Z_{-1}=0$. Note that, though ${\mathcal S}_{k-1}\subset {\mathcal S}_k$ for $k\geq 1$, ${\mathcal S}_{k-1}$ represents only a tiny fraction of ${\mathcal S}_k$ at large $k$.

\vspace{.3cm}
Among the pure sets, some have a simple structure.

\myitem{Russian dolls, or matrioshka\label{i:rdm}}{Russian dolls are defined recursively: $\nun=:\underline{0}$ is a  matrioshka and $\lb \underline{k-1}\rb:=\underline{k}$ is a matrioshka for $k\geq 1$. Stated differently,the Russian doll $\underline{k}$ is a series of $k+1$ opening braces $\lb\right.$ followed by its mirror image made of $k+1$ closing braces $\left.\rb$. Thus $\underline{0}=\nun$, $\underline{1}=\ndeux$, $\underline{2}=\ntrois$, $\underline{3}=\ncinq$, and so on. The depth of $\underline{k}$ is $k$.}

\vspace{.3cm}
As our notation suggests, in early versions of set theory matryoshka were used as set-theoretic versions of the integers. They were later supplanted by the Von Neumann integers.

\myitem{Von Neumann integers\label{i:vni}}{The Von Neumann integers are defined recursively: $\nun=:\overline{0}$ is a Von Neumann integer and $\overline{k-1}\cup \lb \overline{k-1}\rb:=\overline{k}$ is a Von Neumann integer for $k\geq 1$. Thus $\overline{0}=\nun$, $\overline{1}=\ndeux$, $\overline{2}=\bquatre$, $\overline{3}=\btreize$, and so on. In particular, $\overline{k}$ is a set with $k$ elements. The depth of $\overline{k}$ is $k$. By construction, Von Neumann integers are transitive sets (see \autoref{i:ts}). They are the most famous ones, because they are totally ordered by $\in$: is $x,y\in \overline{k}$ then either $x\in y$ or $x=y$ or $y\in x$, They are even well-ordered: every subset of $\overline{k}$ has a smallest element, a property which is obvious for finite totally ordered sets, but with deep implications in general set theory (i.e. with infinite sets).} 

\vspace{.3cm}
However, generic pure sets lack such a simple description. We view a pure set ``in extention'' as a sequence of mixed $\lb \right.$s and $\left.\rb$s consistent with the rules. It is a natural question whether it can be described in a more efficient way, in particular ``in comprehension'' by characterizing its elements. This relates to the theory of complexity and compression. The next section starts with a canonical bijection between pure sets and binary strings, a natural ecosystem for the  concepts of compression. Thus there is nothing specific to learn on pure sets purely from that point of view: uniformly random binary strings of given length cannot be compressed. That's the reason why we try to investigate the intricacies of pure sets by focusing on observables for which the relation $\in$ plays a crucial role. That leads us to the concept of chain, see \autoref{i:cps}. Chains are among the salient features of pure sets, and play the central role in what follows. Before closing this section, we quote a very simple identity that plays a crucial role in what follows.

\myitem{A useful identity\label{i:aui}}{From the definition of the power set of any set $x$ as the set of functions from $x$ to $\lb 0,1 \rb$, we can build an arbitrary element of ${\mathcal S}_k$ by choosing freely for each $y\in {\mathcal S}_{k-1}$ whether $y\in x$ or $y\notin x$. Here is a way to rephrase this fact. Suppose that $(t_y)_{y\in {\mathcal S}_{k-1}}$ is a family of (formal, commuting) variables. Then
\begin{equation}\label{eq:basident}
\sum_{x\in {\mathcal S}_k} \prod_{y\in x} t_y = \prod_{y\in {\mathcal S}_{k-1}} (1+t_y), \end{equation}
an identity which proves useful in both directions.}

\section{Representation of pure sets by familiar objects}\label{sec:rpsc}

In this section we list three ways to represent pure sets: binary strings, Dick paths and trees. 

\myitem{Binary strings\label{i:bs}}{We start with a simple bijection $\varphi_k$ between $\intt{0,Z_k-1}$ and ${\mathcal S}_k$. For $k=0$ there is no choice: $\intt{0,Z_k-1}=\lb 0 \rb$  and ${\mathcal S}_0=\ndeux$ are singletons, so $\varphi_0(0)=\nun$. Then the  definition is recursive. For $k\geq 1$ an elements $n$ of $\intt{0,Z_k-1}$ can be seen in base $2$ as a binary string of length $Z_{k-1}$, i.e. as an arbitrary sequence of zeros and ones: $n=\sum_{m=0}^{Z_{k-1}-1} 2^m\varepsilon_m(n)$ with $\varepsilon_m\in\{0,1\}$ for $m\in \intt{0,Z_{k-1}-1}$. Then $\varphi_k(n)$ is the set whose elements are the sets $\varphi_{k-1}(m)$ for those $m$s such that $\varepsilon_m(n)=1$. If each such $m$ is again converted in binary notation, and the procedure is repeated, one arrives at the expression of $n$ in the so-called hereditary binary representation. The reverse construction is called the Ackermann coding.

The family of bijections $\varphi_k$ has a nice property: For $k\geq 1$ the map $\varphi_{k-1}$ is the restriction of $\varphi_k$ to $\intt{0,Z_{k-1}-1}$.

\myproof The key point is that the ones in the binary expansion of an integer are at the same positions for every ${\mathcal S}_k$ to which it belongs (the difference is only in a string of zeroes for large $m$s). Again, the property is proven by recursion. It holds for $k=1$ because $\varphi_1(0)=\nun=\varphi_0(0)$. Suppose it holds for some $k\geq 1$ and take $n=\sum_{m=0}^{Z_{k-1}-1} 2^m\varepsilon_m(n) \in \intt{0,Z_k-1}$. By definition, $\varphi_{k+1}(n)$ is the set whose elements are the sets $\varphi_{k}(m)$ for those $m$s such that $\varepsilon_m(n)=1$. But if $n \in \intt{0,Z_k-1}$, those $m$s are in   $\intt{0,Z_{k-1}-1}$ and by the recursion hypothesis, they satisfy $\varphi_{k}(m)=\varphi_{k-1}(m)$, implying that $\varphi_{k+1}(n)=\varphi_k(n)$.

\vspace{.3cm}
Thus we may consider a single function $\varphi\colon {\mathbb N}\to {\mathcal S}_{\infty}$, whose restriction to $\intt{0,Z_k-1}$ is $\varphi_k$. Consequently $\varphi$ enumerates pure sets one at a time, starting with $\varphi(0)=\nun$. In this way, pure sets are ordered in a sequence. The enumeration has the natural feature that ${\mathcal S}_{0}$ is enumerated first, then ${\mathcal S}_{1}\setmin {\mathcal S}_{0}$, then ${\mathcal S}_{2}\setmin {\mathcal S}_{1}$ and do on. However, the precise order in which members of  ${\mathcal S}_{k}\setmin {\mathcal S}_{k-1}$ come does not appear to rely on obvious features. 

 As an illustrative example, we compute $\varphi(14)$. The integer $14$ rewrites $14=2^3+2^2+2^1.$ Thus  $\varphi(14)$ is a set with three elements: $\varphi(3)$, $\varphi(2)$ and $\varphi(1)$. Writing $3=2^1+2^0$, $2=2^1$ and $1=2^0$ we get that $\varphi(3)$ has two elements, $\varphi(1)$ and $\varphi(0)$, $\varphi(2)$ has one element, $\varphi(1)$, $\varphi(1)$ has one element, $\varphi(0)$ and $\varphi(0)=\lb\rb$ has no element. We infer successively that $\varphi(1)=\lb\varphi(0)\rb=\lb\lb\rb\rb$, then $\varphi(2)=\lb\varphi(1)\rb=\lb\lb\lb\rb\rb\rb$, then $\varphi(3)=\lb\varphi(1),\varphi(0)\rb=\lb\lb\lb\rb\rb,\lb\rb\rb$. Finally $\varphi(14)=\lb\varphi(3),\varphi(2),\varphi(1)\rb=\lb\lb\lb\lb\rb\rb,\lb\rb\rb,\lb\lb\lb\rb\rb\rb,\lb\lb\rb\rb\rb$. We could dispense with the commas altogether and write  $\varphi(14)=\lb\lb\lb\lb\rb\rb\lb\rb\rb\lb\lb\lb\rb\rb\rb\lb\lb\rb\rb\rb$, that can be read on the fly on the binary hereditary representation $14=2^{2^{2^{0}}+2^{0}}+2^{2^{2^{0}}}+2^{2^{0}}$.

This illustrates a byproduct of the construction: any pure set (except $\nun$) has infinitely many representatives ``in extension'' because its elements, the elements of its elements, and so on, can be repeated (leading to redundancy) and reordered; however the map $\varphi$ yields a well-defined representative of each pure set by a succession of braces. We call this the standard braces representation in the sequel. All other representations without redundancy of a set have the same number of opening and closing braces : a simple recursion argument on the depth shows that they are obtained from the standard braces representation by reordering its elements, the elements of its elements, and so on.  

We give three families of examples of the inverse bijection $\varphi^{-1}$. Starting with $\nun$, the codes for the  matrioshka are $0,1,2,4,\cdots$, and in general the matryoshka $\underline{n}$ has code $Z_{n-1}$ for $n\geq 0$. Starting again with $\nun$, the Von Neumann integers have code $0,1,3,11,\cdots$ and in general the Von Neumann integer $\overline{n}$ has code $c_n$ with $c_0=0$ and $c_{n+1}=2^{c_n}+c_n$, building sequence A$034797$ in \cite{oeis2025}). Finally, starting with $\mathcal{S}_{-1}=\nun$ the codes for the $\mathcal{S}$'s (they that are pure sets themselves) are $0,1,3,15$ and in general $\mathcal{S}_{n}$ has code $Z_{n+1}-1$: this is a string of $1$s of length $Z_n$ so it indeed corresponds to a set with $Z_n$ elements that enumerate $\mathcal{S}_{n}$. As might have been expected, the binary codes for those three simple classes of sets are themselves simple.

The correspondence of pure sets with binary strings allows us to use standard complexity theory but the outcome is in some sense trivial: most binary strings cannot be compressed significantly, so they give generically a minimal representation of pure sets. On the other hand, the standard braces representation turns out to be logarithmically redundant. We shall illustrate this phenomenon in \autoref{i:rsbr} as an early application of the probabilistic model: the computation of the average number of (pairs of) braces of a pure set in ${\mathcal S}_k$. Later in \autoref{i:rsbrv} we shall also estimate the variance, showing that the redundancy of the braces representation is essentially uniform on ${\mathcal S}_k$ at large $k$.

The existence of a global description via $\varphi$ has nice consequences. For instance, suppose that we are given an infinite sequence of bits, $\varepsilon_m\in\{0,1\},\, m=0,1,\cdots$. Setting $x_l:=\varphi\left(\sum_{m<l}2^m\varepsilon_m\right)$ for $\lN$ we obtain an increasing sequence of pure sets starting from $x_0=\nun$ with one element added to $x_{l-1}$ each time $\varepsilon_l=1$. The formal $l\to \infty$ limit leads to a countable set $x_{\infty}$ whose elements are pure sets. However $x_{\infty}$, obviously a subset of ${\mathcal S}_{\infty}$, is a pure set only if all but a finite number of $\varepsilon$s equal $0$. In general $x_{\infty}$ does not belong to the universe of pure sets. 

For $l=Z_{k-1}$, $x_l\in {\mathcal S}_k$. Moreover, anticipating a bit on \autoref{sec:probmod}, if $(\varepsilon_m)_{\mN}$ is chosen with the usual symmetric probability measure on infinite sequences of independant Bernoulli trials\footnote{That is, the $\varepsilon$s are independent and each assumes values $0,1$ with probability $1/2$.} then all elements $x_l\in {\mathcal S}_k$ have the same probability because all finite sequences $(\varepsilon_m)_{m\in \intt{0,l-1}}$ are equally likely.}

\myitem{Dick paths}{We have seen that general pure sets can be represented (in infinitely many ways in general) by a succession of braces. Not every succession of braces represents a set though. For this to be the case, the number of $\lb\right.$s has to match that of $\left.\rb\!$ s. Moreover, reading from left to right, the succession has to start with an opening brace  $\lb\right.$ and from then on, counting braces from left to right, the number of opening braces $\lb\right.$ has to exceed that of closing braces $\left.\rb$ by at least one until the last brace that has to be a closing brace $\left.\rb$.

This leads to a correspondence with a class of paths. Counting from left to right and starting at $0$, we add $1$ for each opening brace $\lb\right.$ and subtract $1$ for each closing brace $\left.\rb$. Then by the rule recalled above the count is always $\geq 1$ until the last step where it returns to $0$.\footnote{We use freely in the sequel that for a pure set of depth $k$ the maximum of the counting function is exactly $k+1$. This is easily seen by recursion.} 

We interpret the evolution of the counting as braces go by as a path. For instance $\nun$ corresponds to \begin{tikzpicture} [scale=.2] \draw [very thick] (0,0)--(1,1)--(2,0); \end{tikzpicture}, and $\lb\lb\rb\lb\lb\rb\rb\rb$ to \begin{tikzpicture} [scale=.2] \draw [very thick] (0,0)--(1,1)--(2,2)--(3,1)--(4,2)--(5,3)--(6,2)--(7,1)--(8,0); \end{tikzpicture}. The number of steps is the number of braces. Removing the first and the last brace, and correspondingly the first and the last step, one obtains the standard notion of a Dyck path, or discrete excursion (see e.g. \cite{flajoletsedgewick2009}). Keeping everything leads to paths that do not come back to level $0$ until the very last step, which we might call strict Dyck paths, or strict discrete excursions. 

To every strict Dyck paths we may associate a pure set by the reverse procedure: a $\lb\right.$ for each step up, and a $\left.\rb$ for each step down. A pair $\lb\rb$ corresponds to a local maximum of the path, and a pair $\left.\rb \lb\right.$ to a local minimum, so for each pure set the number of $\lb\right.\left.\rb$ pairs is one more than the number of $\left.\rb \lb\right.$ pairs, or in a representation using $\eun$s and commas, there is always one  $\eun$ symbol more than there are commas.

However, the vast standard knowledge of (strict) Dyck paths properties does not apply immediately for two reasons:\\
-- We do not constrain the paths to have a given length (or the sets to involve a given number of braces), we fix the maximal height (that is one plus the depth of the associated pure sets). There are infinitely many such paths except of height $1$. \\
-- The infinity is due to multiple counting because the resulting description of sets does not avoid ordering and repetition problems. Once this degeneracy is taken into account, only finitely many distinct pure sets remain for each depth.}

\myitem{Identity trees\label{i:it}}{It is well-known that Dyck paths also code for a class (in fact several classes) of trees (see e.g. \cite{flajoletsedgewick2009}), so pure sets can also be represented as trees. In fact, it is plain to go directly from pure sets to trees.

To any finite set $S=\{s_1,\cdots,s_n\}$ we can associate a rather simplistic tree, with $S$ as root connected to leaves $s_1,\cdots,s_n$. If $S$ is a pure set, then  $s_1,\cdots,s_n$ are pure sets as well and one can iterate, grafting a new (possibly trivial tree) at each leaf. Note the similarity with the recursive procedure to express an integer in hereditary binary representation: indeed the binary representation of a pure set makes the correspondence with the associated tree transparent.

When the iteration steps are over, one ends with a tree whose nodes correspond to pairs of matching braces $\lb \cdots \rb$ --matching means that $\cdots$ stands for nothing or a pure set-- with inner nodes corresponding to a nontrivial $\cdots$ while leaves correspond to adjacent matching braces $\lb\rb$ i.e. to $\emptyset$ in the associated pure set.

If the standard braces representation of pure sets is used, the family of rooted trees one obtains are the so-called identity trees. These are trees without any symmetries, corresponding to the fact that repetitions and reorderings do not lead to different sets.

Notice that a Dick path is recovered as usual by ``traveling around the tree'', counting $+1$ when the step gets farther from the root and $-1$ when it gets closer.

The following remark is obvious but important. In a tree, paths starting at the root are in one to one correspondence with nodes, so that chains in a pure set are in one to one correspondence with nodes in the associated tree and maximal chains with leaves. More precisely a chain of length $n$ in the pure set corresponds to a node (hence to a pair of, in general nonadjacent, matching braces) at distance $n$ from the root. On infers, using the standard braces representation of pure sets, that chains are in natural bijection with pairs of matching braces. In particular, the total number of braces is twice the total number of chains, a fact we shall use repeatedly in the sequel.}

\vspace{.3cm}

The next section gives a list of the pure sets of depth $\leq 3$, a total of $16=1+1+2+12$ sets. Indeed ${\mathcal S}_{k-1}\subset {\mathcal S}_k$ for $k\geq 1$ and ${\mathcal S}_k\setmin {\mathcal S}_{k-1}$ is made of those sets that appear only at depth $k$. There is $1$ such set for $k=0$, $1$ for $k=1$, $2$ for $k=2$, $12$ for $k=3$ and so on.

\section{The four first depths... and beyond}\label{sec:ffdb}

For depth $0,\cdots,3$ we obtain the list
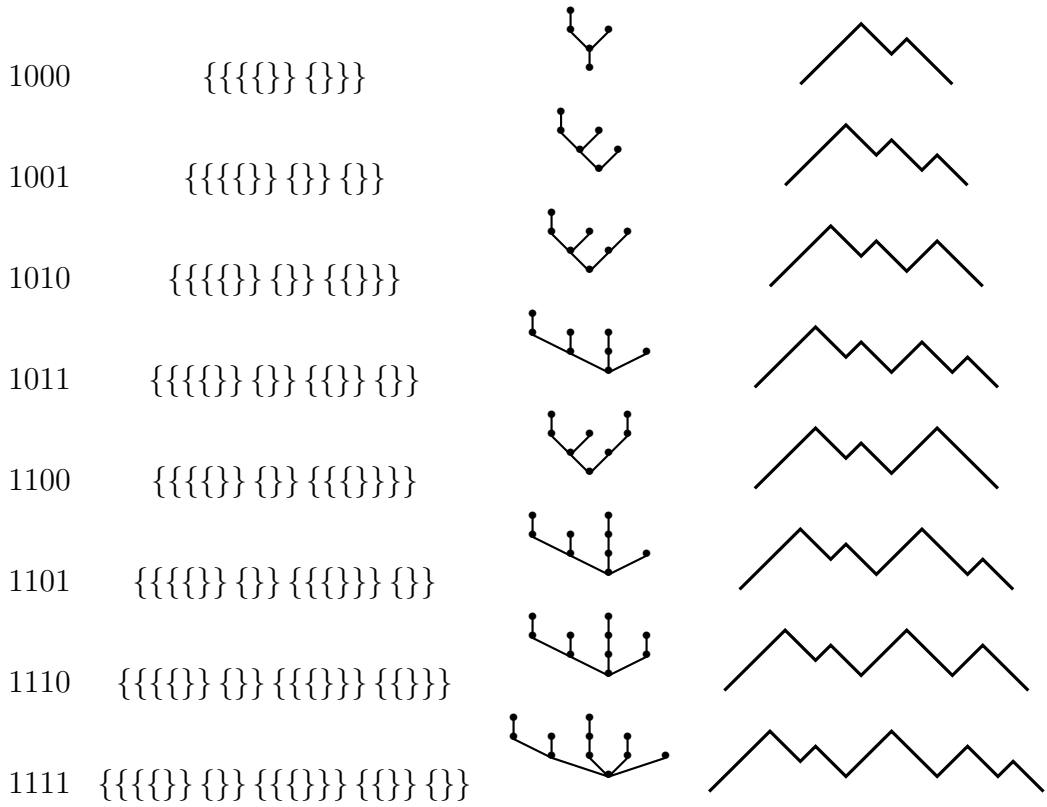
\begin{longtable}{RCCC}
  \text{Strings} & \text{Braces} & \text{Trees} & \text{Strict excursions} \\
   & \nun & \begin{tikzpicture}[scale=.25] \draw (0,-1) node {$\mbullet$}
      ; \end{tikzpicture}  & \begin{tikzpicture} [scale=.2] \draw [very thick] (0,0)--(1,1)--(2,0); \end{tikzpicture} \\
  1 & \ndeux & \begin{tikzpicture}[scale=.25] \draw [thick] (0,-1)--(0,0);  \draw (0,-1) node {$\mbullet$}; \draw (0,0) node {$\mbullet$}; \end{tikzpicture} & \begin{tikzpicture} [scale=.2] \draw [very thick] (0,0)--(1,1)--(2,2)--(3,1)--(4,0); \end{tikzpicture} \\
  10 & \ntrois & \begin{tikzpicture}[scale=.25] \draw [thick] (0,-1)--(0,0)--(0,1);  \draw (0,-1) node {$\mbullet$}; \draw (0,0) node {$\mbullet$}; \draw (0,1) node {$\mbullet$}; \end{tikzpicture}  & \begin{tikzpicture} [scale=.2] \draw [very thick] (0,0)--(1,1)--(2,2)--(3,3)--(4,2)--(5,1)--(6,0); \end{tikzpicture}\\
  11 & \bquatre & \begin{tikzpicture}[scale=.25] \draw [thick] (-1,1)--(-1,0)--(0,-1)--(1,0); \draw (-1,1) node {$\mbullet$}; \draw (-1,0) node {$\mbullet$}; \draw (0,-1) node {$\mbullet$}; \draw (1,0) node {$\mbullet$};  \end{tikzpicture} & \begin{tikzpicture} [scale=.2] \draw [very thick] (0,0)--(1,1)--(2,2)--(3,3)--(4,2)--(5,1)--(6,2)--(7,1)--(8,0); \end{tikzpicture} \\
  100 & \ncinq & \begin{tikzpicture}[scale=.25] \draw [thick] (0,-1)--(0,0)--(0,1)--(0,2);  \draw (0,-1) node {$\mbullet$}; \draw (0,0) node {$\mbullet$}; \draw (0,1) node {$\mbullet$}; \draw (0,2) node {$\mbullet$}; \end{tikzpicture} & \begin{tikzpicture} [scale=.2] \draw [very thick] (0,0)--(1,1)--(2,2)--(3,3)--(4,4)--(5,3)--(6,2)--(7,1)--(8,0); \end{tikzpicture} \\
  101 & \bsept & \begin{tikzpicture}[scale=.25] \draw [thick] (-1,2)--(-1,1)--(-1,0)--(0,-1)--(1,0); \draw (-1,2) node {$\mbullet$}; \draw (-1,1) node {$\mbullet$};  \draw (-1,0) node {$\mbullet$}; \draw (0,-1) node {$\mbullet$}; \draw (1,0) node {$\mbullet$}; \end{tikzpicture} &  \begin{tikzpicture} [scale=.2] \draw [very thick] (0,0)--(1,1)--(2,2)--(3,3)--(4,4)--(5,3)--(6,2)--(7,1)--(8,2)--(9,1)--(10,0); \end{tikzpicture} \\
  110 & \bneuf & \begin{tikzpicture}[scale=.25] \draw [thick] (-1,2)--(-1,1)--(-1,0)--(0,-1)--(1,0)--(1,1); \draw (-1,2) node {$\mbullet$}; \draw (-1,1) node {$\mbullet$}; \draw (-1,0) node {$\mbullet$}; \draw (0,-1) node {$\mbullet$}; \draw (1,0) node {$\mbullet$}; \draw (1,1) node {$\mbullet$}; \end{tikzpicture} & \begin{tikzpicture} [scale=.2] \draw [very thick] (0,0)--(1,1)--(2,2)--(3,3)--(4,4)--(5,3)--(6,2)--(7,1)--(8,2)--(9,3)--(10,2)--(11,1)--(12,0); \end{tikzpicture} \\
  111 & \bdouze & \begin{tikzpicture}[scale=.25] \draw [thick] (-2,2)--(-2,1)--(-2,0)--(0,-1) (0,1)--(0,0)--(0,-1)--(2,0); \draw (-2,2) node {$\mbullet$}; \draw (-2,1) node {$\mbullet$}; \draw (-2,0) node {$\mbullet$}; \draw (0,1) node {$\mbullet$}; \draw (0,0) node {$\mbullet$}; \draw (0,-1) node {$\mbullet$}; \draw (2,0) node {$\mbullet$}; \end{tikzpicture} & \begin{tikzpicture} [scale=.2] \draw [very thick] (0,0)--(1,1)--(2,2)--(3,3)--(4,4)--(5,3)--(6,2)--(7,1)--(8,2)--(9,3)--(10,2)--(11,1)--(12,2)--(13,1)--(14,0); \end{tikzpicture} \\
  1000 & \bsix & \begin{tikzpicture}[scale=.25] \draw [thick] (0,-2)--(0,-1) (-1,1)--(-1,0)--(0,-1)--(1,0); \draw (-1,1) node {$\mbullet$}; \draw (0,-2) node {$\mbullet$}; \draw (-1,0) node {$\mbullet$}; \draw (0,-1) node {$\mbullet$}; \draw (1,0) node {$\mbullet$}; \end{tikzpicture} & \begin{tikzpicture} [scale=.2] \draw [very thick] (0,0)--(1,1)--(2,2)--(3,3)--(4,4)--(5,3)--(6,2)--(7,3)--(8,2)--(9,1)--(10,0); \end{tikzpicture} \\
  1001 & \bhuit & \begin{tikzpicture}[scale=.25] \draw [thick] (-3,-1)--(-2,-2)--(-1,-1) (-4,1)--(-4,0)--(-3,-1)--(-2,0);  \draw (-2,0) node {$\mbullet$}; \draw (-3,-1) node {$\mbullet$}; \draw (-2,-2) node {$\mbullet$}; \draw (-1,-1) node {$\mbullet$}; \draw (-4,0) node {$\mbullet$}; \draw (-4,1) node {$\mbullet$}; \end{tikzpicture} & \begin{tikzpicture} [scale=.2] \draw [very thick] (0,0)--(1,1)--(2,2)--(3,3)--(4,4)--(5,3)--(6,2)--(7,3)--(8,2)--(9,1)--(10,2)--(11,1)--(12,0); \end{tikzpicture} \\
  1010 & \bdix & \begin{tikzpicture}[scale=.25] \draw [thick] (-2,2)--(-2,1)--(-1,0)--(0,-1)--(1,0)--(2,1) (0,1)--(-1,0); \draw (-2,2) node {$\mbullet$}; \draw (-2,1) node {$\mbullet$}; \draw (-1,-0) node {$\mbullet$}; \draw (0,-1) node {$\mbullet$}; \draw (1,0) node {$\mbullet$}; \draw (2,1) node {$\mbullet$}; \draw (0,1) node {$\mbullet$}; \end{tikzpicture} & \begin{tikzpicture} [scale=.2] \draw [very thick] (0,0)--(1,1)--(2,2)--(3,3)--(4,4)--(5,3)--(6,2)--(7,3)--(8,2)--(9,1)--(10,2)--(11,3)--(12,2)--(13,1)--(14,0); \end{tikzpicture} \\
  1011 & \btreize & \begin{tikzpicture} [scale=.25] \draw [thick] (-4,2)--(-4,1)--(-2,0)--(0,-1)--(2,0) (0,1)--(0,0)--(0,-1) (-2,1)--(-2,0); \draw (-4,2) node {$\mbullet$}; \draw (-4,1) node {$\mbullet$}; \draw (-2,0) node {$\mbullet$}; \draw (0,-1) node {$\mbullet$}; \draw (2,0) node {$\mbullet$}; \draw (0,1) node {$\mbullet$}; \draw (0,0) node {$\mbullet$}; \draw (-2,1) node {$\mbullet$}; \end{tikzpicture} & \begin{tikzpicture} [scale=.2] \draw [very thick] (0,0)--(1,1)--(2,2)--(3,3)--(4,4)--(5,3)--(6,2)--(7,3)--(8,2)--(9,1)--(10,2)--(11,3)--(12,2)--(13,1)--(14,2)--(15,1)--(16,0); \end{tikzpicture} \\
  1100 & \bonze & \begin{tikzpicture}[scale=.25] \draw [thick] (-2,2)--(-2,1)--(-1,0)--(0,-1)--(1,0)--(2,1)--(2,2) (0,1)--(-1,0); \draw (-2,2) node {$\mbullet$}; \draw (-2,1) node {$\mbullet$}; \draw (-1,-0) node {$\mbullet$}; \draw (0,-1) node {$\mbullet$}; \draw (1,0) node {$\mbullet$}; \draw (2,1) node {$\mbullet$}; \draw (2,2) node {$\mbullet$}; \draw (0,1) node {$\mbullet$}; \end{tikzpicture} & \begin{tikzpicture} [scale=.2] \draw [very thick] (0,0)--(1,1)--(2,2)--(3,3)--(4,4)--(5,3)--(6,2)--(7,3)--(8,2)--(9,1)--(10,2)--(11,3)--(12,4)--(13,3)--(14,2)--(15,1)--(16,0); \end{tikzpicture} \\
  1101 & \bquatorze & \begin{tikzpicture} [scale=.25] \draw [thick] (-4,2)--(-4,1)--(-2,0)--(0,-1)--(2,0) (0,2)--(0,1)--(0,0)--(0,-1) (-2,1)--(-2,0); \draw (-4,2) node {$\mbullet$}; \draw (-4,1) node {$\mbullet$}; \draw (-2,0) node {$\mbullet$}; \draw (0,-1) node {$\mbullet$}; \draw (2,0) node {$\mbullet$}; \draw (0,2) node {$\mbullet$}; \draw (0,1) node {$\mbullet$}; \draw (0,0) node {$\mbullet$}; \draw (-2,1) node {$\mbullet$}; \end{tikzpicture} & \begin{tikzpicture} [scale=.2] \draw [very thick] (0,0)--(1,1)--(2,2)--(3,3)--(4,4)--(5,3)--(6,2)--(7,3)--(8,2)--(9,1)--(10,2)--(11,3)--(12,4)--(13,3)--(14,2)--(15,1)--(16,2)--(17,1)--(18,0); \end{tikzpicture} \\
  1110 & \bquinze & \begin{tikzpicture} [scale=.25] \draw [thick] (-4,2)--(-4,1)--(-2,0)--(0,-1)--(2,0)--(2,1) (0,2)--(0,1)--(0,0)--(0,-1) (-2,1)--(-2,0); \draw (-4,2) node {$\mbullet$}; \draw (-4,1) node {$\mbullet$}; \draw (-2,0) node {$\mbullet$}; \draw (0,-1) node {$\mbullet$}; \draw (2,0) node {$\mbullet$}; \draw (2,1) node {$\mbullet$}; \draw (0,2) node {$\mbullet$}; \draw (0,1) node {$\mbullet$}; \draw (0,0) node {$\mbullet$}; \draw (-2,1) node {$\mbullet$}; \end{tikzpicture} & \begin{tikzpicture} [scale=.2] \draw [very thick] (0,0)--(1,1)--(2,2)--(3,3)--(4,4)--(5,3)--(6,2)--(7,3)--(8,2)--(9,1)--(10,2)--(11,3)--(12,4)--(13,3)--(14,2)--(15,1)--(16,2)--(17,3)--(18,2)--(19,1)--(20,0); \end{tikzpicture} \\
  1111 & \bseize & \begin{tikzpicture} [scale=.25] \draw [thick] (-5,2)--(-5,1)--(-3,0)--(0,-1)--(3,0) (-1,2)--(-1,1)--(-1,0)--(0,-1)--(1,0)--(1,1) (-3,1)--(-3,0); \draw (-5,2) node {$\mbullet$}; \draw (-5,1) node {$\mbullet$}; \draw (-3,0) node {$\mbullet$}; \draw (0,-1) node {$\mbullet$}; \draw (3,0) node {$\mbullet$}; \draw (-1,2) node {$\mbullet$}; \draw (-1,1) node {$\mbullet$}; \draw (-1,0) node {$\mbullet$}; \draw (1,0) node {$\mbullet$}; \draw (1,1) node {$\mbullet$}; \draw (-3,1) node {$\mbullet$};\end{tikzpicture} & \begin{tikzpicture} [scale=.2] \draw [very thick] (0,0)--(1,1)--(2,2)--(3,3)--(4,4)--(5,3)--(6,2)--(7,3)--(8,2)--(9,1)--(10,2)--(11,3)--(12,4)--(13,3)--(14,2)--(15,1)--(16,2)--(17,3)--(18,2)--(19,1)--(20,2)--(21,1)--(22,0); \end{tikzpicture} \\
\end{longtable}

We may replace each pair of successive matching braces $\lb\rb$ by a $\eun$ and/or add a comma between each pair $\left.\rb \lb\right.$ , i.e. replace it with $\left.\rb , \lb\right.$. These two conventions help readability for the human eye. Here is the representation with commas, with commas and $\eun$, followed by $3$ statistics: the depth, the number of chains and of m(aximal) chains. 

  \begin{longtable}{RLLCCC}
  \text{String} & \text{Commas} & \text{Commas and $\eun$} & \text{Depth} &
  \text{Chains} & \text{M-chains} \\
   & \nun & \eun =\underline{0}=\overline{0} & 0 & 1 & 0 \\
  1 & \ndeux& \edeux =\underline{1}=\overline{1} & 1 & 2 & 1 \\
  10 & \ntrois & \etrois =\underline{2} & 2 & 3 & 1 \\
  11 & \nquatre & \equatre =\overline{2} & 2 & 4 & 2 \\
  100 & \ncinq & \ecinq =\underline{3} & 3 & 4 & 1 \\
  101 & \nsept & \esept & 3 & 5 & 2 \\
  110 & \nneuf & \eneuf & 3 & 6 & 2 \\
  111 & \ndouze & \edouze & 3 & 7 & 3 \\
  1000 & \nsix & \esix & 3 & 5 & 2 \\
  1001 & \nhuit & \ehuit & 3 & 6 & 3 \\
  1010 & \ndix & \edix & 3 & 7 & 3 \\
  1011 & \ntreize & \etreize =\overline{3} & 3 & 8 & 4 \\
  1100 & \nonze & \eonze & 3 & 8 & 3 \\
  1101 & \nquatorze & \equatorze & 3 & 9 & 4 \\
  1110 & \nquinze & \equinze & 3 & 10 & 4 \\
  1111 & \nseize & \eseize & 3 & 11 & 5 \\
\end{longtable}

Why stop the explicit list at depth $3$? The following remarks are just a bit more than numerology.

Remember that $Z_k$ is the number of pure sets of depth $\leq k$. So $Z_0=1$, $Z_1=2$, $Z_2=4$, $Z_3=16$, $Z_4=65536$. To enumerate ${\mathcal S}_4$ takes a fraction of second on a laptop, using the binary representation. For obvious reasons, we do not write $Z_5=2^{65536}$ in full, though any formal computation tool\footnote{The one used for these notes was Maxima\cite{maxima}.} gives $Z_5$ in base $10$ in a fraction of a second, a number of order $2\; 10^{19728}$. Many counting questions related to pure sets lead to integer sequences that grow as a function of the depth $k$ in such a way as to prevent to submit to the OEIS \cite{oeis2025} more than the bare minimum of requested terms, and the sequence $(Z_k)_{\kN}$ is the simplest illustration. We give another one in \autoref{i:cts}. 

Anyway, $Z_5$ dwarfs the present estimates for the number of nucleons in the universe, that in the high hypotheses has $82$ digits. The number of Planck size cells that fit in the observable Universe has about $152$ digits and is no match for $Z_5$ either. It takes a sphere of radius about $10^{6525}$ times that of the observable universe to fit each set of ${\mathcal S}_5$ in a separate Plank cell. Going to spacetime Planck cells does not help much, as the age of the universe has ``only'' about $61$ digits in  Planck times units. A computer able to manipulate in detail ${\mathcal S}_5$ would be able to store (and simulate if we input valid physical laws, that is another matter) the history of the Universe we observe with a wealth of details that we never dreamed of. Needless to say, most members of ${\mathcal S}_5$ will never appear individually in a any useful mathematical. However, the correspondence with binary strings of length $65536$ makes it easy to simulate many samples in ${\mathcal S}_5$  and compute their statistical properties.

\vspace{.3cm}
The number $Z_5=2^{2^{16}}$ is itself dwarfed by several famous integers.

An example from antiquity is Archimedes cattle problem, posed (but not fully solved) by Archimedes. It asks for the size of a cattle by imposing that is satisfies a number of arithmetic conditions. With some effort, the problem can be reduced to a Fermat-Pell equation. Finding solutions remains a challenge, see e.g. \cite{vardi-1998}. The smallest compliant cattle size is about $8\; 10^{206544}$, so lies between $2^{2^{19}}$ and $2^{2^{20}}$. The next one is about $4\; 10^{413090}$ so lies between $2^{2^{20}}$ and $2^{2^{21}}$.

Larger though is the number of books in Jorge Luis Borges' Babel library. This library (see e.g. \cite{borges2017en}) is supposed to contain every possible book of $410$ pages written in an alphabet of $25$ letter were every page has $40$ lines each consisting of \textit{about} $80$ character. The total number of volumes lies between $2^{2^{22}}$ and $2^{2^{23}}$. Without the \textit{about} one is led to a number with $1834097$ digits, a number that written in full glory  would occupy a book and a half or so in the library ($Z_5$ would occupy a bit more that $6$ pages)...but digits are not in the allowed alphabet. Taking the \textit{about} into account would add or subtract a few hundreds (to the number of digits!).

The number of games of go is much larger. It is known to lie somewhere between $2^{2^{163}}$ and $2^{2^{569}}$, see \cite{tromp-farneback-2007}.

All these turn pale compared to $Z_6=2^{2^{65536}}$, a number that goes beyond the limits of (my) imagination: the previous numerology implies in particular that producing a single sample in ${\mathcal S}_6$ is far more challenging than describing in detail each and every spacetime Planck cell of our Universe.

\vspace{.3cm}
Nevertheless, sequences growing much faster than $(Z_k)_{\kN}$ are relevant for some concrete mathematics, for instance Ackermann numbers. The definition is rather simple: $A_{0,n}:=n+1,\ A_{m+1,0}:=A_{m,1},\text{ and } A_{m+1,n+1}:=A_{m,A_{m+1,n}}$. Then $A_{4,3}=Z_6-3$ for instance. The Ackermann numbers are an example of a computable sequence (one can write a program which in principle computes them) requiring inescapably \texttt{while} loops in one guise or another. This is related to the composition of functions in the third item of the definition. See e.g. \cite{conway-guy1996} for a reader-friendly account, or \cite{katz-reimann-2018} for a deeper discussion. While loops can be viewed as the door for the entry of the devil in the theory of computable functions. The $n^{\text{th}}$ Ackermann number is $A(n,n)$.  The fourth Ackermann number is $Z_7-3$ which may not look so large, but the fifth already requires to take special notations to write it down.\footnote{Moreover, even computing $A[2,1]=5$ already requires $35$ calls to the recursive definition. The number of calls for the naive computation of $A[4,3]$ exceeds by a huge amount the size of the stack of present computers.} 

\vspace{.3cm}
To close this excursion into vague numerology, let us note that after six days of hard work to build ${\mathcal S}_0$ to ${\mathcal S}_5$ a day of rest would be well-deserved. Such a day might be suitable to play games.  The chain structure allows to use pure sets for that purpose: $x$ being a pure set, player one chooses (if possible) an $x'\in x$, then player two chooses (if possible) an $x''\in x'$ and so on, the first player who cannot play looses the game. We give a description of the odds in \autoref{i:games}.

\section{Some properties of chain functions}\label{sec:spcf}

Our aim is to study basic statistical properties of chains of length $n$ within elements of ${\mathcal S}_k$. We shall be interested in the large $k$ limit via a recursive procedure on $k$. We start by recalling and expanding the definitions. In the sequel, we often abbreviate ``chain counting function'' simply to ``chain function''. 

Chains were defined formally in \autoref{i:cps} but we repeat the definition for convenience.

\myitem{Chains of pure sets\label{i:cps2}}{A finite sequence $x_0,\cdots,x_n$ of pure sets is called a chain if $x_0\in x_1\in \cdots \in x_n$. Chains never close to make a cycle. We call $n$ the length of the chain.

The set $x_0$ (resp. $x_n$) is called the beginning (resp. the end) end of the chain, and the chain is called maximal if $x_0=\nun$ and $n>0$.

If $x$ is a pure set, a chain in $x$ is, by definition, a chain $x_0,\cdots,x_n$ that ends in $x$, i.e. such that $x_n=x$.}

\myitem{Direct chain counting functions\label{i:dccf}}{Direct chain counting functions are  denoted by  $G^{(n)}_k$ and $H^{(n)}_k$, $\kN$, $n\in {\mathbb Z}$. They map ${\mathcal S}_k$ to  ${\mathbb N}$. For $x\in {\mathcal S}_k$, $G^{(n)}_k(x)$ is the number of maximal chains of length $n$ in $x$ , and $H^{(n)}_k(x)$ the number of chains, maximal or not, of length $n$ in $x$.\footnote{Recall that by convention we do count $\lb\rb$ as a chain, but not as a maximal chain, of length $0$ in $\lb\rb$. Thus $G^{(n)}_k\equiv 0$ unless $0< n \leq k$ and $H^{(n)}_k$ unless $0\leq n \leq k$.}
}

\vspace{.3cm}
The direct chain counting functions have simple recursive properties valid for $\kN^*$, $n\in {\mathbb Z}$ and $x\in {\mathcal S}_k$
\begin{equation} G^{(n)}_k(x)=\sum_{y\in x} G^{(n-1)}_{k-1}(y)+\ind{n=1}\ind{\nun\in x} \qquad H^{(n)}_k(x)=\sum_{y\in x} H^{(n-1)}_{k-1}(y) +\ind{n=0}.\label{eq:recurdir}
\end{equation}
Here both $n$ and $k$ are shifted by $-1$ on the right-hand side. In fact, the relation holds also for $k=0$ with the convention that an empty sum is $0$, without having to specify what chain functions with subscript $-1$ mean. 

Note that in fact $G^{(n)}_k$ is identically $0$ unless $n\in \intt{1,k}$ and $H^{(n)}_k$ is identically $0$ unless $n\in \intt{0,k}$. Also, for $k'\geq k$ and $n\in {\mathbb Z}$ the restriction of $G^{(n)}_{k'}$ (resp. $H^{(n)}_{k'}$ to ${\mathcal S}_k$ is $G^{(n)}_k$ (resp. $H^{(n)}_k$): the number of chains of a given length ending at a pure set $x$ can be computed without reference to a $k$ such that $x \in {\mathcal S}_k$. Equivalently, we can make a global definition

\myitem{Global direct chain counting functions\label{i:gdccf}}{The global direct chain counting functions are  denoted by  $G^{(n)}$ and $H^{(n)}$, $n\in {\mathbb Z}$. They map ${\mathcal S}_{\infty}$ to  ${\mathbb N}$. For $x \in {\mathcal S}_{\infty}$, $G^{(n)}(x)$ is the number of maximal chains of length $n$ in $x$ , and $H^{(n)}(x)$ the number of chains, maximal or not, of length $n$ in $x$.
}

\vspace{.3cm}
The global direct chain functions $G^{(n)}$ (resp. $H^{(n)}$) on ${\mathcal S}_{\infty}$ are such that $G^{(n)}_k$ (resp. $H^{(n)}$) is the restriction of $G^{(n)}$ (resp. $H^{(n)}$) to ${\mathcal S}_k$ for $\kN$.
In terms of global direct chain functions the recursive property rewrites
\begin{equation} G^{(n)}(x)=\sum_{y\in x} G^{(n-1)}(y)+\ind{n=1}\ind{\nun\in x} \qquad H^{(n)}_k(x)=\sum_{y\in x} H^{(n-1)}(y) +\ind{n=0}.\label{eq:recurdirglob}
\end{equation}
for $n\in {\mathbb Z}$ and $x\in {\mathcal S}_{\infty}$.

\vspace{.3cm}
For technical reasons, it is useful to be able to manipulate chain counting functions with a varying $k$ while $k-n$ is kept fixed, leading to the notion of inverse chain counting function.
\myitem{Inverse chain counting functions\label{i:iccf}}{The inverse chain counting functions are defined as $\overline{G}^{(n)}_k:=G^{(k-n)}_k$ and $\overline{H}^{(n)}_k:=H^{(k-n)}_k$, $\kN$, $n\in {\mathbb Z}$.
}

\vspace{.3cm}
In contrast with direct chain functions, due to the fact that inverse chain functions see pure sets ``upside-down'' the subscript $k$ is essential for them. Clearly the families of direct chain functions and inverse chain functions carry the same information ... as long as we do not let $k\to \infty$ carelessly.

For inverse chain counting functions the recursive property rewrites
\begin{equation} \overline{G}^{(n)}_k(x)=\sum_{y\in x} \overline{G}^{(n)}_{k-1}(y)+\ind{k=n+1}\ind{\nun\in x} \qquad \overline{H}^{(n)}_k(x)=\sum_{y\in x} \overline{H}^{(n)}_{k-1}(y) +\ind{n=k},\label{eq:recurinv}
\end{equation}
Here $k$, but not $n$, is shifted by $-1$ on the right-hand side.

\myitem{Trees and pure sets upside-down\label{i:tpsud}}{
Though the use we make of inverse chain functions is mostly of technical nature, looking at pure sets ``upside-down'' is interesting for itself. This viewpoint is perhaps most easily visualized in the tree representation (see \autoref{i:it}). If $x\in {\mathcal S}_k$, hiding the first $k-n$ levels of the tree leaves a collection of identity trees corresponding to pure sets in  ${\mathcal S}_n$. Here is an example with $k=3,n=2$ so $k-n=1$:
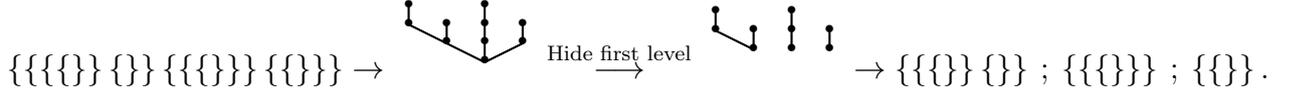
\[ \bquinze \to \begin{tikzpicture} [scale=.25] \draw [thick] (-4,2)--(-4,1)--(-2,0)--(0,-1)--(2,0)--(2,1) (0,2)--(0,1)--(0,0)--(0,-1) (-2,1)--(-2,0); \draw (-4,2) node {$\mbullet$}; \draw (-4,1) node {$\mbullet$}; \draw (-2,0) node {$\mbullet$}; \draw (0,-1) node {$\mbullet$}; \draw (2,0) node {$\mbullet$}; \draw (2,1) node {$\mbullet$}; \draw (0,2) node {$\mbullet$}; \draw (0,1) node {$\mbullet$}; \draw (0,0) node {$\mbullet$}; \draw (-2,1) node {$\mbullet$}; \end{tikzpicture} \stackrel{\tiny \text{Hide first level}}{\longrightarrow} \begin{tikzpicture}[scale=.25]  \draw (0,-1) node {}; \draw [thick] (-4,2)--(-4,1)--(-2,0)--(-2,1) (0,2)--(0,1)--(0,0) (2,1)--(2,0); \draw (-4,2) node {$\mbullet$} (-4,1) node {$\mbullet$} (-2,0) node {$\mbullet$} (-2,1) node {$\mbullet$} (0,2)  node {$\mbullet$} (0,1)  node {$\mbullet$} (0,0) node {$\mbullet$}  (2,1)  node {$\mbullet$} (2,0) node {$\mbullet$}; \end{tikzpicture} \to \bquatre \ ; \ \ntrois \ ; \ \ndeux.\]
The whole procedure can be described in terms of the braces representation as well. Hiding the first level of $x=\lb x_1,\cdots,x_\alpha\rb \in {\mathcal S}_k$ (the elements of $x$ are to be listed without repetition) means removing the opening and closing brace of $x$ leaving $x_1\ ; \ \cdots \ ; \ x_\alpha$ (the order is irrelevant). Hiding the first two levels in $x$ means removing the opening and closing brace of each set in the collection $x_1\ ; \ \cdots \ ; \ x_\alpha$. The procedure can be iterated, removing a shell of braces at each iteration. 
The reader may check that the procedure with $k=3,n=2,1,0$ yields
\[ \bquinze \to \bquatre \ ; \ \ntrois \ ; \ \ndeux \to \ndeux \ ; \ \nun \ ; \ \ndeux \ ; \ \nun \to \nun \ ; \ \nun .\]

As this example shows, for given $n$ and $x\in {\mathcal S}_k$ the resulting  collections of pure sets has multiplicities in general. Formally, we may define a map ${\mathcal S}_k\to {\mathbb N}^{{\mathcal S}_n},\, x \mapsto \sum_{y\in {\mathcal S}_n} m_y(x)y$ where $m_y(x)$ is the number of times $y\in {\mathcal S}_n$ appears when $k-n$ shells of braces have been removed from $x$. Removing shells of braces one by one allows to express $m_y(x)$ in terms of chains: $m_y(x)$ is the number of chains of length $k-n$ starting at $y$ and ending at $x$. This leads to a number of nontrivial counting problems. For example, one might want to compute a generating polynomial like $\sum_{x\in {\mathcal S}_k}  \prod_{y\in {\mathcal S}_n} t_y^{m_{y}(x)}$ where the $t_y$s are formal variables indexed by $y\in {\mathcal S}_n$, or some specializations thereof. Symmetry considerations show that this generating polynomial is symmetric in the $t_y$s, but apart from that little can be said in general. 

For $n=k-1$ there are no multiplicities (for a given $x\in {\mathcal S}_k$, each $y \in {\mathcal S}_{k-1}$ appears at most once when the first level of $x$ is hidden), and one checks that
\[ \sum_{x\in {\mathcal S}_k}  \prod_{y\in {\mathcal S}_{k-1}} t_y^{m_{y}(x)}=\prod_{y\in {\mathcal S}_{k-1}} (1+t_y),\]
which is nothing but \autoref{eq:basident} in disguise. The case $n=k-2$ can also be analyzed (mostly) explicitly. This is done in \autoref{i:sesh}.
For other values of $n$ the situation is more intricate. The case $n=0$, when the procedure ``simply'' yields $m_{\lb\rb}(x)$, i.e. a number of $\nun$s for each $x\in {\mathcal S}_k$, is nontrivial and related to the ``maximal'' maximal chain function, see \autoref{i:mmcf}. We explain the recursive structure of the answer but are able to quote an explicit formula only for $k=0,\cdots,5$.}

\vspace{.3cm}
We shall also have use of total chain counting function.
\myitem{Total chain counting functions\label{i:tccf}}{The sums $G_k:=\sum_n G^{(n)}_k$ (resp. $H_k:=\sum_n H^{(n)}_k$) define the total chain counting functions for $\kN$. They count the number of maximal (rep. arbitrary) chains of arbitrary lengths in an element of ${\mathcal S}_k$. In the tree representation, they count the number of leaves and nodes respectively.
}

\vspace{.3cm}
For total chain counting functions the recursive property rewrites
\begin{equation}
G_k(x)=\sum_{y\in x} G_{k-1}(y)+\ind{\nun\in x} \qquad  H_k(x)=\sum_{y\in x} H^{(n-1)}_{k-1}(y) +1.\label{eq:recurtot}
\end{equation}

If $x,y\in {\mathcal S}_k$ are disjoint, then $G^{(n)}_k(x)+G^{(n)}_k(y)=G^{(n)}_k(x\cup y)$ and $H^{(n)}_k(x)+H^{(n)}_k(y)=H^{(n)}_k(x\cup y)+\ind{n=0}$, so\footnote{With the exception of $H^{(0)}_k=\overline{H}^{(k)}_k$, when there is a simple and explicit additional term $1$.} the chain counting functions of index $k$ are in fact measures on ${\mathcal S}_{k-1}$. In particular, the chain counting functions behave nicely under taking complements: an element $x$ of ${\mathcal S}_k$ is a subset of ${\mathcal S}_{k-1}$ and there is a symmetry under the involution  $x \to {\mathcal S}_{k-1}\setminus x$. For $x\in{\mathcal S}_k$  we have $G^{(n)}_k(x)+G^{(n)}_k({\mathcal S}_{k-1}\setminus x)=G^{(n)}_k({\mathcal S}_{k-1})$ and $H^{(n)}_k(x)+H^{(n)}_k({\mathcal S}_{k-1}\setminus x)=H^{(n)}_k({\mathcal S}_{k-1})+ \ind{n=0}$.

\vspace{.3cm}
We compute the chain counting functions for two simple families of pure sets.

\myitem{The chain counting functions of matryoshka}{We defined the notation for matryoshas in \autoref{i:rdm}  by setting recursively $\underline{0}:=\nun$ and $\underline{k}:=\lb \underline{k-1}\rb$ for $k \geq 1$. It is plain that $\underline{k}\in {\mathcal S}_k$ and $G^{(n)}_{k'}(\underline{k})=\ind{k=n\geq 1}$ and $H^{(n)}_{k'}(\underline{k})=\ind{n\in \intt{0,k}}$ for $k' \geq k \geq 0$.}

\myitem{The chain counting functions of Von Neumann integers}{We defined a notation for Von Neumann integers in \autoref{i:vni} by setting recursively $\overline{0}:=\nun$ and $\overline{k}:=\overline{k-1}\cup \lb \overline{k-1}\rb$. It is plain that $\overline{k}\in {\mathcal S}_k$ and a recursion argument obtains that $G^{(n)}_{k'}(\overline{k})=\binom{k-1}{n-1}$ and $H^{(n)}_{k'}(\overline{k})=\binom{k}{n}$ for $k' \geq k \geq 0$.

\myproof We do the proof for maximal chain functions. For $k=0$ the formula gives $G^{(0)}_{k'}(\overline{0})=0$ for $k'\geq 0$. which is correct. For $k=1$ the formula gives $G^{(n)}_{k'}(\overline{1})=\ind{n=1}$ for $k'\geq 1$ which is correct. Now suppose $k \geq 2$. We note first that $\nun$ is an element both of $\overline{k}$ and $\overline{k-1}$, and second that $y\in \overline{k}$ means $y\in \overline{k-1}$ or (exclusive)  $y=\overline{k-1}$. Hence, using \autoref{eq:recurdir} when needed,
\begin{eqnarray*}
  G^{(n)}_k(\overline{k}) & = & \sum_{y\in \overline{k-1}} G^{(n-1)}_{k-1}(y)+ G^{(n-1)}_{k-1}(\overline{k-1})+\ind{n=1}\ind{\nun\in \overline{k}}  \\ & = & \sum_{y\in \overline{k-1}} G^{(n-1)}_{k-1}(y)+ G^{(n-1)}_{k-1}(\overline{k-1})+\ind{n=1} \\ & = & G^{(n)}_k(\overline{k-1})-\ind{n=1}\ind{\nun\in \overline{k-1}}+ G^{(n-1)}_{k-1}(\overline{k-1})+\ind{n=1} \\ & = & G^{(n)}_k(\overline{k-1})-\ind{n=1}+ G^{(n-1)}_{k-1}(\overline{k-1})+\ind{n=1} \\ & = & G^{(n)}_k(\overline{k-1})+ G^{(n-1)}_{k-1}(\overline{k-1}).
\end{eqnarray*}
If the recursion hypothesis holds for $\overline{k-1}$  we get $G^{(n)}_k(\overline{k})=\binom{k-2}{n-1}+\binom{k-2}{n-2}=\binom{k-1}{n-1}$ so the recursion hypothesis holds for $\overline{k}$, concluding the argument. The case of ordinary chain functions is similar.}

\vspace{.3cm}

These simple results have an immediate application for the proof of the linear independence statement below.

\myitem{Relations among chain counting functions\label{i:linindep}}{The function $G^{(0)}_k$ is identically $0$ and the function $H^{(0)}_k$ is identically $1$ for $\kN$. The linear relations $G^{(k)}_k=H^{(k)}_k$ hold for $k\geq 1$ and the linear relations $G^{(k-1)}_k+G^{(k)}_k=H^{(k-1)}_k$ hold for $k\geq 2$. These are the only linear relations among direct chain counting functions i.e. \\
-- The function $H^{(0)}_0$ is linearly independent on ${\mathcal S}_0$,\\
-- The functions $H^{(0)}_1$ and $G^{(1)}_1$ are linearly independent on ${\mathcal S}_1$,\\
-- The functions $(G^{(n)}_k)_{n\in \intt{1,k}}$ and $(H^{(n)}_k)_{n\in \intt{0,k-2}}$ are linearly independent on ${\mathcal S}_k$ for $k\geq 2$.

\myproof The statements describing the special cases $k=0,1$ result from easy computations, and we concentrate on $k\geq 2$. 

The relation $G^{(k)}_k=H^{(k)}_k$ for $k\geq 1$ has already been noticed and used before.

We show that $G^{(k-1)}_k+G^{(k)}_k=H^{(k-1)}_k$ for $k\geq 2$. If a chain $y\in\cdots \in x$ of length $k-1$ ends at $x\in {\mathcal S}_k$ and starts at $y$ then either $y\in {\mathcal S}_1$ so that $y=\nun$, and then the chain is maximal and counts in $G^{(k-1)}_k$, or $y=\ndeux$ and then the only way to extend it is as the maximal chain $\nun \in y \cdots \in x$ that counts in $G^{(k)}_k$.

To prove the linear independence statement we introduce a compact notation. For $\kN$, set
\[ F^{(n)}_k:=\left\{ \begin{array}{ll} H^{(n)}_k & \text{ for } n\in \intt{0,k},\\ G^{(-n)}_k & \text{ for } n\in\intt{-k,-1}. \end{array}\right. \]
We already know that $F^{(k)}_k=F^{(-k)}_k$ for $k\geq 1$ and $F^{(k-1)}_k=F^{(-k)}_k+F^{(-k+1)}_k$ for $k\geq 2$.
We take for granted that $F^{(0)}_0$ is linearly independent on ${\mathcal S}_0$, and $F^{(-1)}_1,F^{(0}_1$ are linearly independent on ${\mathcal S}_1$. So we need to prove that the functions $F^{(n)}_k$, $n\in \intt{-k,k-2}$ are linearly independent for $k\geq 2$. We argue by recursion. Assume that the linear combination $\sum_{n\in \intt{-k,k-2}}\alpha_n F^{(n)}_k$ vanishes identically on ${\mathcal S}_k$. In particular this linear combination vanishes identically on ${\mathcal S}_{k-1}$ which entails that $\sum_{n\in \intt{-k+1,k-2}}\alpha_n F^{(n)}_{k-1}=0$ on ${\mathcal S}_{k-1}$. For $k-1=1$ we obtain $\alpha_{-1}=\alpha_0=0$, so that $\alpha_{-2}F^{(-2)}_2=0$, which plainly implies that $\alpha_{-2}=0$ and we are done. Suppose that $k\geq 3$ and the result has been established for $k-1$. For the last term, $n=k-2$,  we use $F^{(k-2)}_{k-1}=F^{(-k+1)}_{k-1}+F^{(-k+2)}_{k-1}$. The recursion hypothesis gives $\alpha_{-k+1}+\alpha_{k-2}=\alpha_{-k+2}+\alpha_{k-2}=0$ and $\alpha_n=0$ for $n\in \intt{-k+3,k-3}$. So the linear combination $\sum_{n\in \intt{-k,k-2}}\alpha_n F^{(n)}_k$ collapses to $\alpha_{-k}F^{(-k)}_k+\alpha_{k-2}\left(F^{(k-2)}_k-F^{(-k+1)}_k-F^{(-k+2)}_k\right)$. We evaluate this linear combination on the matryoshka $\underline{k}$ to get $1\cdot \alpha_{-k}+1\cdot \alpha_{k-2}=0$ and on the Von Neumann integer $\overline{k}$ to get $\binom{k-1}{k-1}\cdot \alpha_{-k} +\left(\binom{k}{k-2}-\binom{k-1}{k-2}-\binom{k-1}{k-3}\right)\cdot \alpha_{k-2}=0$, that is $1\cdot \alpha_{-k} +0\cdot \alpha_{k-2}=0$. Thus  $\alpha_{-k}=\alpha_{k-2}=0$ and then $\alpha_{-k+1}=\alpha_{-k+2}=0$ as well, concluding the recursion step.
}

\section{The probabilistic model} \label{sec:probmod}

The growth of the sequence $(Z_k)_{\kN}$ makes an explicit description of each pure set an illusory goal, and we resort to a statistical description. We view ${\mathcal S}_k$ as a sample space where each set has probability $1/Z_k$, that is, ${\mathcal S}_k$ is endowed with the uniform probability measure,  denoted by $\nu_k$ in the sequel.

This means that a sample in ${\mathcal S}_k$ is made of elements of ${\mathcal S}_{k-1}$ chosen with probability $1/2$ independently of each other. We have already explained when we introduced the binary string representation of pure sets (see \autoref{sec:rpsc}) how the usual symmetric probability measure on infinite sequences of independant Bernoulli trials induces the measure $\nu_k$ on each ${\mathcal S}_k$. Thus there is single measure to rule them all. Call it $\nu$. We view $\nu$ as a measure on subsets of ${\mathcal S}_{\infty}$ (not a pure set), i.e. on $2^{{\mathcal S}_{\infty}}=:{\mathcal S}_{\infty+1}$ (not a pure set either): a sample is made by deciding with probability $1/2$ for each pure set independently of the others whether or not it belongs to the sample.\footnote{Just to make things clear, an element of $2^{{\mathcal S}_{\infty}}={\mathcal S}_{\infty+1}$ is an elementary event, and $\nu$ assigns a probability to measurable families of elementary events (the so-called measurable subsets of ${\mathcal S}_{\infty+1}$) that is, only to certain elements of $2^{{\mathcal S}_{\infty+1}}=:{\mathcal S}_{\infty+2}$.} Defining $\Pi_k\colon  {\mathcal S}_{\infty+1}\to {\mathcal S}_k=2^{{\mathcal S}_{k-1}},\; \chi \mapsto \chi \cap {\mathcal S}_{k-1}$, we obtain $\nu_k$ as an image measure: $\nu_k=\nu \circ \Pi_k^{-1}$.

However, beware of two slight traps: \\
-- First, ${\mathcal S}_{k-1}$ is a subset of ${\mathcal S}_k$ but the probability of a pure set in ${\mathcal S}_{k-1}$ (i.e. under $\nu_{k-1}$) is not its probability as an element of ${\mathcal S}_k$ (i.e. under $\nu_k$). In terms of random processes, one usually sees, say, $1$ and $01$ as distinct with $1$ being interpreted  as the (non-elementary) event ``anything on the left''$1$ and $01$ as ``anything on the left''$01$, but the map to pure sets $\varphi$ views both as an elementary event, a left-infinite strings of $0$s with a final $1$ on the right.\\
-- Second, we have observed after  \autoref{i:gdccf} that the direct chain functions $G^{(n)}_k$ (resp. $H^{(n)}_k$) are restrictions to ${\mathcal S}_k$ of well-defined global chain functions $G^{(n)}$ (resp. $H^{(n)}$) defined on ${\mathcal S}_{\infty}$. However, the global chain functions cannot be extended as useful random variables defined ${\mathcal S}_{\infty+1}$. They would take infinite values almost everywhere. 

Nevertheless, we dispense (most of the time) with the notation $\nu_k$ and use $\nu$ (almost) systematically, with associated expectation symbol ${\mathbb E}$. Any function $f$ from  ${\mathcal S}_k$ to some set $X$ defines a random element $f\circ \Pi_k$ on ${\mathcal S}_{\infty+1}$ with values in $X$. Accordingly, we turn the chain counting functions at fixed $k$ (\rv s on ${\mathcal S}_k$) into random variables on ${\mathcal S}_{\infty+1}$ by composition with $\Pi_k$ but keep the same notation. For instance, according to the context, $G^{(n)}_k$ stands either for $\text{(former)}G^{(n)}_k$ or $\text{(new)}G^{(n)}_k:=\text{(former)}G^{(n)}_k\circ \Pi_k$, this should cause no confusion.

\vspace{.3cm}
A word on speed of convergence. We shall have many occasions to meet $k$-sequences involving random variables on ${\mathcal S}_k$ and whose probabilistic properties (averages for instance) have a large $k$ limit with corrections of order $1/Z_{k-1}$ (resp. $1/Z_{k-2}$). As functions of $k$ these behaviors  amount to a very fast decay. But from the point of view of statistical mechanics, one should think of ${\mathcal S}_k$ as the configuration space of a system of size $Z_{k-1}$, and then the above corrections are just like usual (resp. logarithmic) finite size corrections.

\vspace{.2cm}

We illustrate the probabilistic model with a few examples.

\myitem{Games\label{i:games}}{Recall that, $x$ being a pure set, player one chooses (if possible) an $x'\in x$, then player two chooses (if possible) an $x''\in x'$ and so on. The first player who cannot play looses the game.  If $Z_k^{\text{one}}$ (resp. $Z_k^{\text{two}}$) denotes the number of pure sets in ${\mathcal S}_k$ for which player one (resp. player two) has a winning strategy, then $Z_k^{\text{one}}+Z_k^{\text{two}}=Z_k$ and $Z_k^{\text{two}}=2^{Z_{k-1}^{\text{one}}}=Z_k2^{-Z_{k-1}^{\text{two}}}$ for $\kN^*$. Indeed, $a)$ no draw is possible, $b)$ $\nun$, the only element of ${\mathcal S}_0$, is winning for player two, giving a winning strategy for two on $\nun$, and $c)$ $x\in {\mathcal S}_k$ is winning for the second player to play if and only if every element $x'$ of $x$ is winning for the first player to play (in $x'$) so each element of ${\mathcal S}_k$ can be tagged as winning or loosing for the first player to play once this assignment has been done for elements of ${\mathcal S}_{k-1}$, and finally $d)$ the corresponding strategies are also computed recursively: if $x\in {\mathcal S}_k$ is tagged as winning for the first player to play, he should choose any element of $x'$ of $x$ that is loosing for the first player to play.

The uniform probability measure on ${\mathcal S}_k$ is perfect for player two for $k=0$, fair for $k=1,2$ and then the odds quickly get extremely small for the second player. Indeed, precisely because  $Z_{k-1}^{\text{two}}$ grows rapidly with $k$, the winning probability for player two,  $Z_k^{\text{two}}/Z_k=2^{-Z_{k-1}^{\text{two}}}$, soon gets miserable ($1/4$ for $k=3$, $1/16$ for $k=4$, $1/2^{4096}$ for $k=5$ and the worse is to come).\footnote{Of course, this could be compensated by ajusting the bets of the two players.}} 

\myitem{Being both an element and a subset}{Let $k\geq 1$ and fix $y\in {\mathcal S}_k$. Take $x$ in  ${\mathcal S}_k$ with the uniform probability measure. The probability that $y \subset x$ is $\frac{1}{2^{\# y}}$. Hence, if $y\in {\mathcal S}_{k-1}$, the probability that $y\cup\lb y\rb \subset x$ is $\frac{1}{2^{1+\# y}}$. Thus the average number of subsets of $x$ that are also elements of $x$ is $\sum_{y\in {\mathcal S}_{k-1} } \frac{1}{2^{1+\# y}}=\sum_{y\subset {\mathcal S}_{k-2} } \frac{1}{2^{1+\# y}}=\sum_n  \binom{Z_{k-2}}{n} \frac{1}{2^{1+n}}=\frac{1}{2}\left(\frac{3}{2}\right)^{Z_{k-2}}$.}

\myitem{Counting transitive pure sets\label{i:cts}}{This is related to the previous example. Recall from \autoref{i:ts} that a set $x$ is called transitive if every element of $x$ is also a subset of $x$. From the explicit description of the first levels in \autoref{sec:ffdb} we read that ${\mathcal S}_0$ contains one transitive set, $\nun$, in ${\mathcal S}_1$ there is a new one, $\ndeux$, then a new one is ${\mathcal S}_2$, $\bquatre$, and three new ones in ${\mathcal S}_3$, $\bdouze$, $\btreize$, and $\bseize$. So, if $T_k$ is the number of transitive sets in ${\mathcal S}_k$,
\[ T_0=1\ T_1=2\ T_2=3\ T_3=6.\]
Clearly, transitive sets have a recursive nature, but to reveal it requires to keep track of more information. We assume for now that $k\geq 2$.

We partition ${\mathcal S}_{k-1}$ as $\left({\mathcal S}_{k-1}\setminus {\mathcal S}_{k-2}\right) \cup {\mathcal S}_{k-2}$. Accordingly, if $z\in {\mathcal S}_k$, that is, if $z\subset {\mathcal S}_{k-1}$, we can decompose $z=y\cup x$, where the elements of $y$ have depth exactly $k-1$ and those of $x$ depth at most $k-2$. Let us denote by $T_{k,n,q}$ the number of transitive sets $z$ in ${\mathcal S}_k$ such that $\# y=n$ ($y$ has cardinal $n$) and  $\# x=q$.

An element of an element of $x$, if any, has depth at most $k-3$, hence cannot be an element of y. Thus,\\
-- If $z$ is transitive, so must be $x$. \\
An element of an element of $y$, if any, has depth at most $k-2$ hence cannot be an element of $y$. Thus, \\
-- If $x$ is transitive, then $z$ is transitive if and only if any element of an element of $y$ is an element of $x$. This can be rephrased as the fact that $y$ is a subset of $2^x$.\\

Assuming that $x \in {\mathcal S}_{k-1}$ is transitive, it is easy to count the number of sets $y\in {\mathcal S}_k\setminus {\mathcal S}_{k-1}$ such that  $\# y=n$ and $z:=y\cup x$ is transitive. But one needs to take into account that $y$ does not intersect ${\mathcal S}_{k-1}$.  We decompose $x$ as $x=w\cup v$ where $w\in {\mathcal S}_{k-1}\setminus {\mathcal S}_{k-2}$ and $v\in {\mathcal S}_{k-2}$. Set $m:=\# w$ and $p:=\# v$.  If $y$ is nonempty, i.e. if $n >0$ then any element of $y$ has depth exactly $k-1$ so must contain at least one an element of $w$. Thus for given $x$ there are $\binom{(2^{m+p}-2^p}{n}$ matching $y$s. Consequently, the following recursion relation holds:
\begin{equation} T_{k,n,q}=\sum_{m+p=q} \binom{2^{m+p}-2^p}{n}T_{k-1,m,p} \text{ for } k\geq 2.\label{eq:recurtrans} \end{equation}
The initial conditions are $T_{1,0,0}=T_{1,1,0}=1$ (and the other $T_{1,\cdot,\cdot}=0$).

Then one can sum over $n$ and $q$ to get $T_k$. This leads to $T_4=4131$. An immediate consequence of \autoref{eq:recurtrans} is
$\sum_n T_{k,n,q}=\sum_{m+p=q}  2^{2^q-2^p}T_{k-1,m,p} \text{ for } k\geq 2$. This little trick allows to compute $T_5$, a number of order $3\; 10^{19723}$, in a fraction of a second.

If $x$ is as large as possible, i.e. $x={\mathcal S}_{k-2}$, the above discussion shows that any choice of $y\in {\mathcal S}_k\setminus {\mathcal S}_{k-1}$ is such that $y\cup x$ is a transitive set in ${\mathcal S}_k$. Thus a lower bound for $T_k$ is $Z_k/Z_{k-1}$ for $k\geq 1$. This approximation is excellent at large $k$ : for $k=0,\cdots,5$, $T_kZ_{k-1}/Z_k-1$ is $-1,0,1/2,1/2,< 10^{-2},< 10^{-9863}$.

To summarize, at depth $\leq k$, $k$ large, the transitive sets form only a tiny fraction, of order $1/Z_{k-1}$, of the total. However, as already noticed, in terms of statistical mechanics and entropy ${\mathcal S}_k$ is the configuration space for a system of size $Z_{k-1}$: taking logarithms (in base $2$) $\log T_k-\log Z_k = \log T_k-Z_{k-1} \simeq -\log Z_{k-1}$, a mere logarithmic correction.

The sequence $T_k$ was submitted to, and inserted as A$381080$ by, the OEIS \cite{oeis2025}. 

}
  
\myitem{Simple conditioning}{Choose a random element $x$ in  ${\mathcal S}_k$ with the uniform probability measure $\nu_k$ conditioned  on $x\neq \eun$ and choose an element $y$ in $x$ with the uniform probability measure on $x$. Then $y$, an element of ${\mathcal S}_{k-1}$, is distributed with $\nu_{k-1}$.

\myproof This result is natural because there is nothing that singles out $y$ from other members of ${\mathcal S}_{k-1}$. Here is the explicit combinatorics.  The joint law for $x,y$ is $\frac{1}{Z_k-1} \frac{1}{\# x}\ind{y\in x}$. Thus the probability to pick $y$ is
\[ \sum_{x\in {\mathcal S}_k\, x\neq \eun} \frac{1}{Z_k-1} \frac{1}{\# x}\ind{y\in x}=\frac{1}{Z_k-1} \sum_{h=1}^{Z_{k-1}} \frac{1}{h} \sum_{x\in {\mathcal S}_k,\, \# x=h}\ind{y\in x}.\]
The last sum is simply the combinatorial factor for choosing $h-1$ distinct elements in ${\mathcal S}_{k-1}\setmin \lb y \rb$, that is $\binom{Z_{k-1}-1}{h-1}=\frac{h}{Z_{k-1}}\binom{Z_{k-1}}{h}$. To summarize, the probability to pick $y$ is
\[ \frac{1}{Z_k-1} \sum_{h=1}^{Z_{k-1}} \frac{1}{Z_{k-1}} \binom{Z_{k-1}}{h}= \frac{1}{Z_k-1}\frac{1}{Z_{k-1}}  \left(2^{Z_{k-1}}-1\right)=\frac{1}{Z_{k-1}} \]
as announced.}

\myitem{Walks to leaves\label{i:wtl}}{This is related to the previous item. Consider the following procedure to build a chain. Take $k\geq 1$. Choose $x_k$  uniformly in ${\mathcal S}_k$ . If $x_k=\eun$ stop, else (this alternative is possible only if $k\geq 1$) choose  $x_{k-1}$ uniformly in $x_k$. If $x_{k-1}=\eun$ stop, else (this alternative is possible only if $k\geq 2$) choose  $x_{k-2}$ uniformly in $x_{k-1}$, and so on. This procedure might be easier to visualize in terms of trees. Choose an identity tree of depth at most $k$ at random uniformly. Then choose a descendant of the root, if any, uniformly. Then a descendant of that node, if any,  uniformly and so on until a leaf is reached.\footnote{Here, we break one of our conventions and take the root as a leaf in the trivial tree.\label{ft:root}} By the above lemma, the probability to survive for $j$ steps, i.e. that $x_{k-j}\neq \eun$ is $\prod_{0\leq i \leq j}\left(1-\frac{1}{Z_{k-i}} \right)$. There is a limit when $k\to \infty$ with $k-j=\overline{\jmath}$ fixed: in this limit, the probability to survive for $k-\overline{\jmath}$ steps is $\prod_{\overline{\jmath}\leq \overline{\imath}}\left(1-\frac{1}{Z_{\overline{\imath}}}  \right)$. It is not difficult to show that this number is transcendental, a Liouville number in fact. We shall not give an explicit proof here, but see the end of this section.}\addtocounter{footnote}{-1}

\myitem{Walks to the leaves, again}{There is a related question. Taking  $x$ in  ${\mathcal S}_k$ with the uniform probability measure $\nu_k$, what is the probability that all leaves are at distance at least $l$ from the root.\footnotemark\ If $Z_{k,l}$ denotes the number of elements in ${\mathcal S}_k$ such that all leaves are at distance at least $l$ from the root, then $Z_{k,0}=Z_k$ and $Z_{k,l}=2^{Z_{k-1,l-1}}-1$ for $k\geq l\geq 1$. This is because to build an $x$ in  ${\mathcal S}_k$ counting for $Z_{k,l}$, we have to chose the elements of $x$ freely among the members of ${\mathcal S}_{k-1}$ contributing to $Z_{k-1,l-1}$, but $x=\nun$ is not allowed. Then $Z_{k,1} =Z_k-1$, $Z_{k,2}=\frac{1}{2}Z_k-1$, $Z_{k,3}=\frac{1}{2}Z_k^{1/2}-1$ and so on. Starting with $l=3$, $Z_{k,l}/Z_k$ goes to $0$ at large $k$.\footnote{However $Z_{k,l}$ still grows quite rapidly at large $k$ for fixed $l$. For instance, starting at $k=0$, the sequence $Z_{k,4}$ begins with $0,0,0,0,1,2^{127}-1,\cdots$ ($2^{127}-1=170141183460469231731687303715884105727$ turns out to be a prime), while the sequence $Z_{k,3}$ starts with $0,0,0,1,127,2^{32767}-1,\cdots$. Note that $Z_{k,k}=1$ for every $k$, corresponding to the sequence of matrioshka.} Hence at large $k$, about half of the samples in ${\mathcal S}_k$ have all there leaves at distance at least $1$ from the root and a leaf at distance $1$, while about half of the samples have all there leaves at distance at least $2$ from the root and a leaf at distance $2$. In fact they have many leaves at distance $2$.

The fact that the story ends there, i.e. that  $Z_{k,3}/Z_k$ goes to $0$ (very fast) at large $k$, illustrates an important feature of our model, a phenomenon that we shall meet repeatedly in the sequel. This leads us to our next example.}

\myitem{The second shell\label{i:sesh}}{While the first shell is really random, namely if $y\in {\mathcal S}_{k-1}$ is given  and $x$ is a sample in ${\mathcal S}_k$ then $y$ belongs to $x$ with probability $1/2$ (this applies to $y=\nun$, explaining what happens at distance $1$ in the previous example), the second shell obeys a law of large numbers. We assume in this item that $k\geq 2$.

Before turning to probabilities, we start with generating functions, a preparation for the approach we follow in the sequel for more general questions. Recall from \autoref{i:tpsud} that, for $x\in {\mathcal S}_k$ and $z\in {\mathcal S}_{k-2}$, $m_z(x)$ counts the number of times $z$ appears in the second shell of $x$, or quivalently the number of chains $z\in y \in x$. Letting $(\tau_z)_{z\in {\mathcal S}_{k-2}}$ be a family of formal variable and applying \autoref{eq:basident} to $t_y:=\prod_{z\in y}\tau_z$ we infer that 
\[ \sum_{x\in {\mathcal S}_k} \prod_{z\in {\mathcal S}_{k-2}} \tau_z^{m_z(x)}=\sum_{x\in {\mathcal S}_k}  \prod_{z\in y\in x} \tau_z = \prod_{y\in {\mathcal S}_{k-1}} \left(1+\prod_{z\in y} \tau_z \right).\]
This formula is still complicated, but simplifies drastically in the special case when all $\tau_{\sbullet}$s but one are specialized to $1$. Thus we fix $z\in {\mathcal S}_{k-2}$ and set $\tau_z:=u$, $\tau_{z'}=1$ for $z'\in {\mathcal S}_{k-2} \setmin \lb z \rb$. Then  $\prod_{z'\in y\in x} \tau_{z'}= u^{\# \lb y | \, y \in x \text{ and } z \in y\rb}=u^{m_z(x)}$ leading to  
\[ \sum_{x\in {\mathcal S}_k} u^{m_z(x)}=\prod_{y\in {\mathcal S}_{k-1}} \left(1+u^{\ind{z\in y}} \right).\]
Splitting $\prod_{y\in {\mathcal S}_{k-1}}= \prod_{y\in {\mathcal S}_{k-1},\, z\in y} \prod_{y\in {\mathcal S}_{k-1},\, z\notin y}$ we finally get
\[ \sum_{x\in {\mathcal S}_k} u^{m_z(x)}=2^{Z_{k-1}/2}(1+u)^{Z_{k-1}/2}.\]
Returning to the probabilistic interpretation, that is, viewing $m_z$ (a function of $x$) as a random variable on ${\mathcal S}_k$, we obtain
\[ \expec{u^{m_z}}=\frac{1}{Z_k} \sum_{x\in {\mathcal S}_k} u^{m_z(x)} =2^{-Z_{k-1}/2}\left(1+u\right)^{Z_{k-1}/2}.\]
This means that the number of times $z$ appear as an element of the second shell of members of ${\mathcal S}_k$ follows a symmetric binomial distribution, the same for every $z$, the exponent of which is $Z_{k-1}/2$: informally, by the (oldest) law of large numbers, for large $k$ the fraction of $x$s in ${\mathcal S}_k$ for whom $z$ appears as an element of an element of $x$ about $Z_{k-1}/4$ times is close to $1$.

Keeping the formal variables $\tau_{\sbullet}$ arbitrary, we would obtain the joint distribution of all the random variables $m_z$, $z\in {\mathcal S}_{k-2}$. But this is not a simple object. 

We can refine our result for a single $m_z$ by conditioning on the size of the sample $x$ in ${\mathcal S}_k$.
Conditioned on $\# x =i$, a sample $x$ in ${\mathcal S}_k$ has probability $1/\binom{Z_{k-1}}{i}$. Half of the $Z_{k-1}$ elements of ${\mathcal S}_{k-1}$ contain $z$ as an element, and half don't. To make samples where $z$ appears $j$ times as an element of an element of $x$ we choose $j$ among the ones that do, and $i-j$ among the ones that don't. Thus the probability is
\[\frac{\binom{Z_{k-1}/2}{j}\binom{Z_{k-1}/2}{i-j}}{\binom{Z_{k-1}}{i}}.\]
Again, the law does not depend on $z$. Equivalently, introducing the counting variable $u_1$ for $\# x$ (the index $1$ is for the first shell) and substituting $u_2$ for $u$ (the index $2$ is for the second shell), we get 
\[ \frac{1}{Z_k} \sum_{x\in {\mathcal S}_k} u_1^{\# x} u_2^{\sum_{y\in x} \ind{z\in y}}=\left(\frac{1}{4}(1+u_1)(1+u_1u_2)\right)^{Z_{k-1}/2}.\]
One computes that, conditioned on $\# x =i\in \intt{0,Z_{k-1}}$, the mean of $\sum_{y\in x} \ind{z\in y}$ is $\frac{i}{2}$ and the variance is $\frac{i(Z_{k-1}-i)}{4(Z_{k-1}-1)}$.
  This leads to the following observation. If no size condition if imposed on the sample $x$, its average size is $Z_{k-1}/2$ and its variance $Z_{k-1}/4$, while the mean of $\sum_{y\in x} \ind{z\in y}$ is $Z_{k-1}/4$ and its variance is $Z_{k-1}/8$. If we plug the mean of $\# x$ for $\# x$ in the formula with size conditions imposed, we find that the mean of $\sum_{y\in x} \ind{z\in y}$ coincides with that of the unconditional case, while the variance is reduced by a factor about $2$ (exactly $2$ in the large $k$ limit). This gives a quantitative measure of the contributions of the fluctuations of $\# x$ to the fluctuations of  $\sum_{y\in x} \ind{z\in y}$. The penultimate example in this section (see \autoref{i:h1h2}), as well as most of the rest of these notes, illustrates that the fluctuations of  $\# x$ play the dominant role in the fluctuations of all chain functions, and account for all the fluctuations of chain functions in a precise sense in the large depth limit.}  

\myitem{Average number of chains\label{i:anc}}{The following application is directly related to our main interest. We compute the average number of chains of a given (non-negative) length. We start with maximal chains (of length $\geq 1$). The computation is based on the identities quoted at the end of \autoref{i:tccf}: for $x\in {\mathcal S}_k$, $G^{(n)}_k(x)+G^{(n)}_k({\mathcal S}_{k-1}\setminus x)=G^{(n)}_k({\mathcal S}_{k-1})$ and, for $x\in {\mathcal S}_k$ $k\geq 1$, $G^{(n)}_k(x)=\sum_{y\in x} G^{(n-1)}_{k-1}(y)+\ind{n=1}\ind{\nun \in x}$.\footnote{Note that in this sentence $G^{(n)}_k$ has its initial meaning: a function on ${\mathcal S}_k$.}

The first identity entails that $\expec{G^{(n)}_k}=\frac{1}{2} G^{(n)}_k({\mathcal S}_{k-1})$ and the second that  $G^{(n)}_k({\mathcal S}_{k-1})=Z_{k-1} \expec{G^{(n-1)}_{k-1}}+\ind{n=1}$ for $k\geq 1$.\footnote{In this sentence, the situation is more tricky: in $G^{(n)}_k({\mathcal S}_{k-1})$, $G^{(n)}_k$ keeps its initial meaning, but in $\expec{G^{(n)}_k}$  (resp. $\expec{G^{(n-1)}_{k-1}}$) composition on the right with $\Pi_k$ (resp. $\Pi_{k-1}$) is implied. The same yoga applies in the sequel.} The outcome is a family of recursion relations that are easily solved:
\[ G^{(1)}_k({\mathcal S}_{k-1})=\ind{k\geq 1} \qquad  G^{(n)}_k({\mathcal S}_{k-1})=\ind{k\geq n} \prod_{l=1}^{n-1} \frac{Z_{k-l}}{2} \text{ for } n\geq 2,\]
or equivalently
\[ \expec{G^{(1)}_k}=\frac{1}{2}\ind{k\geq 1} \qquad \expec{G^{(n)}_k}=\frac{1}{2}\ind{k\geq n} \prod_{l=1}^{n-1} \frac{Z_{k-l}}{2} \text{ for } n\geq 2.\]

We continue with arbitrary chains (of length $\geq 0$), starting from the identities quoted at the end of \autoref{i:tccf}: for $x\in {\mathcal S}_k$, $H^{(n)}_k(x)+H^{(n)}_k({\mathcal S}_{k-1}\setminus x)=H^{(n)}_k({\mathcal S}_{k-1})+ \ind{n=0}$ and, for $x\in {\mathcal S}_k$ $k\geq 1$, $H^{(n)}_k(x)=\sum_{y\in x} H^{(n-1)}_{k-1}(y)+\ind{n=0}$.

We infer from the first identity that $\expec{H^{(n)}_k}=\frac{1}{2} \left(H^{(n)}_k({\mathcal S}_{k-1})+\ind{n=0}\right)$ and from the second that $H^{(n)}_k({\mathcal S}_{k-1})=Z_{k-1} \expec{ H^{(n-1)}_{k-1}}+\ind{n=0}$ for $k\geq 1$. The outcome is:
\[ H^{(0)}_k({\mathcal S}_{k-1})=1 \qquad H^{(n)}_k({\mathcal S}_{k-1})=2\,\ind{k\geq n} \prod_{l=1}^{n} \frac{Z_{k-l}}{2} \text{ for } n\geq 1\]
or equivalently
\[ \expec{H^{(0)}_k}=1 \qquad \expec{H^{(n)}_k}=\ind{k\geq n} \prod_{l=1}^{n} \frac{Z_{k-l}}{2} \text{ for } n\geq 1.\]

As  $G^{(k)}_k=H^{(k)}_k$ for $k\geq 1$ it is reassuring that our expressions for evaluation on ${\mathcal S}_{k-1}$ or equivalently for  expectations do match, due to $Z_0=1$. It is convenient to set $\tilde{Z}_k:=\prod_{l=0}^{k-1}\frac{Z_l}{2}$ that is exactly $\expec{G^{(k)}_k}=\expec{H^{(k)}_k}$.\footnote{The first values of $\tilde{Z}_k$, starting at $k=0$, are $1,1/2,1/2,1,8,262144=2^{18},2^{65553},\cdots$.} For instance we can rewrite
\[ \expec{G^{(n)}_k}=\ind{k\geq n}\frac{1}{2}\frac{\tilde{Z}_k}{\tilde{Z}_{k-n+1}} \text{ for }n\geq 1\qquad \expec{H^{(n)}_k} = \ind{k\geq n}\frac{\tilde{Z}_k}{\tilde{Z}_{k-n}} \text{ for }n\geq 0. \]
Note that $\tilde{Z}_k$ grows very rapidly, but only logarithmically, with $Z_k$, the size of the sample space.}

\myitem{Average total number of chains}{Recall that the total chain functions are $G_k:=\sum_{n\in \intt{1,k}} G^{(n)}_k$ and  $H_k:=\sum_{n\in \intt{0,k}} H^{(n)}_k$. Then $\expec{H_k}=2\expec{G_k}+1$ and both expectations are  of the order of the number of chains of length $k$:
\[ \frac{\expec{G_k}}{\tilde{Z}_k}=\frac{1}{2}\sum_{l=1}^k \frac{1}{\tilde{Z}_l} \qquad \frac{\expec{H_k}}{\tilde{Z}_k}=\sum_{l=0}^k \frac{1}{\tilde{Z}_l}.\]

The two series converge very rapidly, and the limits, say $\gamma$ and $\eta$ satisfy $\eta=2\gamma+1$ and are transcendental. They are already well approximated by their value at $k=5$, namely $\gamma\simeq 2.5625019073486328125$ and $\eta\simeq 6.125003814697265625$.\footnote{For the last one, the difference between the approximation and the limit starts with $19733$ zeroes after the decimal point.} The transcendence is easily established because the series are of Liouville type: writing $\tilde{Z}_k=2^{n_k}$ we find that $n_k$ is an integer, positive for $k\geq 3$ and such that $n_{k+1}/n_k$ goes to $+\infty$ at large $k$, so that transcendence follows by the Liouville criterion (a really elementary sufficient condition, see e.g. \cite{baker2012} for a general reference or \cite{kac-ulam1968} p.9 to go straight to the matter). A slight refinement of this argument shows that the number $\prod_{\overline{\imath}\geq \overline{\jmath}}\left(1-\frac{1}{Z_{\overline{\imath}}}  \right)$ encountered in \autoref{i:wtl} is a Liouville number (hence transcendental) for each $\overline{\jmath}$.

Notice that another tree model, with a root having $\frac{Z_{k-1}}{2}$ descendants in average (as for random pure sets) but with independent descendants (not as for random pure sets) each with $\frac{Z_{k-2}}{2}$ descendants in average and so on would predict the same  average number of nodes in the tree (which is the total number of chains in the random pure set model).
}

\myitem{Redundancy of the standard braces representation\label{i:rsbr}}{We can now estimate the redundancy of the standard braces representation. Recall that in that representation chains are in bijections with pairs of matching braces, i.e. the number of braces is $2H_k$. Thus a pure set in ${\mathcal S}_k$, encoded by a binary string of length $Z_{k-1}$, has a standard braces representation of average length $2\tilde{Z}_k\sum_{l=0}^k \frac{1}{\tilde{Z}_l}$. Rewriting $2\tilde{Z}_k=Z_{k-1}\prod_{l=0}^{k-2}\frac{Z_l}{2}$, we infer that the redundancy is logarithmic: though very large as a function of $k$ even for moderate $k$, the redundancy $\prod_{l=0}^{k-2}\frac{Z_l}{2}$ is only logarithmic in $Z_{k-1}$, the length of a generically incompressible representation of a pure set in ${\mathcal S}_k$.}

\vspace{.3cm}
In the next section we shall use generating functions and recursion relations to have a flexible access also to (co)variances of the number of chains of different lengths. One of the consequences will be that the number of braces in the standard braces representation is generically (i.e. not only in average) logarithmically redundant compared to the binary string representation, see \autoref{i:rsbrv}.

To show the usefulness of generating functions, we go to our next two examples.

\myitem{Chains of length $1$ and maximal chains of length $1,2$\label{i:hungdeux}}{Counting chains of length $1$ (ending at $x \in {\mathcal S}_k$) is the same as counting elements of $x$: $H^{(1)}_k(x) =\sum_{y\in x} 1$. If one assigns a weight $u_1$ to each such chain in each member of ${\mathcal S}_k$, the generating function is simply $(1+u_1)^{Z_{k-1}}$, an obvious fact already used before. Maximal chains of length $1$ are simple: $\nun$ is an element of half the samples of ${\mathcal S}_k$, $k\geq 1$. Maximal chains of length $2$ (ending at $x$) occur once for each $y$ such that $\nun \in y \in x$. Thus for $x\in {\mathcal S}_k$, $k\geq 2$ we have
\[
\begin{aligned}
H^{(1)}_k(x) =&\sum_{y\in x} 1=\ind{\nun\in x}+\sum_{y\in x,\, y\neq \nun} 1,\\
G^{(1)}_k(x) =&\ind{\nun\in x},\\
G^{(2)}_k(x) =&\sum_{y\in x} \ind{\nun\in y}=\sum_{y\in x,\, y\neq \nun} \ind{\nun\in y},\\
(H^{(1)}_k-G^{(2)}_k-G^{(1)}_k)(x) =&\sum_{y\in x,\, y\neq \nun} \ind{\nun\notin y}.  
\end{aligned}
\]
The last line is a linear combination of the previous ones. Among the elements of ${\mathcal S}_{k-1}$, $k\geq 2$, one is $\nun$, $Z_{k-1}/2-1$ differ from $\nun$ but do not contain $\nun$ and $Z_{k-1}/2$ contain $\nun$ (hence differ from $\nun$). Hence, taking $v_1,v_2$ as the weights of each maximal chain of length $1,2$, and $u_1$ as  the weight of each chain of length $1$, one finds for $k\geq 2$ (note that $k=0,1$ do no fit the pattern)
\begin{eqnarray*}
\sum_{x\in {\mathcal S}_k} u_1^{(H^{(1)}_k-G^{(2)}_k-G^{(1)}_k)(x)}(u_1v_1)^{G^{(1)}_k(x)}(u_1v_2)^{G^{(2)}_k(x)} & = & \\ & & \hspace{-5cm} (1+(u_1v_1))(1+u_1)^{Z_{k-1}/2-1}(1+(u_1v_2))^{Z_{k-1}/2}.
\end{eqnarray*}

This is a counting function, and to go to probabilistic expectations we just have to divide by $Z_k$.
The analysis is totally straightforward: $G^{(1)}_k$, $G^{(2)}_k$ and $H^{(1)}_k-G^{(2)}_k-G^{(1)}_k$ are independent under $\nu$ with $G^{(1)}_k$ a symmetric (i.e. parameter $1/2$) Bernoulli \rv\  and $G^{(2)}_k$, $H^{(1)}_k-G^{(2)}_k-G^{(1)}_k$ binomial \rv s of parameters $(1/2,Z_{k-1}/2)$ and $(1/2,Z_{k-1}/2-1)$ respectively. Concentrating on expectations and (co)variances we obtain
\[ \expec{G^{(2)}_k}=\frac{1}{2}\frac{Z_{k-1}}{2} \qquad \var{G^{(2)}_k}=\frac{1}{4}\frac{Z_{k-1}}{2},\]
\[ \expec{H^{(1)}_k-G^{(2)}_k-G^{(1)}_k}=\frac{1}{2}\left(\frac{Z_{k-1}}{2}-1\right) \qquad \var{H^{(1)}_k-G^{(2)}_k-G^{(1)}_k}=\frac{1}{4}\left(\frac{Z_{k-1}}{2}-1\right),\]
and
\[\cov{H^{(1)}_k-G^{(2)}_k-G^{(1)}_k,G^{(2)}_k}=0 \qquad \cov{H^{(1)}_k,G^{(2)}_k}=\frac{1}{4}\frac{Z_{k-1}}{2}. \]
In particular, the correlation between $H^{(1)}_k$ and $G^{(2)}_k$ is exactly $1/\sqrt{2}$ for $k\geq 2$. 

In the large $k$ limit, nothing happens to $G^{(1)}_k$ and the normalized\footnote{See \autoref{i:nrv}. In words normalizing a \rv\ means subtracting its mean and dividing by its standard deviation, i.e. the square root of its variance.} versions of $G^{(2)}_k$ and $H^{(1)}_k-G^{(2)}_k-G^{(1)}_k$ become independent $N(0,1)$ (that is, Gaussian with mean $0$ and variance $1$) \rv s.}

\vspace{.3cm}
We can also deal with chains of length $2$. 

\myitem{Chains of length $1$ and $2$\label{i:h1h2}}{Recall from the previous example that the generating function for chains of length $1$ in $x \in {\mathcal S}_k$) is simply $(1+u_1)^{Z_{k-1}}$. It is possible to take chains of length $2$ into account: for each $x \in {\mathcal S}_k$, each $y\in x$ accounts for $\# y$ chains of length $2$ in $x$, and $x$ is build by choosing among the $\binom{Z_{k-2}}{l}$ elements of ${\mathcal S}_{k-1}$ with size $l$.  Thus, assigning a weight  $u_2$ to each chain of length $2$ in each member of ${\mathcal S}_k$,  each $y\in x \in {\mathcal S}_k$ contributes a factor $u_2^{\# y}$ in the generating function, which henceforth reads in full $\prod_l(1+u_1u_2^l)^{\binom{Z_{k-2}}{l}}$ for $k\geq 2$ (though $k=0,1$ also fit, with a convention on binomial numbers). 

  Taking the logarithm yields a cumulant generating function for the joint law of the number of chains of length $1$ and of length $2$. Recall that we denote the corresponding random variables $H^{(1)}_k$ and $H^{(2)}_k$. Translated in probabilistic terms, the generating function above reads
\begin{equation} \expec{u_1^{H^{(1)}_k}u_2^{H^{(2)}_k}}=\frac{1}{Z_k} \prod_l(1+u_1u_2^l)^{\binom{Z_{k-2}}{l}}\label{eq:cgf12} \end{equation}
  We compute readily by taking the appropriate derivatives that
\[ \expec{H^{(1)}_k}=\frac{Z_{k-1}}{2} \qquad \expec{H^{(2)}_k}=\frac{Z_{k-1}}{2}\frac{Z_{k-2}}{2},\] 
\[\var{H^{(1)}_k}=\frac{1}{2}\frac{Z_{k-1}}{2} \qquad \var{H^{(2)}_k}=\frac{1}{2}\frac{Z_{k-1}}{2}\frac{Z_{k-2}}{2}\frac{Z_{k-2}+1}{2}\] and
\[\cov{H^{(1)}_k,H^{(2)}_k}=\frac{1}{2}\frac{Z_{k-1}}{2}\frac{Z_{k-2}}{2} .\]
In particular, the correlation between $H^{(1)}_k$ and $H^{(2)}_k$ is $\sqrt{\frac{Z_{k-2}}{Z_{k-2}+1}}$. It  goes to $1$ at large $k$. Thus, though $H^{(1)}_k$ and $H^{(2)}_k$ have wildly different orders of magnitude, the fluctuations of their normalized partners are very close, and as $H^{(1)}_k$ has Gaussian fluctuations, so must $H^{(2)}_k$.

In this simple instance we can do better. Introducing the normalized variables  $\widehat{H}^{(i)}_k:=\frac{H^{(i)}_k-\expec{H^{(i)}_k}}{\sqrt{\var{H^{(i)}_k}}}$ for $i=1,2$ we find that (after some changes of notation)
\begin{eqnarray*}
 \log \expec{e^{u_1 \widehat{H}^{(1)}_k+u_2 \widehat{H}^{(2)}_k}} & = & -u_1\frac{\expec{H^{(1)}_k}}{\sqrt{\var{H^{(1)}_k}}}-u_2\frac{\expec{H^{(2)}_k}}{\sqrt{\var{H^{(2)}_k}}}\\ & + & \sum_l \binom{Z_{k-2}}{l} \log \frac{1+e^{\frac{u_1}{\sqrt{\var{H^{(1)}_k}}}+l\frac{u_2}{\sqrt{\var{H^{(2)}_k}}}}}{2}
\end{eqnarray*}
and straightforward computations gives
\[ \lim_{k\to \infty} \log \expec{e^{u_1 \widehat{H}^{(1)}_k+u_2 \widehat{H}^{(2)}_k}}=\frac{(u_1+u_2)^2}{2}.\]
The previous results guarantees that this is correct up to second order, and a power counting argument is enough to show that the higher cumulants vanish in the large $k$ limit.

Thus, the joint law of the \rv s$\widehat{H}^{(i)}_k$ for $i=1,2$ has a degenerate Gaussian limit at large $k$.}

\vspace{.3cm}

One can obtain such explicit generating function taking into account chains up to length $2$ and maximal chains up to length $3$, see \autoref{i:mc3c2}. Chains of length $\geq 3$ and maximal chains of length $\geq 4$ do not lead to totally explicit generating functions. It is our aim in the following to compute the simplest features of their joint law in the large $k$ limit despite the lack of simple closed expressions.

In fact the previous two examples prepare for the general pattern. The penultimate example introduced the ``main actors'': $G^{(1)}_k$, $G^{(2)}_k$ and $H^{(1)}_k-G^{(2)}_k-G^{(1)}_k$. The last example showed that $H^{(2)}_k$ aligns with $H^{(1)}$: $H^{(2)}_k$ and $H^{(1)}$ are correlated strongly enough that their normalized versions coincide in the large $k$ limit.

Our central result, presented in \Autoref{sec:mr}, is that, in the large $k$ limit,  all chain counting functions except those involving $H^{(0)}_k$ (deterministic and equal to $1$), $G^{(1)}_k$ and $G^{(2}_k$ exhibit the ``alignment along $H^{(1)}$'' phenomenon. The proof will occupy most of the rest of this work.

\vspace{.3cm}

We finish this long list of examples with an independence result.

\myitem{Independence with respect to $G^{(1)}_k$\label{i:gunindep}}{For $k\geq 1$ there is a simple but non-trivial involution
\[ \jmath \colon {\mathcal S}_k \to  {\mathcal S}_k,\ x \mapsto \lb \begin{aligned}x \setmin \ndeux & \text{ if } \nun \in x\\ x \cup  \ndeux & \text{ if } \nun \notin x \end{aligned} \right. \]
with the property that $G^{(1)}_k$ takes values in $\{0,1\}$ with $G^{(1)}_k+G^{(1)}_k\circ \jmath=1$ (the constant function $1$) while the chain counting functions $G^{(n)}_k,H^{(n)}_k$, $n\geq 2$ are plainly invariant under $\jmath$. As $\nu_k$ is the uniform measure on ${\mathcal S}_k$, that is, is proportional to the counting measure, this proves that under $\nu_k$ the chain counting function $G^{(n)}_k,H^{(n)}_k$, $n\geq 2$ are independent of $G^{(1)}_k$. We have seen before that $H^{(1)}_k-G^{(1)}_k$ is also independent of $G^{(1)}_k$.}

\section{Weights and their generating function's}\label{sec:wgf}

To build a set $x\in {\mathcal S}_k$, we decide for each $y\in {\mathcal S}_{k-1}$ whether or not it will be an element of $x$. We want to use this simple yes-no picture to construct weight functions with nice properties.

\myitem{Weight functions\label{i:wf}}{We use collections of indeterminates $u_j,\overline{u}_j$, $\jN$ and $v_j,\overline{v}_j$, $\jN ^*$.

A weight function is the assignment of a monomial in these indterminates to every couple $(k,x)$ where $\kN$ and $x\, \in {\mathcal S}_k$.

A particular case is when the weight depends only on $x$, not on $k$, leading to the notion of uniform weight. Thus a uniform weight is simply an assignment of a monomial to every member of ${\mathcal S}_{\infty}$.}
    
Let us illustrate the notion.

\myitem{Uniform weight functions for maximal chains\label{i:uwfmc}}{For each $\lN$ we define a uniform weight function $w'_l$: for $x\in {\mathcal S}_{\infty}$,
\[ w'_l(x):=\prod_{\jN^*} v_{l+j}^{G^{(j)}(x)},\]
where the global chain counting function $G^{(j)}(x):=\#\left\{\text{Maximal chains of length } j \text{ ending at } x\right\}$ was introduced in \autoref{i:gdccf}.

Note that $w'_l$ is obtained from $w'_0$ simply by shifting the label of the $u$-indeterminates by $l$. This may seem redundant, but the usefulness stems from the recursion formula
\begin{equation}
\label{eq:mfdg}
w'_l(x)=\prod_{y\in x} w'_{l+1}(y) \text{ if } \nun \notin x \qquad w'_l(x)=v_{l+1}\prod_{y\in x} w'_{l+1}(y) \text{ if } \nun \in x,
\end{equation}
the translation at the level of weights of the recursion formula \autoref{eq:recurdirglob} for the chain functions $G^{(\cdot)}$. 
  
}

\myitem{Uniform weight functions for chains\label{i:uwfc}}{For each $\lN$ we define a uniform weight function $w_l$: for $x\in {\mathcal S}_{\infty}$,
\[ w_l(x):=\prod_{\jN} u_{l+j}^{H^{(j)}(x)},\]
where the global chain counting function $H^{(j)}(x):=\#\left\{\text{Chains of length } j \text{ ending at } x\right\}$ was introduced in \autoref{i:gdccf}.

Again, the recursion formula \autoref{eq:recurdirglob} for the chain functions $H^{(\cdot)}$ leads to a recursion formula
 a weight function $\overline{w}'_k$: for $x\in {\mathcal S}_k$,
\begin{equation}
\label{eq:mfdh}
w_l(x):=u_l\prod_{y\in x} w_{l+1}(y).
\end{equation}
}

\vspace{.3cm}
We observe that this framework is convenient to study chains of length $0,1,2,\cdots$ as $k$ varies. But it is also interesting to have control of the chains of length $k,k-1,k-2,\cdots$ as $k$ varies. To do that requires the use of non-uniform weights.
\myitem{Weight functions for inverse maximal chain functions\label{i:wfimcf}}{For each $\kN$ we define a weight function $\overline{w}'_k$: for $x\in {\mathcal S}_k$,
\begin{equation}
\label{eq:mfig}
\overline{w}'_k(x)=\prod_{\jN^*} \overline{v}_{j}^{\overline{G}^{(j)}_k(x)},\footnote{We do not introduce an independent indeterminate $\overline{v}_0$, equivalently we set $\overline{v}_0:=1$,  because it would be redundant with $\overline{u}_0$: in ${\mathcal S}_{k}$ a chain of length $k$ is automatically maximal.}  
\end{equation}
where the inverse chain counting function $\overline{G}^{(j)}_k(x)$ was introduced in \autoref{i:iccf}.

The recursion formula \autoref{eq:recurinv} translates into 
\[ \overline{w}'_k(x):=\prod_{y\in x} \overline{w}'_{k-1}(y) \text{ if } \nun \notin x \qquad \overline{w}'_k(x):=\overline{v}_k\prod_{y\in x} \overline{w}'_{k-1}(y) \text{ if } \nun \in x\]
for $\kN^*$.}

\myitem{Weight functions for inverse chain functions\label{i:wficf}}{For each $\kN$ we define a weight function $\overline{w}_k$: for $x\in {\mathcal S}_k$,
\[ \overline{w}_k(x)=\prod_{\jN} \overline{u}_{j}^{\overline{H}^{(j)}_k(x)},\]
where the inverse chain counting function $\overline{H}^{(j)}_k(x)$ was introduced in \autoref{i:iccf}.

The recursion formula \autoref{eq:recurinv} translates into
\begin{equation}
 \label{eq:mfih}
\overline{w}_k(x):=\overline{u}_k\prod_{y\in x} \overline{w}_{k-1}(y)  
\end{equation}
for $\kN^*$.
}

\vspace{.3cm}
The recursion relations for inverse weight functions are also valid for $k=0$ with the convention that an empty product is $1$ and that $\overline{v}_0=1$, without having to specify what inverse weight functions of index $-1$ mean.

We have now the basic building blocks needed to define the most general weight function that we need to study the statistics of chains and maximal chains.

\myitem{General weight function for chains and its generating function\label{i:gwfc}}{For $k,\lN$ we define $W_{k,l}$ on ${\mathcal S}_{k}$ by  $W_{k,l}:=w'_l w_l \overline{w}'_k\overline{w}_k$. We denote by $W_l$ the family $(W_{k,l})_{\kN}$ and by $W$ the family $(W_l)_{\lN}=(W_{k,l})_{k,\lN}$.

The generating function is defined as
\[ {\mathcal Z}_{k,l}:= \sum_{x\in {\mathcal S}_{k}} W_{k,l}(x). \]
} 

\vspace{.3cm}
The multiplicative structure of the weight function $W$ leads to factorization properties of the corresponding generating functions
\myitem{The factorization formula\label{i:ff}}{For $\kN^*$ 
\[ {\mathcal Z}_{k,l}=u_l\overline{u}_k (1+u_{l+1}\overline{u}_{k-1}v_{l+1}\overline{v}_{k-1}) \prod_{y\in {\mathcal S}_{k-1},\,y\neq \nun} (1+W_{k-1,l+1}(y)).\]
In case all the indeterminates for maximal chains, $v,\overline{v}$, are specialized to $1$, it specializes to
\[ {\mathcal Z}_{k,l}=u_l\overline{u_k}\prod_{y\in {\mathcal S}_{k-1}} (1+w_{l+1}(y)\overline{w}_{k-1}(y)).\]

\myproof Using \Autoref{eq:mfdg,eq:mfdh,eq:mfig,eq:mfih} we get 
\[ {\mathcal Z}_{k,l}= \sum_{x\in {\mathcal S}_{k}} (u_l\overline{u}_k) (v_{l+1}u_{l+1}\overline{v}_{k-1} \overline{u}_{k-1})^{\ind{\nun \in x}}\prod_{y\in x,y \neq \nun} W_{k-1,l+1}(y), \]
where we have used the explicit form of the weight functions on the empty set $\nun$. Now we recall a fact observed early on in this work: to build a set $x\in {\mathcal S}_{k}$ one simply chooses, for each $y\in {\mathcal S}_{k-1}$ whether $y$ belongs to $x$ or not. This leads directly to the factorization formula.

The specialized formula follows directly.
}

\vspace{.3cm}
Note that when all the indeterminates are specialized to $1$ the generating function ${\mathcal Z}_{k,l}$ reduces to the partition function $Z_k$ as it should.

\vspace{.3cm}
The factorization formula has an (implicit) recursive structure. We try to make it a bit less implicit.
\myitem{Recursive structure of the factorization formula\label{i:rsff}}{Define, for each monomial $\mathfrak{m}$ in the $u,v,\overline{u},\overline{v}$ indeterminates, a bookkeeping function $D_{k,l;\mathfrak{m}}:= \# \{x\in {\mathcal S}_k, W_{k,l}(x)=\mathfrak{m}\}$. Then ${\mathcal Z}_{k,l}=\sum_{\mathfrak{m}} D_{k,l;\mathfrak{m}} \mathfrak{m}$, and the factorization formula rewrites
\[ \sum_{\mathfrak{m}} D_{k,l;\mathfrak{m}} \mathfrak{m} = u_l\overline{u}_k\frac{1+u_{l+1}\overline{u}_{k-1}v_{l+1}\overline{v}_{k-1}}{1+u_{l+1}\overline{u}_{k-1}} \prod_{\mathfrak{m}} (1+{\mathfrak{m}})^{D_{k-1,l+1;\mathfrak{m}}},\]
were we need to divide by the contribution of $\nun$, $1+u_{l+1}\overline{u}_{k-1}$, to compensate for the fact that the product over $m$ includes this contribution. Knowing  $D_{k-1,l+1;\mathfrak{m}}$ for every $\mathfrak{m}$ we can expand the right-hand side using the binomial formula, collect the monomials and infer $D_{k,l;\mathfrak{m}}$ for every $\mathfrak{m}$. In particular, knowing $D_{k-1,l+1;\mathfrak{m}}$ for every $l$ and $\mathfrak{m}$, we infer $D_{k,l;\mathfrak{m}}$ for every $l$ and $\mathfrak{m}$. This is recursive in $k$, starting from the knowledge, for every $l$ and $\mathfrak{m}$, of $D_{0,l;\mathfrak{m}}$. But plainly $D_{0,l;\mathfrak{m}}=\ind{\mathfrak{m}=u_l\overline{u}_0 }$, as ${\mathcal S}_0$ contains nothing but $\nun$. Of course, getting explicit expressions for the $D_{k,l;\mathfrak{m}}$s is another matter.}

\myitem{The special case of inverse weight  functions\label{i:scicf}}{Suppose that only the inverse weight functions are taken into account, i.e. that all the $u,v$ variables are specialized to $1$.  Set $D_{k;\mathfrak{m}}:= \# \{x\in {\mathcal S}_k\, | \overline{w}'_k(x)\overline{w}_k(x)=\mathfrak{m}\}$ where $\mathfrak{m}$ now stands for an arbitrary monomial in the $\overline{u},\overline{v}$ variables. Then
\[\sum_{x\in {\mathcal S}_{k}} \overline{w}'_k(x)\overline{w}_k(x)=\sum_{\mathfrak{m}} D_{k;\mathfrak{m}} \mathfrak{m}=\overline{u}_k\frac{1+\overline{u}_{k-1}\overline{v}_{k-1}}{1+\overline{u}_{k-1}} \prod_{\mathfrak{m}} (1+{\mathfrak{m}})^{D_{k-1;\mathfrak{m}}},\]
which is directly recursive in $k$ by expansion via the binomial formula. This formula codes for all inverse chain functions, hence for all chain functions in fact ... as long as we do not let $k\to \infty$ carelessly.}

\vspace{.3cm}
The generic formul\ae\ are a bit complicated, so as a warm-up let us see the recursive structure on a simple case.
\myitem{Maximal chains up to length $3$ and/or chains up to length $2$\label{i:mc3c2}}{Specialize all indeterminates except $u_0,u_1,u_2,v_1,v_2,v_3$ to $1$. For $k\geq 3$\ \footnote{The values of ${\mathcal Z}_{k,0}$ for $k=0,1,2$ are easily computed by hand, and they do not always match with the generic formula, but our interest is in the large $k$ limit anyway.}
\begin{eqnarray} \label{eq:cgf123} {\mathcal Z}_{k,0} & = & u_0(1+u_1v_1)\prod_{(i,j)\neq (0,0)} (1+u_1u_2^iv_3^j)^{\binom{Z_{k-2}/2-1}{i-j}\binom{Z_{k-2}/2}{j}} \nonumber \\ & & \prod_{(i,j)}(1+u_1u_2^iv_2v_3^j)^{\binom{Z_{k-2}/2-1}{i-j-1}\binom{Z_{k-2}/2}{j}}.\end{eqnarray}
The value of $ {\mathcal Z}_{k,l}$ is obtained by shifting the indices of the indeterminates by $l$.

\myproof $W_{k,l}$ is identically $1$ for $l\geq 3$, leading via the factorization formula for the generating function to ${\mathcal Z}_{k,3}=Z_k$ for $k\geq 0$.  Thus ${\mathcal Z}_{k,2}=u_2(1+v_3)Z_k/2$ for $k\geq 1$ so $W_{k,2}$ takes value $u_2$ on $Z_k/2$ members of ${\mathcal S}_{k}$ (including $\nun$) and $u_2v_3$ on $Z_k/2$ members of ${\mathcal S}_{k}$. The factorization formula leads to ${\mathcal Z}_{k,1}=u_1(1+u_2v_2)(1+u_2)^{Z_{k-1}/2-1}(1+u_2v_3)^{Z_{k-1}/2}$ for $k\geq 2$ so $W_{k,1}$ takes value $u_1u_2^iv_3^j$ on $\binom{Z_{k-1}/2-1}{i-j}\binom{Z_{k-1}/2}{j}$ members of ${\mathcal S}_{k}$ ($i=j=0$ is for $\nun$) and $u_1u_2^iv_2v_3^j$ on $\binom{Z_{k-1}/2-1}{i-j-1}\binom{Z_{k-1}/2}{j}$ members of ${\mathcal S}_{k}$. A final use of the factorization formula leads to the announced value for ${\mathcal Z}_{k,0}$.

The consequence for the evaluation of ${\mathcal Z}_{k,l}$ is plain. 
}

\vspace{.3cm}
It seem that this is as far as one can go using only explicit and elementary combinatorial functions. This formula is still simple enough to derive a detailed result for the joint law of maximal chains up to length $3$ and/or chains up to length $2$ under $\nu_k$ in the large $k$ limit. However this is a bit cumbersome and we omit the discussion because we shall get more general results later by another approach. An explicit analysis of simple special cases ($u_0=u_2=v_3=1$ and $u_0=v_1=v_2=v_3=1$) was given in the last two examples at the end of \autoref{sec:probmod}.

\vspace{.3cm}
As another, alas less explicit, computation, we apply the factorization formula to the case of chains of length $k$ in members of ${\mathcal S}_k$ i.e. to the chain counting functions $H^{(k)}_k=\overline{H}^{(0)}_k$.
\myitem{The ``maximal'' maximal chain function\label{i:mmcf}}{Specialize all indeterminates except $\overline{u}_0$ to $1$. For $h\in{\mathbb N}$ set $D_{k;h}:=\# \lb x\in {\mathcal S}_k \, \left|  \, \overline{H}^{(0)}_k(x)=h\rb \right.$. Then $D_{0;h}=\ind{h=1}$ and for $k\geq 1$
\[ D_{k;h} = \sum_{(l_{\tilde h})\in \prod_{{\tilde h} \in {\mathbb N}} \intt{0,D_{k-1;{\tilde h}} },\, \sum_{{\tilde h}l_{\tilde h}=h} }\prod_{\tilde h} \binom{D_{k-1;{\tilde h}}}{l_{\tilde h}}. \]

\myproof This is just the awkward rewriting of 
\[ \sum_h D_{k;h} \overline{u}_0^h=\prod_{\tilde h} \left(1+\overline{u}_0^{\tilde h}\right)^{D_{k-1;{\tilde h}}}.\]
Iteration leads to ${\mathcal Z}_{1,l}=1+\overline{u}_0$, ${\mathcal Z}_{2,l}=2(1+\overline{u}_0)$, ${\mathcal Z}_{3,l}=4(1+\overline{u}_0)^2$, ${\mathcal Z}_{4,l}=2^4(1+\overline{u}_0)^8(1+\overline{u}_0^2)^4$,
\begin{eqnarray*} {\mathcal Z}_{5,l} & = & 2^{16}(1+x)^{128}(1+x^2)^{512}(1+x^3)^{1408}(1+x^4)^{3008}(1+x^5)^{5248} \\ & & (1+x^6)^{7680}(1+x^7)^{9600}(1+x^8)^{10336}(1+x^9)^{9600}(1+x^{10})^{7680} \\ & & (1+x^{11})^{5248}(1+x^{12})^{3008}(1+x^{13})^{1408}(1+x^{14})^{512}(1+x^{15})^{128}(1+x^{16})^{16},\end{eqnarray*}  and so on. Though the principle is straightforward, obviously no simple closed formula for generic $k$ is to be hoped, either for ${\mathcal Z}_{k,l}$ itself, or for the $D_{k;h}$s. It seems that ${\mathcal Z}_{k,l}$ is the $Z_{k-2}$-th power of a polynomial. The symmetry sending a pure set in  ${\mathcal S}_k$ to is complement is apparent.}

\vspace{.3cm}
As a last illustration, we consider the total chain counting functions 
\myitem{The case of total chain counting functions\label{ctcf}}{To concentrate on the total chain counting functions $G_k:=\sum_n G^{(n)}_k=\sum_n \overline{G}^{(n)}_k$ and $H_k:=\sum_n H^{(n)}_k=\sum_n \overline{H}^{(n)}_k$, we specialize to $\overline{u}_j=\overline{u}$ and $\overline{v}_j=\overline{v}$ for every $j$, and specialize all other indeterminates to $1$. Then ${\mathcal Z}_{0,l}=\overline{u}$ and
\[ {\mathcal Z}_{k,l}:= \sum_{x\in {\mathcal S}_k}\overline{v}^{\overline{G}_k(x)}\overline{u}^{\overline{H}_k(x)} =\overline{u}(1+\overline{v}\,\overline{u})\prod_{y\in {\mathcal S}_{k-1},\,y\neq \nun} \left(1+\overline{v}^{\overline{G}_{k-1}(x)}\overline{u}^{\overline{H}_{k-1}(x)}\right)\]
for $k\geq 1$. Again, the $l$ dependence drops out. 

This time we set $D_{k;g,h}:= \# \lb x\in {\mathcal S}_k \, | \, G_k(x)=g,\, H_k(x)=h\rb$ and find  $D_{0;g,h}=\ind{(g,h)=(0,1)}$ and for $k\geq 1$
\[\sum_h D_{k;g,h} \overline{v}^g\overline{u}^h=\overline{u}(1+\overline{v}\,\overline{u})\prod_{(g,h)\neq (0,1)}\left(1+ \overline{v}^g\overline{u}^h\right)^{D_{k-1;g,h}}\]
which we refrain from expanding via the binomial formula. By iteration ${\mathcal Z}_{1,l}=\overline{u}(1+\overline{v}\,\overline{u})$, ${\mathcal Z}_{2,l}=\overline{u}(1+\overline{v}\,\overline{u})(1+\overline{v}\,\overline{u}^2)$, ${\mathcal Z}_{3,l}=\overline{u}(1+\overline{v}\,\overline{u})(1+\overline{v}\,\overline{u}^2)(1+\overline{v}\,\overline{u}^3)(1+\overline{v}^2\,\overline{u}^4)$ and so on, when again no obvious general formula is to be expected.

\vspace{.3cm}
Every pure set with $\leq k+1$ pairs of braces shows up in ${\mathcal S}_k$ and taking the limit $k \to \infty$ in the above generating function is legitimate, leading to the enumeration of pure sets (of arbitrary depth) weighted by the number of chains and maximal chains. In the representation of pure sets by identity trees, maximal chains correspond to leaves and chains to nodes. The equation for $k\to \infty$, \[\sum_h D_{\infty,g,h} \overline{v}^g\overline{u}^h=\overline{u}(1+\overline{v}\,\overline{u})\prod_{(g,h)\neq (0,1)}\left(1+ \overline{v}^g\overline{u}^h\right)^{D_{\infty,g,h}},\]
has a unique solution as a formal power series in $\overline{u}$. The coefficient of each power of $\overline{u}$ is a polynomial in $\overline{v}$. The enumeration is, as expected, the one of identity trees, as can be found in \cite{oeis2025}, see A$055327$, or A$004111$ for the specialization to $\overline{v}=1$.
}

\vspace{.3cm}
The important lesson is that, despite the absence of closed formul\ae\ in general, the factorization formula allows to control by recursion the moments and cumulants for the chain counting functions and derive limit results at large $k$. This is the subject to which we turn next, starting with an illustrative example.

\section{An illustrative example} \label{sec:ie}

This section gives an application of generating functions to probabilistic computations relevant for our purpose. It illustrates the general strategy that we apply in section \autoref{sec:mcvcf} to compute the covariance of two generic chain functions. Along the way, we introduce a simple recursion relation that plays an important role, see \Autoref{i:rrc,i:spr}.

\myitem{Illustrative example\label{i:ie}}{For $\kN$ large enough, we aim to compute the correlations between the simplest chain function $H^{(1)}_k$, that counts the size of elements of  ${\mathcal S}_k$ and a function on ${\mathcal S}_k$, that we denote by $\varphi_k$. We assume that with respect to the parameter $k$ the functions $\varphi_k$ satisfy a recursion relation that ensures a recursive formula for the weights: for some $\nN$, $\varphi_k(x)=\sum_{y\in x}\varphi_{k-1}(y)$ for $k>n$ and $x\in {\mathcal S}_k$.\footnote{The functions $\varphi_k$, $k<n$ may well be defined, but our interest is in the large $k$ limit anyway.} This assumption is relevant for our discussion because, by \autoref{eq:recurinv}, any linear combination of inverse chain counting functions  $\overline{G}^{(m)}_k$ with $m<n$ and $\overline{H}^{(m)}_k$ with $m\leq n$ satisfies the property assumed of $\varphi_k$.}

\vspace{.3cm}

At the end of the section, we shall modify this example slightly to deal with $G_k$, the function counting every maximal chain (of any length) in a sample of  ${\mathcal S}_k$.
\myitem{Single step computation\label{i:ssc}}{We first concentrate on a single recursion step, so we simplify the notation. We let ${\mathcal S}$ be any finite non empty set with cardinal $Z$ endowed with the uniform probability measure, that we denote by $\nu_{\mathcal S}$ (expectation symbol ${\mathbb E}_{\mathcal S}$). We also endow $2^{\mathcal S}$, the set of subsets of $S$,  with the uniform probability measure, that we denote by $\nu_{2^{\mathcal S}}$ (expectation symbol ${\mathbb E}_{2^{\mathcal S}}$). Let $\varphi$ be any numerical function on ${\mathcal S}$ and define a numerical function $\Phi$ on $2^{\mathcal S}$ by $\Phi(x):=\sum_{y\in x}\varphi(y)$. For $x\in 2^{\mathcal S}$ we denote by  $\Gamma(x)$ the cardinal of $x$.

To make contact with the original questions, for some given $k>n$ ${\mathcal S}$ is to be thought of as ${\mathcal S}_{k-1}$ (then $2^{\mathcal S}$ as ${\mathcal S}_k$), $\varphi$ is as $\varphi_{k-1}$ (then $\Phi(x)$ as $\varphi_k$), and finally $\Gamma$ as $H^{(1)}_k$. 

Viewing $\varphi$ (resp. $\Phi$, $\Gamma$) as a \rv\ on ${\mathcal S}$ (resp. $2^{\mathcal S}$) we compute that 
\[ \expb{2^{\mathcal S}}{e^{u\Gamma+v\Phi}}=\frac{1}{2^{Z}}\sum_{x\subset\mathcal S}  e^{u\Gamma(x)+v\Phi(x)}=\prod_{y\in {\mathcal S}} \frac{1+e^{u+v\varphi(y)}}{2},\]
so that
\begin{eqnarray*}
\log \expb{2^{\mathcal S}}{e^{u\Gamma+v\Phi}} & = & Z\; \expb{\mathcal S}{\log\frac{1+e^{u+v\varphi}}{2}}\\ & = & Z \left( \frac{u}{2}+ \frac{v}{2}\expb{\mathcal S}{\varphi} + \expb{\mathcal S}{\log \cosh \frac{u+v\varphi}{2}}\right).
\end{eqnarray*}
Setting $v=0$ one retrieves the expected result that $\Gamma$ is the sum of $Z$ independent symmetric Bernoulli \rv s. In particular  $\expb{2^{\mathcal S}}{\Gamma}=Z/2$ and  $\varb{2^{\mathcal S}}{\Gamma}=Z/4$. Expanding in both variables up to second order, we find
\begin{equation} \label{eq:ssc} \expb{2^{\mathcal S}}{\Phi}=\frac{Z}{2}\expb{\mathcal S}{\varphi} \quad \varb{2^{\mathcal S}}{\Phi}=\frac{Z}{4}\expb{\mathcal S}{\varphi^2} \quad \covb{2^{\mathcal S}}{\Gamma,\Phi}=\frac{Z}{4} \expb{\mathcal S}{\varphi}. \end{equation}

In particular the correlation coefficient is
\[ \frac{\covb{2^{\mathcal S}}{\Gamma,\Phi}}{\sqrt{\varb{2^{\mathcal S}}{\Gamma}\varb{2^{\mathcal S}}{\Phi}}}=\frac{\expb{\mathcal S}{\varphi}}{\sqrt{\expb{\mathcal S}{\varphi^2}}}=\frac{\expb{\mathcal S}{\varphi}}{\sqrt{\expb{\mathcal S}{\varphi}^2+\varb{\mathcal S}{\varphi}}}=:\frac{1}{\sqrt{1+c_v^2}}.\]

The important feature of this result is that $Z$, the size of ${\mathcal S}$ does not appear explicitly. It expresses the correlation of $\Gamma$ and $\Phi$ in terms of $c_v=\frac{\sqrt{\var{\varphi}}}{\expec{\varphi}}$ and nothing else. 

\myitem{Variation coefficient\label{i:vc}}{If $A$ is a \rv\ on some probability space, with finite variance and non-zero expectation,  the variation coefficient of $A$ is defined as the ratio $\frac{\sqrt{\var{A}}}{\expec{A}}$.
}

The variation coefficient is a standard numerical characteristic of \rv s in statistics, and $c_v$ is nothing but the variation coefficient of $\varphi$. 

\vspace{.3cm}
In a situation when $\expb{\mathcal S}{\varphi}\neq 0$ (then $\expb{2^{\mathcal S}}{\Phi}=\frac{Z}{2}\expb{\mathcal S}{\varphi}$ is nonzero of the same sign) and the variation coefficient $c_v$ of $\varphi$ is small, we find that the correlation of $\Gamma$ and $\Phi$ is close to $\sgn (\expb{\mathcal S}{\varphi})$ so that the normalized fluctuations of $\Gamma$ and $\Phi$ are close (modulo a sign).

We keep the assumption that $c_v$ is small. If $Z$ is large, the fluctuations of $\Gamma$ are close to Gaussian, and so must be those of $\Phi$. This is a very standard fact, but let's make it explicit to explain the meaning we assign to this statement. Denoting as usual by $\widehat{\Gamma}$, $\widehat{\Phi}$ the normalized versions of $\Gamma$ and $\Phi$ we have 
\[ \expb{2^{\mathcal S}}{\widehat{\Gamma}^2}=\expb{2^{\mathcal S}}{\widehat{\Phi}^2}=1 \quad \expb{2^{\mathcal S}}{\widehat{\Gamma}\; \widehat{\Phi}}=(1+c_v^2)^{-1/2}\]  so $\expb{2^{\mathcal S}}{\left(\widehat{\Gamma}-\widehat{\Phi}\right)^2}=2(1-(1+c_v)^{-1/2})\simeq c_v^2$ and in terms of characteristic functions
$\left|\expb{2^{\mathcal S}}{e^{iu\widehat{\Phi}}}-\expb{2^{\mathcal S}}{e^{iu\widehat{\Gamma}}}\right|\lesssim 2|u|c_v$.\footnote{In detail,
$\left|\expb{2^{\mathcal S}}{e^{iu\widehat{\Phi}}}-\expb{2^{\mathcal S}}{e^{iu\widehat{\Gamma}}}\right| \leq \expb{2^{\mathcal S}}{\left|e^{iu\widehat{\Phi}}-e^{iu\widehat{\Gamma}}\right|}=\expb{2^{\mathcal S}}{2\left|\sin u(\widehat{\Phi}-\widehat{\Gamma})/2\right|} \leq 2|u| \expb{2^{\mathcal S}}{\left|\widehat{\Gamma}-\widehat{\Phi}\right|} \leq  2|u| \expb{2^{\mathcal S}}{\left(\widehat{\Gamma}-\widehat{\Phi}\right)^2}^{1/2}\simeq 2|u|c_v$.}
Thus, $\widehat{\Gamma}$ and $\widehat{\Phi}$ are close in quadratic mean, and their characteristic functions are close at least for moderate $u$.

We end the Item with a remark. There is a simple relation between the variation coefficient $C_v$ of $\Phi$ and that of $\varphi$. Indeed, from \autoref{eq:ssc} we get
\[ \frac{\varb{2^{\mathcal S}}{\Phi}}{\expb{2^{\mathcal S}}{\Phi}^2}=\frac{1}{Z}\left(1+ \frac{\varb{\mathcal S}{\varphi}}{\expb{\mathcal S}{\varphi}^2}\right),\]
that is $C_v^2=(1+c_v^2)/Z$. We have made heavy use of the assumption that $c_v$ was small. But if $Z$ is large, it is enough for the variation coefficient of $c_v$ of  $\varphi$ to be  small compared to $\sqrt{Z}$ to enforce that the variation coefficient $C_v$ of $\Phi$ is small.
}

\vspace{.3cm}
These considerations (in particular the remark closing the previous Item) become effective if we return to the intended interpretation. We assume that $\expec{\varphi_n}\neq 0$ and then $\expec{\varphi_k}$ is nonzero and of the same sign for $k\geq n$. Denoting by ${c_v}_k$ the variation coefficient of $\varphi_k$, the relation between the successive variation coefficients becomes ${c_v}_k^2=(1+ {c_v}_{k-1}^2)/Z$ for $k>n$. This motivates to introduce a recursion relation.

\myitem{The $R$ recursion relation\label{i:rrc}}{For given $\nN$ and $r$ (initial conditions), the sequence $(R_{n,k}(r))_{k\geq n}$ is defined by $R_{n,n}(r)=r$ and $R_{n,k}(r)=(1+R_{n,k-1}(r))/Z_{k-1}$ for $k>n$. We call this an $R$ recursion formula in the sequel.}

\vspace{.3cm}
Then
\[ \frac{\varb{\nu_k}{\varphi_k}}{\expb{\nu_k}{\varphi_k}^2}=R_{n,k}\left(\frac{\varb{\nu_n}{\varphi_n}}{\expb{\nu_n}{\varphi_n}^2}\right) \quad \text{ for } k>n.\]
We make a detour to list some simple but useful properties of $R$. As this is the only situation of interest for us we assume $r\geq 0$ from now on. 

\myitem{Some properties of $R$\label{i:spr}}{The $R$ recursion formula is easily solved by ``variation of the constant'', yielding\footnote{In the following formul\ae\ we make the usual convention that an empty product has value $1$ and an empty sum value $0$. Beware of boundary effects!}
  \[ R_{n,k}(r)= \frac{r}{\prod_{n\leq  j<k} Z_j}+ R_{n,k}(0)=\frac{r}{\prod_{n\leq  j<k} Z_j} +\sum_{n\leq  i <k} \frac{1}{\prod_{i\leq  j<k} Z_j} \text{ for } k\geq n. \]
 It is sometimes useful to rewrite this result in terms of the $\tilde{Z}_{\sbullet}$ (as defined in the end of \autoref{i:anc}) as
\[ R_{n,k}(r)= \frac{1}{2^k\tilde{Z}_k} \left( 2^n\tilde{Z}_nr +\sum_{n\leq j < k}2^j \tilde{Z}_j \right) \text{ for } k\geq n. \] 
  
We start with the case $r=0$, $n=0$. For $k\geq 1$, $R_{0,k}(0)$ is a sum of $k$ terms that are all $\leq 1/Z_{k-1}$ so $R_{0,k}(0)\leq k/Z_{k-1}\leq 1$ for $\kN$ (with $0/0=0$ for $k=0$). But $R_{n,k}(0)$ is a positive decreasing function of $n$ for fixed $k$, and we conclude
\[ 0 \leq R_{n,k}(0) \leq 1 \quad \text{ for } k\geq n \geq 0.\]
We use this to prime the pump. For $k\geq n+1$, $R_{n,k}(0)=(1+R_{n,k-1}(0))/Z_{k-1}$. Pursuing, for $k\geq n+2$, $R_{n,k}(0)=1/Z_{k-1}+(1+R_{n,k-2}(0))/(Z_{k-2}Z_{k-1})$ so $0\leq Z_{k-1}R_{n,k}(0)-1\leq 2/Z_{k-2}$. But $Z_{k-1}R_{n,k}(0)-1=0$ for $k=n+1$ so we conclude that
\begin{equation}
  \label{eq:Ras0} 0\leq Z_{k-1}R_{n,k}(0)-1\leq 2/Z_{k-2} \quad \text{ for } k\geq \max\{n+1,2\}.
\end{equation}
Returning to the general case $r\geq 0$, using the formula $R_{n,k}(r)= r/\prod_{n\leq  j<k} Z_j+ R_{n,k}(0)$ we infer
\begin{equation}
  \label{eq:Rasr}
  0\leq Z_{k-1}R_{n,k}(r)-1\leq (2+r)/Z_{k-2} \quad \text{ for } k\geq n+2.
\end{equation}
An important feature is that the bounds are uniform in $n$. In particular, the solution of an $R$ recursion goes to $0$ at large $k$ for fixed $n$.}

\vspace{.3cm}
Thus, the variation coefficient of ${c_v}_k$ of $\varphi_k$ goes to zero at large $k$. This forces $\widehat{H}^{(1)}_k-\widehat{\varphi}_k$ to go to $0$ in mean square because $\expec{\left(\widehat{H}^{(1)}_k-\widehat{\varphi}_k\right)^2}\simeq {c_v}_{k-1}^2$, and then also in probability and in law. In particular, as $\widehat{\Gamma}_k$ converges in law to a normalized Gaussian \rv , $\widehat{\varphi}_k$ will have to do the same. Beware that we do not claim that $\widehat{\Gamma}_k$ or $\widehat{\varphi}_k$ have a limit in mean square, in fact $\widehat{\Gamma}_k$ doesn't (a classical fact in the closely related context of the simple symmetric random walk, see also \autoref{i:rfmaineq}) and then $\widehat{\varphi}_k$ cannot either.

The previous discussion is enough to show that, in the various modes of convergence recalled above (in mean square, and then in probability and in law) the fluctuations of any inverse chain counting function align at large $k$ with the fluctuations of the simplest direct chain counting function, namely $H^{(1)}_k$. This implies that in fact all the fluctuations of all inverse chain functions align at large $k$, which is one of our central results.

\vspace{.3cm}
We close  this section with the case of the total number of chains, counted by $H_k=\sum_{n\in \intt{0,k}}  H^{(n)}_k=\sum_{n\in \intt{0,k}}  \overline{H}^{(n)}_k$, and of maximal chains, counted by $G_k=\sum_{n\in \intt{1,k}}  G^{(n)}_k=\sum_{n\in \intt{0,k-1}}  \overline{G}^{(n)}_k$. Dealing with the normalized fluctuations of $H_k$ and $G_k$ to show that they align with those of $H^{(1)}_k$ requires a minor adaptation of \Autoref{i:ie,i:ssc} to which the reader is referred for notations. 

\myitem{Illustrative example, continued\label{i:iee}}{To deal with the total number of chains, we replace the homogeneous recursion relation satisfied by the functions $\varphi_k$ by the inhomogeneous relation $\varphi_k(x)=1+\sum_{y\in x}\varphi_{k-1}(y)$ for $\kN$ and $x\in {\mathcal S}_k$. Setting $\varphi_0=1$, it is easily seen by recursion that  $\varphi_k=H_k$ for $\kN$.

For a single recursion step, $\Phi(x):=1+\sum_{y\in x}\varphi(y)$, leading to
\begin{eqnarray*}
\log \expb{2^{\mathcal S}}{e^{u\Gamma+v\Phi}} & = & v+ Z\; \expb{\mathcal S}{\log\frac{1+e^{u+v\varphi}}{2}}\\ & = & v+ Z \left( \frac{u}{2}+ \frac{v}{2}\expb{\mathcal S}{\varphi} + \expb{\mathcal S}{\log \cosh \frac{u+v\varphi}{2}}\right).
\end{eqnarray*}
The appearance of a lonely $v$ is the consequence of the inhomogeneous term in the recursion relation for the $\varphi_k$s. Expansion to second order in $u,v$ leads to
\[ \expb{2^{\mathcal S}}{\Phi}=1+ \frac{Z}{2}\expb{\mathcal S}{\varphi} \quad \varb{2^{\mathcal S}}{\Phi}=\frac{Z}{4}\expb{\mathcal S}{\varphi^2} \quad \covb{2^{\mathcal S}}{\Gamma,\Phi}=\frac{Z}{4} \expb{\mathcal S}{\varphi},\]
together with the familiar $\expb{2^{\mathcal S}}{\Gamma}=Z/2$ and  $\varb{2^{\mathcal S}}{\Gamma}=Z/4$

The formula for the correlation between $\Gamma$ and $\Phi$ is unchanged and reads
\[ \frac{\covb{2^{\mathcal S}}{\Gamma,\Phi}}{\sqrt{\varb{2^{\mathcal S}}{\Gamma}\varb{2^{\mathcal S}}{\Phi}}}=\frac{\expb{\mathcal S}{\varphi}}{\sqrt{\expb{\mathcal S}{\varphi^2}}}=\frac{\expb{\mathcal S}{\varphi}}{\sqrt{\expb{\mathcal S}{\varphi}^2+\varb{\mathcal S}{\varphi}}}.\]
Using $\expb{2^{\mathcal S}}{\Phi}\geq  \frac{Z}{2}\expb{\mathcal S}{\varphi}$ we retrieve an inequality for the variation coefficients
\[ \frac{\varb{2^{\mathcal S}}{\Phi}}{\expb{2^{\mathcal S}}{\Phi}^2} \leq \frac{1}{Z}\left(1+ \frac{\varb{\mathcal S}{\varphi}}{\expb{\mathcal S}{\varphi}^2}\right).\]

We translate these results in terms of our original problem. We substitute $\varphi_{k-1}=H_{k-1}$ for $\varphi$, which goes along with the substitution of $\varphi_k=H_k$ for $\Phi$ and $H^{(1)}_k$ for $\Gamma$.  We retrieve that the correlation between $H^{(1)}_k$ and $H_k$ equals $1/\sqrt{1+\var{H_{k-1}}/\expec{H_{k-1}}^2}$ while
\[\frac{\var{H_k}}{\expec{H_k}^2} \leq  \frac{1}{Z_{k-1}}\left(1+\frac{\var{H_{k-1}}}{\expec{H_{k-1}}^2}\right).\]
The last relation implies by recursion that  $\var{H_k}/\expec{H_k}^2 \leq R_{0,k}(1)$ for $\kN$ so that $\var{H_k}/\expec{H_k}^2$ goes to $0$ at large $k$. We conclude that the correlation between $H^{(1)}_k$ and $H_k$ goes to $1$, meaning that the difference $\widehat{H}_k-\widehat{H}^{(1)}_k$ of normalized fluctuations goes to $0$ in mean square as announced.

\vspace{.3cm}
To deal with the total number of maximal chains, we replace the homogeneous recursion relation satisfied by the functions $\varphi_k$ by the inhomogeneous relation $\varphi_k(x)=\ind{\nun \in x}+\sum_{y\in x}\varphi_{k-1}(y)$ for $\kN$ and $x\in {\mathcal S}_k$. Setting $\varphi_0=0$, it is easily seen by recursion that  $\varphi_k=G_k$ for $\kN$.

For a single recursion step, $\Phi(x):=\ind{\nun \in x}+\sum_{y\in x}\varphi(y)$, leading to
\begin{eqnarray*}
\log \expb{2^{\mathcal S}}{e^{u\Gamma+v\Phi}} & = & \log\frac{1+e^v}{2}+ Z\; \expb{\mathcal S}{\log\frac{1+e^{u+v\varphi}}{2}}\\ & = & v/2+ \log \cosh \frac{v}{2}+ Z \left( \frac{u}{2}+ \frac{v}{2}\expb{\mathcal S}{\varphi} + \expb{\mathcal S}{\log \cosh \frac{u+v\varphi}{2}}\right).
\end{eqnarray*}
The appearance of lonely $v$ terms is the consequence of the inhomogeneous term in the recursion relation for the $\varphi_k$s (and signs that $G^{(1)}_k$ is a Bernoulli \rv ). Expansion to second order in $u,v$ leads to
\[ \expb{2^{\mathcal S}}{\Phi}=\frac{1}{2}(1+ Z\expb{\mathcal S}{\varphi}) \quad \varb{2^{\mathcal S}}{\Phi}=\frac{1}{4}(1+Z\expb{\mathcal S}{\varphi^2}) \quad \covb{2^{\mathcal S}}{\Gamma,\Phi}=\frac{Z}{4} \expb{\mathcal S}{\varphi},\]
together with the familiar $\expb{2^{\mathcal S}}{\Gamma}=Z/2$ and  $\varb{2^{\mathcal S}}{\Gamma}=Z/4$.

The formula for the correlation between $\Gamma$ and $\Phi$ reads
\[ \frac{\covb{2^{\mathcal S}}{\Gamma,\Phi}}{\sqrt{\varb{2^{\mathcal S}}{\Gamma}\varb{2^{\mathcal S}}{\Phi}}}=\frac{\expb{\mathcal S}{\varphi}}{\sqrt{\expb{\mathcal S}{\varphi^2}+1/Z}}=\frac{\expb{\mathcal S}{\varphi}}{\sqrt{\expb{\mathcal S}{\varphi}^2+\varb{\mathcal S}{\varphi}+1/Z}}.\]

Thus there are some correction terms to the usual equations, that we have not found a quick and elegant way to deal with. Setting $\delta:=(\varb{\mathcal S}{\varphi}+1/Z)/\expb{\mathcal S}{\varphi}^2$, the correlation between $\Gamma$ and $\Phi$ rewrites $1/\sqrt{1+\delta}$. The analog of $\delta$ but for $\Phi$ is $\Delta:=(\varb{2^{\mathcal S}}{\Phi}+1/2^Z)/\expb{2^{\mathcal S}}{\Phi}^2$. We compute that
\[ \Delta = \frac{1}{Z}\frac{(Z\expb{\mathcal S}{\varphi})^2(1+\delta)+4Z/2^Z}{(1+ Z\expb{\mathcal S}{\varphi})^2}.\]
If $Z,\expb{\mathcal S}{\varphi},\delta$ are $\geq 0$ (as is the case in the application we have in mind), we obtain that for $Z \geq 4$ (i.e. $4Z/2^Z\leq 1$)
\[ (Z\expb{\mathcal S}{\varphi})^2(1+\delta)+4Z/2^Z\leq (1+\delta)(1+ Z\expb{\mathcal S}{\varphi})^2.\]
Consequently, for $Z\geq 4$, $\Delta \leq (1+\delta)/Z$.

We translate these results in terms of our original problem. We substitute $\varphi_{k-1}=G_{k-1}$ for $\varphi$, which goes along with the substitution of $\varphi_k=G_k$ for $\Phi$ and $H^{(1)}_k$ for $\Gamma$. In the same spirit we set $\delta_k:=(\var{\varphi_k}+1/Z_k)/\expec{\varphi_k}^2$. Then the correlation between  $G_k$ and $H^{(1)}_k$ rewrites $1/\sqrt{1+\delta_{k-1}}$ and for $k\geq 3$ (so $Z_{k-1} \geq 4$) the inequality  $\delta_k \leq (1+\delta_{k-1})/Z_{k-1}$ holds. This is the analog, with an inequality, of the $R$ recursion relation and we infer that $\delta_k \leq R_{2,k}(\delta_2)$ for $k\geq 2$. As the limit of an $R$ recursion relation is always $0$, in fact is a $O(1/Z_{k-1})$, we conclude that $\delta_k$ goes to $0$ at large $k$ and that the correlation between  $G_k$ and $H^{(1)}_k$ goes to $1$ with corrections that are $O(1/Z_{k-2})$. This ensures that $\widehat{G}_k-\widehat{H}^{(1)}_k$ converges to $0$ at large $k$ in mean square, as announced. 
}

\vspace{.3cm}
The result for $H_k$ has the following corollary, strengthening \autoref{i:rsbr}. 

\myitem{Redundancy of the standard braces representation, variance\label{i:rsbrv}}{We just established in \autoref{i:iee} that $\var{H_k}/\expec{H_k}^2 \leq R_{0,k}(1)$ for $\kN$. Recall $R_{0,k}(1)\sim 1/Z_{k-1}$ at large $k$. Rewriting this inequality in terms $B_k$, the number of braces in the standard braces representation of a pure set in ${\mathcal S}_k$ which satisfies $B_k=2H_k$, we get $\var{B_k}/\expec{B_k}^2 \leq R_{0,k}(1)$ so $\var{B_k}/\expec{B_k}^2 \lesssim 1/Z_{k-1}$ at large $k$. This implies a variety of statements exploiting that the fluctuations of $B_k$ are much smaller than $\expec{B_k}$ at large $k$. To give but the most elementary, we use the Markov inequality: for $\varepsilon >0$, the probability that $|B_k-\expec{B_k}|$  exceeds $\varepsilon \expec{B_k}$ is bounded by $\var{B_k}/(\varepsilon^2\expec{B_k}^2\lesssim 1/(\varepsilon^2Z_{k-1})$. Stated informally, at large $k$ most samples ${\mathcal S}_k$ have $|B_k -\expec{B_k}|$ no larger than $\sim \expec{B_k}/\sqrt{Z_{k-1}}$.

The binary string representation of a pure set in ${\mathcal S}_k$ has $Z_{k-1}$ bits and is generically incompressible. We can rephrase our result as follows using \autoref{i:rsbr}. At large $k$, the redundancy of the standard braces representation is logarithmic uniformly over ${\mathcal S}_k$: $B_k/Z_{k-1}\simeq \prod_{l=0}^{k-2}\frac{Z_l}{2}$ on most of ${\mathcal S}_k$.}

\section{Mean and (co)variance of chain functions}\label{sec:mcvcf}

The main goals of this section are to give explicit formul\ae\ for the (co)variances of the various chain functions. These formul\ae\ stem from the multiplicative formula for the generating functions, which yield recursion relations related to the nested structure of pure sets. The explicit computations are a bit repetitive and their essence transpires already in the illustrative example  of the previous section. Adding that they are technical in nature, the reader is advised to look at \autoref{i:ecvccf}, which lists the results (repeating the formul\ae\ already established for the means) and to skip the rest of this section (giving the detailed justifications) on a first reading.

\myitem{Formul\ae\ for expectations and (co)variances of chain counting function\label{i:ecvccf}}{\\
The (co)variances involve solutions of the $R$ recursion, which was defined in \autoref{i:rrc}. 

\vspace{.3cm}
\monemph{Expectations}
\begin{equation}
\label{eq:mainexp}  
\begin{aligned}\expec{H^{(n)}_k} & =\ind{k\geq n} \frac{\tilde{Z}_k}{\tilde{Z}_{k-n}} \text{ for } n\geq 0 \\ \expec{G^{(n)}_k} & =\ind{k\geq n}\frac{\tilde{Z}_k}{2\tilde{Z}_{k-n+1}} \text{ for } n\geq 1 .
\end{aligned}
\end{equation}
\monemph{Variances}
\begin{equation}
\label{eq:mainvar}
\begin{aligned}  
\var{H^{(n)}_k} & = \expec{H^{(n)}_k}^2R_{k-n,k}(0) \quad \text{ for } k\geq n\geq 0,\\
\var{G^{(n)}_k} & = \expec{G^{(n)}_k}^2R_{k-n+1,k}(1) \quad \text{ for }k\geq n\geq 1 .
\end{aligned}                        
\end{equation}
\monemph{Covariances}
\begin{equation}
\label{eq:maincov}
\begin{aligned}
 \cov{H^{(n)}_k,H^{(n')}_k} & = \expec{H^{(n)}_k}\expec{H^{(n')}_k}R_{k-\min \{n,n'\},k}(0) \\
 \text{ for } & k \geq n,n' \geq 0 ,\\
 \cov{G^{(n)}_k,G^{(n')}_k} & = \expec{G^{(n)}_k}\expec{G^{(n')}_k}R_{k-\min \{n,n'\}+1,k}(0)\\
 \text{ for } & k \geq n\neq n' \geq 1,\\
\cov{H^{(n)}_k,G^{(n')}_k}  & = \expec{H^{(n)}_k}\expec{G^{(n')}_k} R_{k-\min \{n,n'\}+\ind{n'<n},k}(0) \\
\text{ for } & k \geq n\geq 0 \text{ and } k\geq n'\geq 1.
\end{aligned}
\end{equation}

\vspace{-.5cm}
}

\vspace{.3cm}
These expressions cover the case of direct chain functions but for the whole range of parameters, so they can be translated straightforwardly to cover both direct and inverse chain functions. In fact, they are established below via computations for inverse chain function. 

We start by rewriting the consequences of the multiplicative formula for the generating function in probabilistic terms, as in \autoref{i:ssc}. For our purpose, we only need to keep track of the ``inverse'' parameters and we set all direct parameters $u_j$, $\jN$ and $v_j$, $\jN ^*$ to $1$. The multiplicative formula specializes to
\[
\sum_{x\in {\mathcal S}_{k}} \prod_{\jN}\overline{u}_{j}^{\overline{H}^{(j)}_k(x)} \prod_{\jN ^*} \overline{v}_{j}^{\overline{G}^{(j)}_k(x)} = \overline{u}_k (1+\overline{u}_{k-1}\overline{v}_{k-1}) \prod_{y\in {\mathcal S}_{k-1},\,y\neq \nun} \left(1+\prod_{\jN}\overline{u}_{j}^{\overline{H}^{(j)}_{k-1}(y)} \prod_{\jN ^*} \overline{v}_{j}^{\overline{G}^{(j)}_{k-1}(y)}\right)  
\]
for $k\geq 1$. This formula rewrites 
\[
\sum_{x\in {\mathcal S}_{k}} \prod_{\jN}\overline{u}_{j}^{\overline{H}^{(j)}_k(x)} \prod_{\jN ^*} \overline{v}_{j}^{\overline{G}^{(j)}_k(x)} = \overline{u}_k \frac{(1+\overline{u}_{k-1}\overline{v}_{k-1})}{(1+\overline{u}_{k-1})} \prod_{y\in {\mathcal S}_{k-1}} \left(1+\prod_{\jN}\overline{u}_{j}^{\overline{H}^{(j)}_{k-1}(y)} \prod_{\jN ^*} \overline{v}_{j}^{\overline{G}^{(j)}_{k-1}(y)}\right)  
\]

This formula is still rather complicated. As we focus on expectations and (co)variances, we shall be able to further specialize but one or two ``inverse'' parameters to $1$. The strategy is then to expand to second order to get recursion relations, that will in general mix cumulants and moments. We include a derivation for the expectations, which comes from the first order, though they have already been computed by a different method.

In the discussion that follows, the $R$ recursion appears repeatedly. The reader is referred to \Autoref{i:rrc,i:spr} for its definition and basic properties, that we use freely in what follows. 

We introduce an ad-hoc terminology for convenience. It generalizes the notion of variation coefficient, recalled in \autoref{i:vc} and which is standard in statistics.

\myitem{Covariation coefficient\label{i:cvc}}{If $A,B$ are square integrable \rv s on some probability space, with $\expec{A},\expec{B}\neq 0$ we call the ratio $\frac{\cov{A,B}}{\expec{A}\expec{B}}=:{\mathbb C}\text{vc}(A,B)$ the covariation coefficient  of $A$ and $B$. Note that ${\mathbb C}\text{vc}(A,A)$ is the square of the variation coefficient of $A$.}

\vspace{.3cm}
The usefulness for us of this notion is that the covariation coefficient of two (maximal or not) chain functions satisfies the R recursion relation in the variable $k$ indexing the depth (as in ${\mathcal S}_k$). Only the initial condition makes the difference. The relation between the covariation coefficient and $R$ for our problem is obvious from the formul\ae\ for (co)variances recalled in \Autoref{eq:mainvar,eq:maincov}.

\myitem{Expectation for chains\label{i:efc}}{
We start with a single chain function, say $\overline{H}^{(n)}_k=:Y_k$ ($n$ is fixed). Introduce a new parameter $u$ with $\overline{u}_n=:e^u$ and specialize all other indeterminates to $1$. We shall use freely that $Y_k=\ind{k=n}$ for $k\in \intt{0,n}$. The factorization identity becomes
\[ \sum_{x\in {\mathcal S}_{k}} e^{uY_k(x)}=e^{u\ind{k=n}}\prod_{y\in {\mathcal S}_{k-1}}(1+e^{uY_{k-1}(y)}).\]
We divide both sides by $Z_k=2^{Z_{k-1}}$ and take the logarithm to get
\begin{eqnarray} \log \expec{e^{uY_k}} & = & u\ind{k=n}+\expec{\log \frac{1+e^{uY_{k-1}}}{2}} \nonumber \\ & = & u(\ind{k=n}+Z_{k-1}\expec{Y_{k-1}}/2)+Z_{k-1}\expec{\log \cosh uY_{k-1}/2}. \label{eq:fic1}
\end{eqnarray}
The right-hand side is even but for a linear term in the $u$ variable as expected: this is due to the symmetry of $\overline{H}^{(n)}_k$ under the exchange of an element of ${\mathcal S}_k$, i.e. a subset of ${\mathcal S}_{k-1} $, with its complement in ${\mathcal S}_{k-1} $. This property was noticed  at the end of \autoref{sec:psbasics} and used to compute first moments in section \autoref{sec:probmod}.

Anyway, expansion to first order in $u$ yields the recursion relation
\[ Y_0=\ind{n=0} \quad \expec{Y_k}=\ind{k=n}+\frac{Z_{k-1}}{2}\expec{Y_{k-1}}.\]
We infer that $\expec{Y_k}=\ind{k\geq n}\tilde{Z}_k/\tilde{Z}_n$. This gives $\expec{\overline{H}^{(n)}_k}=\ind{k\geq n}\tilde{Z}_k/\tilde{Z}_n$ and taking $n\to k-n$ leads to $\expec{H^{(n)}_k}=\ind{k\geq n} \tilde{Z}_k/\tilde{Z}_{k-n}$.\footnote{The pre-factor $\ind{k\geq n}$ in this relation comes from a hidden $\ind{n\geq 0}$ in the formula for $\expec{Y_k}$.} We retrieve a result already known from \autoref{sec:probmod}, 
\[ \expec{H^{(n)}_k} =\ind{k\geq n} \frac{\tilde{Z}_k}{\tilde{Z}_{k-n}} \text{ for } n\geq 0.\]
We have thus (re)established the first formula in \autoref{eq:mainexp}. 
}

\myitem{Variance for chains\label{i:vfc}}{The second order in $u$ in \autoref{eq:fic1} yields
\[\var{Y_k}=\frac{Z_{k-1}}{4}\expec{Y_{k-1}^2} \text{ i.e. } \var{Y_k}=\frac{Z_{k-1}}{4}\left(\var{Y_{k-1}}+ \expec{Y_{k-1}}^2\right)
.\]
For $k>n$ divide both sides by $\expec{Y_{k}}^2=\frac{Z_{k-1}^2}{4}\expec{Y_{k-1}}^2$ (that is non-zero) to get
\[ \frac{\var{Y_k}}{\expec{Y_{k}}^2}=\frac{1}{Z_{k-1}} \left(\frac{\var{Y_{k-1}}}{\expec{Y_{k-1}}^2}+1\right).\]
This is the $R$ recursion relation for the covariation coefficient $\cvc{Y_k,Y_k}$, with initial condition ($k=n$) $\cvc{Y_{n},Y_{n}}=\frac{\var{Y_n}}{\expec{Y_{n}}^2}=0$ and we infer that
\[ \var{\overline{H}^{(n)}_k}=  \expec{\overline{H}^{(n)}_k}^2R_{n,k}(0) \qquad \var{H^{(n)}_k}=  \expec{H^{(n)}_k}^2R_{k-n,k}(0).
\]
For $n=0,1$ we retrieve $\var{H^{(0)}_k}=0$ and $\var{H^{(1)}_k}=Z_{k-1}/4=\expec{H^{(1)}_k}/2$, as expected for a symmetric binomial \rv . However, for $n>1$ we find that $\var{H^{(n)}_k}\sim \expec{H^{(n)}_k}^2\frac{1}{Z_{k-1}} = \expec{H^{(n)}_k} \frac{\tilde{Z}_{k-1}}{2\tilde{Z}_{k-n}}$ so the variance, though much smaller than the square of the expectation, is much larger than the expectation itself.}

\myitem{Covariance for chains\label{i:cfc}}{We consider two chain functions $\overline{H}^{(n)}_k=:Y_k$, $\overline{H}^{(n')}_k=:Y'_k$ ($n,n'$ are fixed, we may and shall assume $n<n'$). Introduce two parameters $u,u'$ with $\overline{u}_n=:e^u$,$\overline{u}_{n'}=:e^{u'}$  and specialize all other indeterminates to $1$. We follow the same steps as before to get
\begin{eqnarray} \log \expec{e^{uY_k+u'Y'_k}} & = & u\ind{k=n}+u'\ind{k=n'}+\expec{\log \frac{1+e^{uY_{k-1}+u'Y'_{k-1}}}{2}}\\ & = & u(\ind{k=n}+Z_{k-1}\expec{Y_{k-1}}/2)+ u'(\ind{k=n'}+Z_{k-1}\expec{Y'_{k-1}}/2)\\ & &  + Z_{k-1}\expec{\log \cosh (uY_{k-1}+u'Y'_{k-1})/2}.\label{eq:fic2}
\end{eqnarray}
Expansion to first order in  both $u$ and $u'$ leads to $\cov{Y_k,Y'_k}=\frac{Z_{k-1}}{4}\expec{Y_{k-1}Y'_{k-1}}$, that is 
\[\cov{Y_k,Y'_k}=\frac{Z_{k-1}}{4}\left(\cov{Y_{k-1},Y'_{k-1}}+\expec{Y_{k-1}}\expec{Y'_{k-1}}\right).\]
Taking into account again that $n<n'$ and $Y'_k=\ind{k=n'}$ for $k\in \intt{0,n'}$, we take $k>n'$ and divide both sides by $\expec{Y_k}\expec{Y'_k} = \frac{Z_{k-1}^2}{4}\expec{Y_{k-1}}\expec{Y'_{k-1}}$. We recover the $R$ recursion relation, only this time starting at $n'$ and with initial condition $\cvc{Y_{n'},Y'_{n'}}=0$ (just as for $\cvc{Y'_k,Y'_k}$). We infer that for $k\geq n'$
\[ \frac{\cov{Y_k,Y'_k}}{\expec{Y_k}\expec{Y'_k}}=R_{n',k}(0)=\frac{\var{Y'_k}}{\expec{Y'_{k}}^2}.\]
The equality of the two extreme terms calls for a more illuminating explanation, but we have not found one. Note that the formula holds also for $n=n'$. In terms of direct chain functions the first equality reads 
\[ \frac{\cov{H^{(k-n)}_k,H^{(k-n')}_k}}{\expec{H^{(k-n)}_k}\expec{H^{(k-n')}_k}}=R_{n',k}(0) \quad \text{ for } k \geq n' \geq n \geq 0.\]
This is readily seen to be equivalent to the first formul\ae\ in \Autoref{eq:mainvar,eq:maincov}.

}

\vspace{.3cm}
We now turn to the case of maximal chains.

\myitem{Expectation for maximal chains\label{i:efmc}}{We start with a single maximal chain function, say $\overline{G}^{(n)}_k=:X_k$ ($n$ is fixed). Introduce a new parameter $v$ with $\overline{v}_n=:e^v$ and specialize all other indeterminates to $1$. We note that $X_k=0$ for $k\in \intt{0,n}$ and that $X_{n+1}$ is a symmetric Bernoulli random variable. The factorization identity becomes
\begin{eqnarray*} \sum_{x\in {\mathcal S}_{k}} e^{vX_k(x)} & = & \left(1+e^{v\ind{k=n+1}}\right)\prod_{y\in {\mathcal S}_{k-1}\,y\neq \nun}(1+e^{vX_{k-1}(y)})\\ & = & \frac{1+e^{v\ind{k=n+1}}}{2}\prod_{y\in {\mathcal S}_{k-1}}(1+e^{vX_{k-1}(y)}).
\end{eqnarray*}
It follows that 
\begin{eqnarray} \log \expec{e^{vX_k}} & = & \log \frac{1+e^{v\ind{k=n+1}}}{2}+Z_{k-1}\expec{\log \frac{1+e^{vX_{k-1}}}{2}}\\ & = & \frac{v}{2}(\ind{k=n+1}+ Z_{k-1}\expec{X_{k-1}}) \\ & & + \log \cosh \frac{v}{2}\ind{k=n+1} +Z_{k-1}\expec{\log \cosh \frac{v}{2}X_{k-1}}.\label{eq:fimc1}
\end{eqnarray}
Solving the recursion relation for the expectation (first order in $v$) yields $\expec{X_k}=\frac{1}{2}\ind{k\geq n+1}\tilde{Z}_k/\tilde{Z}_{n+1}$. This gives $\expec{\overline{G}^{(n)}_k}=\frac{1}{2}\ind{k\geq n+1}\tilde{Z}_k/\tilde{Z}_{n+1}$ and $\expec{G^{(n)}_k}=\frac{1}{2}\ind{k\geq n}\tilde{Z}_k/\tilde{Z}_{k-n+1}$ (valid for $n\geq 1$).}

\myitem{Variance for maximal chains\label{i:vfmc}}{The second order in $v$ in \autoref{eq:fimc1} gives the initial conditions $\var{X_k}=\frac{1}{4}\ind{k=n+1}$ for $k\leq n+1$ and the by now familiar recursion relation $\var{X_k}=\frac{Z_{k-1}}{4}\expec{X_{k-1}^2}$ for $k>n+1$, that translates into the $R$ recursion relation for the covariation coefficient. The differences with the previous cases are a different starting point and initial condition, $\cvc{X_{n+1},X_{n+1}}=1$ (not $0$). We obtain
\[ \var{\overline{G}^{(n)}_k}=  \expec{\overline{G}^{(n)}_k}^2R_{n+1,k}(1) \qquad \var{G^{(n)}_k}=  \expec{G^{(n)}_k}^2R_{k-n+1,k}(1).\]
The second equation is readily seen to be equivalent to the second formul\ae\ in \autoref{eq:mainvar}.
}  

\myitem{Covariance for maximal chains\label{i:cfmc}}{We consider two maximal chain functions $\overline{G}^{(n)}_k=:X_k$, $\overline{G}^{(n')}_k=:X'_k$ ($n,n'$ are fixed, we may and shall assume $n<n'$). Introduce two parameters $v,v'$ with $\overline{v}_n=:e^v$, $\overline{v}_{n'}=:e^{v'}$  and specialize all other indeterminates to $1$. We follow the same steps as before to get
\begin{eqnarray} \log \expec{e^{vX_k+v'X'_k}} & = & \log \frac{1+e^{v\ind{k=n+1}}}{2}+\log \frac{1+e^{v'\ind{k=n'+1}}}{2}+\expec{\log \frac{1+e^{vX_{k-1}+v'X'_{k-1}}}{2}}\\ & = & \frac{v}{2}(\ind{k=n+1}+ Z_{k-1}\expec{X_{k-1}})+\frac{v'}{2}(\ind{k=n'+1}+ Z_{k-1}\expec{X'_{k-1}})
\\ & & + \log \cosh \frac{v}{2}\ind{k=n+1} +\log \cosh \frac{v'}{2}\ind{k=n'+1} \\ & & + Z_{k-1}\expec{\log \cosh (vX_{k-1}+v'X'_{k-1})/2}.\label{eq:ficm2}
\end{eqnarray}
Expanding to first order in both $v$ and $v'$ and rewriting in terms of the covariation coefficient leads again to the $R$ recursion formula. The initial condition is $0$ but at $k=n'+1$ i.e. there is a shift. The outcome can be written
\begin{eqnarray*} 
\cov{\overline{G}^{(n)}_k,\overline{G}^{(n')}_k} & = & \expec{\overline{G}^{(n)}_k}\expec{\overline{G}^{(n')}_k}R_{n'+1,k}(0) \\ \cov{G^{(n)}_k,G^{(n')}_k} & = & \expec{G^{(n)}_k}\expec{G^{(n')}_k}R_{k-n'+1,k}(0),
\end{eqnarray*}
where the first equation is for $n<n'$ and the second for $n > n'$. 

The second formula is readily seen to be equivalent to the first formul\ae\ in \Autoref{eq:mainvar,eq:maincov}.

We retrieve that the correlation between $G^{(n)}_k$, $n >1$ and $G^{(1)}_k$ is $0$ (we have already seen that the two \rv s are independent in \autoref{sec:wgf}).
}

It remains to compute correlations between chain and maximal chain counting functions.

\myitem{Mixed chains covariance\label{i:mccc}}{We consider two chain functions $\overline{H}^{(n)}_k=:Y_k$, $\overline{G}^{(n')}_k=:X'_k$ ($n,n'$ are fixed). Introduce two parameters $u,v'$ with $\overline{u}_n=:e^u$, $\overline{v}_{n'}=:e^{v'}$  and specialize all other indeterminates to $1$. The factorization identity becomes 
\begin{eqnarray*}
\sum_{x\in {\mathcal S}_{k}} e^{uY_k(x)+v'X'_k(x)} & = & e^{u\ind{k=n}}\left(1+e^{u\ind{k=n+1}+v'\ind{k=n'+1}}\right)\prod_{y\in {\mathcal S}_{k-1}\,y\neq \nun}(1+e^{uY_{k-1}(y)+v'X'_{k-1}(y)})\\ & = & e^{u\ind{k=n}}\frac{1+e^{u\ind{k=n+1}+v'\ind{k=n'+1}}}{1+e^{u\ind{k=n+1}}}\prod_{y\in {\mathcal S}_{k-1}}(1+e^{uY_{k-1}(y)+v'X'_{k-1}(y)}). 
\end{eqnarray*}

We start with the case when $n'=n$. It leads to
\begin{eqnarray} \log \expec{e^{uY_k+v'X'_k}}  & = & u(\ind{k=n}+Z_{k-1}\expec{Y_{k-1}}/2)+ \frac{v'}{2}(\ind{k=n+1}+ Z_{k-1}\expec{X'_{k-1}})\\ & &  +\log \frac{\cosh \frac{u+v'}{2}\ind{k=n+1}}{\cosh \frac{u}{2}\ind{k=n+1}}+ Z_{k-1}\expec{\log \cosh (uY_{k-1}+v'X'_{k-1})/2}.\label{eq:fice}
\end{eqnarray}
Expanding to first order in both $u$ and $v'$  yields $\covb{\nu_{n+1}}{Y_{n+1},X_{n+1}}=1/4$ and a recursion relation that rewritten in terms of the covariation coefficient leads again to that of $R$. As $\expec{Y_{n+1}}=Z_n/2$ and $\expec{X_{n+1}}=1/2$ the initial condition is $r=1/Z_n$ at $k=n+1$. But $R_{n+1,k}(1/Z_n)=R_{n,k}(0)$ for $k\geq n+1$ so that $\cvc{Y_k,X'_k}=R_{n,k}(0)$.

Turning to the case  $n \neq n'$, we find 
\begin{eqnarray} \log \expec{e^{uY_k+v'X'_k}}  & = & u(\ind{k=n}+Z_{k-1}\expec{Y_{k-1}}/2)+ \frac{v'}{2}(\ind{k=n'+1}+ Z_{k-1}\expec{X'_{k-1}})\\ & &  +\log \cosh \frac{v'}{2}\ind{k=n'+1}+ Z_{k-1}\expec{\log \cosh (uY_{k-1}+v'X'_{k-1})/2}.\label{eq:ficd}
\end{eqnarray}
We follow the same steps as before to get an $R$ recursion formula for the covariation coefficient. For $n <n'$ we find $\cvc{Y_k,X'_k}=R_{n'+1,k}(0)$ if $k\geq n'+1$ while for $n' <n$ we find $\cvc{Y_k,X'_k}=R_{n,k}(0)$ if $k\geq n$.

Putting everything together, we obtain
\begin{eqnarray*}
 \cov{\overline{H}^{(n)}_k,\overline{G}^{(n')}_k} & = & \expec{\overline{H}^{(n)}_k}\expec{\overline{G}^{(n')}_k} \left\{\begin{array}{lcr} R_{n'+1,k}(0) & \text{for} & n < n' \\ R_{n,k}(0) & \text{for} & n' \leq  n \end{array} \right. \\
  \cov{H^{(n)}_k,G^{(n')}_k} & = & \expec{H^{(n)}_k}\expec{G^{(n')}_k}\left\{\begin{array}{lcr} R_{k-n'+1,k}(0) & \text{for} & n' < n \\ R_{k-n,k}(0) & \text{for} & n \leq  n' \end{array} \right.
\end{eqnarray*}
The second formula is readily seen to be equivalent to the third formula in \autoref{eq:maincov}.

}

This concludes the verification of \Autoref{eq:mainexp,eq:mainvar,eq:maincov}.

\myitem{Exact versus approximate normalization\label{i:evan}}{Normalizing random variables by subtracting the exact mean and dividing by the exact standard deviation is customary, but replacing the two first cumulants by appropriate approximations would not change the conclusions. In our context, we observe that according to \autoref{eq:mainvar} $\widehat{H}^{(n)}_k$ and $\widehat{G}^{(n)}_k$ rewrite 
  \[ \widehat{H}^{(n)}_k=\frac{1}{\sqrt{R_{k-n,k}(0)}}\left(\frac{H^{(n)}_k}{\expec{H^{(n)}_k}}-1\right) \quad  \widehat{G}^{(n)}_k=\frac{1}{\sqrt{R_{k-n+1,k}(0)}}\left(\frac{G^{(n)}_k}{\expec{G^{(n)}_k}}-1\right).\]
  The point is that, according to \autoref{eq:Rasr}, all the $R$ prefactors have the same asymptotics: for large $k$ all $R$s equal $1/Z_{k-1}(1+O(1/Z_{k-2}))$. Thus, to compare chain functions, instead of considering, say, $\widehat{H}^{(n')}_k-\widehat{H}^{(n)}_k$, we could work with the simpler $\sqrt{Z_{k-1}}\left(\frac{H^{(n')}_k}{\expec{H^{(n')}_k}}- \frac{H^{(n)}_k}{\expec{H^{(n)}_k}}\right)$. For instance, using \Autoref{eq:mainvar,eq:maincov} leads to
\begin{eqnarray*}
Z_{k-1}\expec{\left(\frac{H^{(n')}_k}{\expec{H^{(n')}_k}}- \frac{H^{(n)}_k}{\expec{H^{(n)}_k}}\right)^2} & = &Z_{k-1} (R_{k-\max \{n,n'\},k}(0)-R_{k-\min \{n,n'\},k}(0)) \\ & = & O(1/Z_{k-2})    
\end{eqnarray*}
a bound of quality comparable to 
\begin{eqnarray*}\expec{\left(\widehat{H}^{(n')}_k-\widehat{H}^{(n)}_k\right)^2}& =& 2\left(1-\sqrt{\frac{R_{k-\min \{n,n'\},k}(0)}{ R_{k-\max \{n,n'\},k}(0)}}\right)\\ & = & O(1/Z_{k-2}).
\end{eqnarray*}
The last formula is again a direct consequence of \autoref{eq:maincov}, and show sthat $\widehat{H}^{(n')}_k-\widehat{H}^{(n)}_k$ goes to $0$ in mean square at large $k$, a special case of the results established in \autoref{sec:lka}.

The other possible combinations of chain functions (an $H$ with a $G$, or two $G$s) lead to similar results. Obviously $\widehat{H}^{(n')}_k-\widehat{H}^{(n)}_k$ for instance is not normalized, and it is more convenient to work with the closely related $\frac{H^{(n')}_k}{\expec{H^{(n')}_k}}- \frac{H^{(n)}_k}{\expec{H^{(n)}_k}}$. We shall make use of this combination in \autoref{sec:ld}.}

\section{Large $k$ asymptotics}\label{sec:lka}

We start with a few simple consequences of \autoref{i:ecvccf}.

-- The formul\ae\ from  \autoref{i:ecvccf} are consistent with our earlier observations that $H^{(0)}_k$ is deterministic, and that $G^{(1)}_k$, $G^{(2)}_k$,  $H^{(1)}_k-G^{(1)}_k-G^{(2)}_k$ are independent \rv s with $G^{(1)}_k$ a symmetric Bernoulli \rv\ and $G^{(2)}_k$, $H^{(1)}_k-G^{(2)}_k-G^{(1)}_k$ binomial \rv s of parameters $(1/2,Z_{k-1}/2)$ and $(1/2,Z_{k-1}/2-1)$ respectively. In particular, $H^{(1)}_k$ is a binomial \rv s of parameters $(1/2,Z_{k-1})$. They are also consistent with the observation that $G^{(1)}_k$ is independent of all other chain counting functions (with the same $k$). These simple facts were established in \autoref{i:hungdeux} and \autoref{i:gunindep}. We conclude that normalized version of the pair $(G^{(2)}_k,H^{(1)}_k-G^{(2)}_k-G^{(1)}_k)$ converges in law at large $k$ towards a Gaussian pair of independent $N(0,1)$ (i.e. normalized Gaussian) \rv s, and $\sqrt{2}$ time the normalized version of $H^{(1)}_k$ converges to their sum. This can be rephrased as follows: the couple  $(\widehat{G}^{(2)}_k,\widehat{H}^{(1)}_k)$ converges in law at large $k$ towards a Gaussian vector $(U,V)$ of $N(0,1)$ \rv s with $\cov{U,V}=1/\sqrt{2}$ and the normalized version of $H^{(1)}_k-G^{(2)}_k-G^{(1)}_k)$ converges in law at large $k$ towards $\sqrt{2}V-U$.

-- As a consequence of the asymptotics of the R recursion relation, \autoref{i:spr}, the formul\ae\ for (co)variances show that, with the exception of $G^{(1)}_k$ (a Bernoulli \rv\ for any $k$) all the (co)variation coefficients go to $0$ at large $k$. This will play an important role when we discuss the Gaussian limit.

-- The various chain functions have generically different orders of magnitude at large $k$, be they measured by expectation or variance. There is one family of exception, the couples $(H^{(n)}_k,G^{(n+1)}_k)_{\nN}$. Indeed, for $n\geq 1$,  $\expec{H^{(n)}_k}$ grows without bounds at large $k$ and $\var{H^{(n)}_k} = \expec{H^{(n)}_k}^2R_{k-n,k}(0) \sim \expec{H^{(n)}_k}^2/Z_{k-1}$ at large $k$. However, one finds after a few simplifications that
\[ \expec{H^{(n)}_k-2G^{(n+1)}_k}=0 \qquad \var{H^{(n)}_k-2G^{(n+1)}_k}=\frac{\expec{H^{(n)}_k}^2}{\prod_{k-n\leq j <k} Z_j}
\] for $k \geq 1$. 

\vspace{.3cm}
We henceforth concentrate on consequences of \autoref{i:ecvccf} that we do not already know and require a more thorough explanation. 

Using \Autoref{eq:mainvar,eq:maincov} leads to the following formul\ae\ for the normalized versions of the chain counting functions:\footnote{Recall that by definition normalized versions have zero mean and unit variance, so covariance and correlation are one and the same.}
\begin{equation}
\label{eq:maincor}
\begin{aligned}
 \cov{\widehat{H}^{(n)}_k,\widehat{H}^{(n')}_k} & = \frac{R_{k-\min \{n,n'\},k}(0)}{\sqrt{R_{k-n,k}(0)R_{k-n',k}(0)  }} \text{ for } k \geq n,n' \geq 1 ,\\
 \cov{\widehat{G}^{(n)}_k,\widehat{G}^{(n')}_k} & =  \frac{R_{k-\min \{n,n'\}+1,k}(0)}{\sqrt{R_{k-n+1,k}(1)R_{k-n'+1,k}(1)}}\text{ for } k \geq n\neq n' \geq 1,\\
\cov{\widehat{H}^{(n)}_k,\widehat{G}^{(n')}_k}  & =  \frac{R_{k-\min \{n,n'\}+\ind{n'<n},k}(0)}{\sqrt{R_{k-n,k}(0)R_{k-n'+1,k}(1)}} \text{ for }  k \geq n,n'\geq 1.
\end{aligned}
\end{equation}

We return to the  compact notation. Recall that we have defined, for $\kN$,
\[ F^{(n)}_k:=\left\{ \begin{array}{ll} H^{(n)}_k & \text{ for } n\in \intt{0,k},\\ G^{(-n)}_k & \text{ for } n\in\intt{-k,-1}. \end{array}\right. \]
Using \Autoref{eq:Ras0,eq:Rasr} (plus elementary properties of the square root) we infer bounds that hold for $k\geq 2$,
\begin{equation}
\label{eq:maincoras}
\begin{array}{cl} \left|\cov{\widehat{F}^{(n)}_k,\widehat{F}^{(n')}_k} -1\right| \leq \frac{3}{Z_{k-2}} &\text{ for } n,n' \in \intt{-k,k}\setmin \{-2,-1,0\},\\
\left|\cov{\widehat{F}^{(n)}_k,\widehat{F}^{(-2)}_k} -\frac{1}{\sqrt{2}}\right| \leq \frac{3}{Z_{k-2}}  &\text{ for } n \in \intt{-k,k}\setmin \{-2,-1,0\}.\end{array}
\end{equation}
The ``$3$'' in $3/Z_{k-2}$ could be improved for specific values of $n,n'$ (for instance, but not only, if $n=n'$ obviously) but it is a simple uniform constant\footnote{Only the existence of a uniform constant matters.} in these inequalities. They can be completed by the information that $F^{(0)}_k$ is the constant $1$ for $k\geq 0$, and $F^{(-1)}_k$ is a symmetric Bernoulli \rv\ independent of  $F^{(n)}_k$ for $k\geq 1$ and $n \in \intt{-k,k'}\setmin \{-1\}$.

It is important to compute the consequences of the above bounds on linear combinations of chain functions. This leads to introduce
\myitem{The vector space of chain functions}{We let ${\mathfrak F}_k$ denote the vector subspace of ${\mathbb R}^{{\mathcal S}_k}$ generated by chain counting functions, that is ${\mathfrak F}_k:=\sum_{n\in \intt{-k,k}} {\mathbb R}F^{(n)}_k$. The sum is not direct in general: as seen in \autoref{i:linindep}, ${\mathfrak F}_0 ={\mathbb R}F^{(0)}_0$, ${\mathfrak F}_1 ={\mathbb R}F^{(-1)}_1 \oplus {\mathbb R}F^{(0)}_1 \oplus {\mathbb R}F^{(1)}_1$ and ${\mathfrak F}_k =\bigoplus_{n\in \intt{-k,k-2}} {\mathbb R}F^{(n)}_k$ for $k\geq 2$.

  We also introduce  $\widehat{\mathfrak F}_k:=\sum_{n\in {\mathbb I}_k} {\mathbb R}\widehat{F}^{(n)}_k$, $k\geq 2$ (recall  ${\mathbb I}_k:=\intt{-k,k}\setminus \{-2,-1,0\}$).

}

\vspace{.3cm}
The subspace $\widehat{\mathfrak F}_k$ plays an important role, as might be expected from \autoref{eq:maincoras}. It consists of centered \rv s.\footnote{Remember, ${\mathfrak F}_k$ contains the constant function $F^{(0)}_k$, that can be used to subtract expectations.} For $F:=\sum_{n\in {\mathbb I}_k}\alpha_n \widehat{F}^{(n)}_k\in \widehat{\mathfrak F}_k$ we have $\expec{F}=0$ and \autoref{eq:maincoras} yields immediately that
\begin{equation}|\expec{F^2}- (\sum_{n\in {\mathbb I}_k} \alpha_n)^2|\leq \frac{3}{Z_{k-2}} (\sum_{n\in {\mathbb I}_k} |\alpha_n|)^2.\label{eq:maineq} \end{equation}

Notice that $\expec{F}$ and $\expec{F^2}$ are intrinsic characteristics of $F$, but due to linear dependencies, the $\alpha_n$s are not. In particular, one can always arrange, for given $F$, that $\sum_{n\in {\mathbb I}_k} \alpha_n=0$ by using the relation $F^{(1-k)}_k+F^{(-k)}_k=F^{(k-1)}_k$.\footnote{The relation $F^{(-k)}_k=F^{(k)}_k$ does not help, though.} However, the price to pay is that $\sum_{n\in {\mathbb I}_k} |\alpha_n|$ will overwhelm  $Z_{k-2}$. Indeed, rewritten for normalized \rv s, the linear relation turns into $r_k(\widehat{F}^{(1-k)}_k+\widehat{F}^{(-k)}_k)=\widehat{F}^{(k-1)}_k$ where  $r_k:=\frac{1}{2}\sqrt{\frac{R_{2,k}(1)}{R_{1,k}(0)}}=\frac{1}{2}\sqrt{\frac{R_{1,k}(1)}{R_{1,k}(0)}}$.\footnote{The second equality holds because  $R_{1,2}(1)=R_{2,2}(1)$ which by the $R$ recursion relation implies $R_{1,k}(1)=R_{2,k}(1)$ for each $k\geq 2$.} Hence, adding $0=\alpha (\widehat{F}^{(k-1)}_k-r_k(\widehat{F}^{(1-k)}_k+\widehat{F}^{(-k)}_k)$ to $F$ adds $\alpha (1-r_k)$ to $\sum_{n\in {\mathbb I}_k} \alpha_n$. But the explicit solution of the $R$ recursion relation yields $1-r_k\sim \frac{1}{2} \frac{1}{\prod_{1\leq  j\leq k-2} Z_j}\ll  \frac{1}{Z_{k-2}}$ at large $k$ and it takes a huge $\alpha$ to counterbalance even a modest deviation of $\sum_{n\in {\mathbb I}_k} \alpha_n$ from $0$.

\autoref{eq:maineq} can be rephrased as expressing the mean square convergence (and the convergence in law) of a whole zoo of random processes. This requires to introduce some notation.

\myitem{The spaces  ${\mathfrak A}^1$, ${\mathfrak M}^1$}{Set ${\mathbb I}_{\infty}:={\mathbb Z}\setminus \{-2,-1,0\}$.

Let ${\mathfrak A}^1$ be the space of arbitrary real arrays $\alpha=(\alpha_{k,m})_{k\geq 2,\, m\in {\mathbb I}_{k}}$. We define a collection of linear maps ${\mathfrak p}_k\colon {\mathfrak A}^1 \to  \widehat{\mathfrak F}_k, \, \alpha \mapsto \sum_{m\in {\mathbb I}_k} \alpha_{k,m}  \widehat{F}^{(m)}_k$ and a collection of linear forms $s_k\colon {\mathfrak A}^1 \to  {\mathbb R}, \, \alpha \mapsto \sum_{m\in {\mathbb I}_k} \alpha_{k,m}$. We say that the array is positive if $\alpha_{k,m}\geq 0$ for $k\geq 2,\ m\in {\mathbb I}_{k}$, and denote by  $|\alpha|$ the positive array $(|\alpha_{k,m}|)_{k\geq 2,\, m\in {\mathbb I}_{k}}$.By construction, $\expec{{\mathfrak p}_k(\alpha)}$ is always zero, so that $\expec{{\mathfrak p}_k(\alpha)^2}$ is the variance. 

We introduce some relevant subspaces of ${\mathfrak A}^1$.
\begin{eqnarray*}
{\mathfrak M}^1 & := & \{ \alpha \in {\mathfrak A}^1| \lim_{k\to \infty} s_k(\alpha) \text{ exists and } \lim_{k\to \infty} s_k(|\alpha|)/Z_{k-2}=0 \},\\
  {\mathfrak M}^1_{\sigma} & := & \{ \alpha \in {\mathfrak A}^1| \lim_{k\to \infty} s_k(\alpha)=\sigma \text{ and } \lim_{k\to \infty} s_k(|\alpha|)/Z_{k-2}=0 \}.                         
  \end{eqnarray*}
If $\alpha \in {\mathfrak M}^1$ we set $s_{\infty}(\alpha)=\lim_{k\to \infty} s_k(\alpha)$. 
}

\vspace{.3cm}
With these definitions, we obtain immediately
\myitem{Reformulation of \autoref{eq:maineq}\label{i:rfmaineq}}{Let $\alpha$ be a member of ${\mathfrak A}^1$. Then for each $k\geq 2$
\begin{equation} \label{eq:rfmaineq}
|\expec{{\mathfrak p}_k(\alpha)^2}-s_k(\alpha)^2|\leq \frac{3s_k(|\alpha|)^2}{Z_{k-2}}.
\end{equation}
In particular, if $\alpha \in {\mathfrak M}^1$ then 
\[ \lim_{k\to +\infty} \expec{{\mathfrak p}_k(\alpha)^2}=s_{\infty}(\alpha)^2.\]
We do not claim that ${\mathfrak p}_k(\alpha)$ has a large $k$ limit in mean square for $\alpha \in {\mathfrak M}^1$. It does not generically. As a simple illustration, take $\alpha_{k,m}:=\ind{k\geq 2}\ind{m=1}$ so that  ${\mathfrak p}_k(\alpha)=\widehat{F}^{(1)}_k=\widehat{H}^{(1)}_k$. One checks that, for $k \geq 1$, $\expec{(\widehat{H}^{(1)}_k-\widehat{H}^{(1)}_{k-1})^2}=2(1-\sqrt{Z_{k-2}/Z_{k-1}})$ which goes to $2$ at large $k$,\footnote{A quick way to do this computation is to note that, as a process indexed by $k$, $2H^{1}_k-Z_{k-1}$ is a simple symmetric random walk sampled at time $Z_{k-1}$.} so that $((\widehat{H}^{(1)}_k)_{k\geq 0}$ is not a Cauchy sequence in mean square. But the fact that $\lim_{k\to +\infty} \expec{{\mathfrak p}_k(\alpha)^2}$ factorizes as the square of the linear form $s_{\infty}$ is stricking.

Moreover, if $\alpha \in {\mathfrak M}^1_0$  then $\lim_{k\to +\infty} \expec{{\mathfrak p}_k(\alpha)^2}=0$, that is, the process ${\mathfrak p}_k(\alpha)$ converges to $0$ in mean square at large $k$. As ${\mathfrak M}^1_0$ is a subspace of co-dimension one in ${\mathfrak M}^1$, this fact and the factorization noted above give a precise meaning to \autoref{enumitem:gen} of \autoref{i:lbcf} and can be rephrased informally as ``$\widehat{\mathfrak F}_k$ becomes effectively 1-dimensional in the large $k$ limit''.
}

\vspace{.3cm}

\myitem{Convergence in law\label{i:rfmaineqcor}}{If $\alpha \in {\mathfrak M}^1_{\sigma}$, then ${\mathfrak p}_k(\alpha)$ converges in law to an $N(0,\sigma^2)$ \rv\ that can be identified with $\sigma V$ where $V$ is the limit in law $V$ of $\widehat{F}^{1}_k=\widehat{H}^{1}_k$. More precisely, the couple $(\widehat{F}^{(-2)}_k,{\mathfrak p}_k(\alpha))$ converges in law to the Gaussian vector $(U,\sigma V)$ where $U$ the limit in law of $\widehat{F}^{-2}_k=\widehat{G}^{2}_k$.

\myproof This is because convergence in mean square implies convergence in law. Indeed the array $\alpha':=(\alpha_{k,m}-\sigma \ind{m=1})_{k\geq 2,\, m\in {\mathbb I}_{k}}$ belongs to ${\mathfrak M}^1_0$ so  ${\mathfrak p}_k(\alpha')={\mathfrak p}_k(\alpha)-\sigma \widehat{F}^{1}_k$ converges to $0$ in mean square at large $k$ by \autoref{i:rfmaineq}. In particular, at large $k$, the couple $(\widehat{F}^{(-2)}_k,{\mathfrak p}_k(\alpha))$ has the same limit in law as $(\widehat{F}^{(-2)}_k,\sigma \widehat{F}^{1}_k)$. \autoref{enumitem:undgauss} in \autoref{i:lbcf} follows.

As an important special case, take $(\sigma_i)_{i\in \intt{1,m}}\in {\mathbb R}^m$ and  $n_{k,i} \in {\mathbb I}_{k}$ for $k \geq 2$ and $i\in \intt{1,m}$. Set $\alpha_{k,l}:=\sum_{i\in \intt{1,m}} \sigma_i\ind{l=n_{k,i}}$ . As  $(\widehat{F}^{(-2)}_k,{\mathfrak p}_k(\alpha))$ converges in law to the Gaussian vector $(U,\sigma V)$ with $\sigma:=\sum_{i\in \intt{1,m}}\sigma_i$ for any $m$-tuple $(\sigma_i)_{i\in \intt{1,m}}$, we can infer that the $(m+1)$-tuple $(\widehat{F}^{(-2)}_k,(\widehat{F}^{(n_{k,i})}_k)_{i\in \intt{1,m}})$ converge in law to the degenerate $(m+1)$-dimensional Gaussian vector $(U,V,\cdots V)$.  
}

\vspace{.3cm}

\myitem{Almost sure convergence of processes\label{i:ascp0}}{Suppose that both $\sum_k s_k(|\alpha|)^2/Z_{k-2}$ and $\sum_k s_k(\alpha)^2$ are finite. Then ${\mathfrak p}_k(\alpha)$ converges to $0$ almost surely and in mean square.

\myproof The convergence in mean square has already be dealt with. Set $\delta_k:=s_k(\alpha)^2+ 3s_k(|\alpha|)^2/Z_{k-2}$. We get from \autoref{eq:rfmaineq} that, for $k\geq 2$, $\expec{{\mathfrak p}_k(\alpha)^2} \leq \delta_k$ with $\sum_{k\geq 2} \delta_k < +\infty$. Then almost sure convergence of ${\mathfrak p}_k(\alpha)$ to $0$ is a consequence of the following classical computation (take $\Theta_k:={\mathfrak p}_k(\alpha)^2$). This concludes the proof of \autoref{enumitem:asconv} in \autoref{i:lbcf}.
}

\myitem{A classical computation\label{i:acc}}{Let $(\Theta_k)_{k\geq 2}$ be a sequence of \rv s (on some unspecified probability space) such that $\expec{|\Theta_k|}\leq \delta_k$ with $\sum_{k\geq 2} \delta_k < +\infty$. Then $\Theta_k$ converges almost surely to $0$. Setting $\varepsilon_k:=\sum_{l\geq k} \delta_l$, the rate of convergence is (at least) arbitrarily close to that of $\varepsilon_k$ (in a sense made precise below). 

\myproof If $\delta_k=0$ for some $k$, then the corresponding $\Theta_k$ is almost surely $0$ as well, so we may and shall assume that $\delta_k>0$ for every $k$. For $k\geq 2$, set $\varepsilon_k:=\sum_{l\geq k} \delta_l$ which is finite by assumption and decreases to $0$. Let $f\colon [0,+\infty[ \to [0,+\infty[$ be (strictly) increasing, continuous with $f(0)=0$, and be differentiable on $]0,+\infty[$ with $\lim_{0^+} f' =+\infty$. Set  $\tilde{\varepsilon}_k:=\delta_k/(f(\varepsilon_k)-f(\varepsilon_{k+1}))$. From  $\delta_k=\varepsilon_k-\varepsilon_{k+1}$ we infer by the intermediate value theorem that the sequence $\tilde{\varepsilon}_k$ converges to $0$.

Let ${\mathbb P}$ denote the probability measure. By the Markov inequality, ${\mathbb P}(|\Theta_k|\geq \tilde{\varepsilon}_k)\leq \delta_k/\tilde{\varepsilon}_k=f(\varepsilon_k)-f(\varepsilon_{k+1})$, so that $\sum_{k\geq 2} {\mathbb P}(|\Theta_k|\geq \tilde{\varepsilon}_k)\leq f(\varepsilon_2)<+\infty$. The Borel lemma implies that, with probability $1$, $|\Theta_k|< \tilde{\varepsilon}_k$ from some (sample dependent in general) $k$ onward. As $\tilde{\varepsilon}_k$ converges to $0$, this yields  the almost sure convergence of $|\Theta_k|$ to $0$ with asymptotic rate at least that of $\tilde{\varepsilon}_k$.

For instance, taking $\theta$ in $]0,1[$ and $f(x):=x^{\theta}/\theta$ we get $\tilde{\varepsilon}_k< \varepsilon_k^{1-\theta}$ by the intermediate value theorem. Thus the rate of convergence of $|\Theta_k|$ to $0$ is at least $\varepsilon_k^{1-\theta}$ for every $\theta>0$, whatever small. This is what we mean by saying that the rate of convergence is arbitrarily close to that of $\varepsilon_k$.
}

\vspace{.3cm} 
Note that by taking other functions $f$ we could improve the rate a little bit. We included the discussion on the rate as a preparation for the natural examples of almost sure convergence below. We start with a simple fact.

\myitem{Convergence from positivity\label{i:cp}}{Let $\alpha \in {\mathfrak A}^1$ be positive and such that, from some $k$ on, ${\mathfrak p}_k(\alpha)$ has unit variance. Then $|1-s_k(\alpha)|\leq 3/Z_{k-2})$ at large $k$.\footnote{In fact, the proof shows that any number $>3/2$ would do instead of $3$ but this is of little use.} In particular, $\alpha\in {\mathfrak M}^1_1$. 

\myproof Indeed, we have from \autoref{eq:rfmaineq} that $|1-s_k(\alpha)^2|\leq 3s_k(|\alpha|)^2/Z_{k-2}$ for large enough $k$. As $\alpha$ is positive by assumption,  $s_k(\alpha)=s_k(|\alpha|)$ is a positive increasing sequence, hence has a limit. As $Z_{k-2}$ goes to infinity at large $k$, an infinite limit or a limit different from $1$ yield a contradiction, so $s_k(\alpha)$ must go to $1$ at large $k$. On the left-hand side $|1-s_k(\alpha)^2|\sim 2|1-s_k(\alpha)|$ while on the right-hand side $3s_k(|\alpha|)^2/Z_{k-2} \sim 3/Z_{k-2}$, leading to the announced bound. That $\alpha\in {\mathfrak M}^1_1$ is a plain consequence.
} 

\vspace{.3cm}

We use this to establish
\myitem{Almost sure convergence of processes\label{i:ascp}}{Let $\alpha, \beta  \in {\mathfrak A}^1$ be positive and such that, from some $k$ on, ${\mathfrak p}_k(\alpha)$  and ${\mathfrak p}_k(\beta)$ have variance $1$. Then $\alpha-\beta \in {\mathfrak M}^1_0$ and the difference ${\mathfrak p}_k(\alpha)-{\mathfrak p}_k(\beta)$ goes to $0$ at large $k$, in quadratic mean and almost surely.

\myproof Indeed, ${\mathfrak p}_k(\alpha)-{\mathfrak p}_k(\beta)={\mathfrak p}_k(\alpha-\beta)$. Using \autoref{i:cp} we observe that $s_k(|\alpha-\beta|)\leq s_k(\alpha)+s_k(\alpha)$ which goes to $1+1=2$ while $s_k(\alpha-\beta)$ goes to $1-1=0$, in fact $|s_k(\alpha-\beta)|\leq  6/Z_{k-2})$, at large $k$. So first  $\alpha-\beta \in  {\mathfrak M}^1_0$ and second, using \autoref{eq:rfmaineq}, $\expec{{\mathfrak p}_k(\alpha-\beta)^2}\leq 12/Z_{k-2}$. This gives the convergence in mean square of $\alpha-\beta$ to $0$ (which already followed from \autoref{i:rfmaineq}), but the rate allows to apply \autoref{i:acc}. Observe that $\sum_{k\geq 2} 12/Z_{k-2} <+\infty$ and in fact $\sum_{l\geq k} 12/Z_{l-2}\sim 12/Z_{k-2}$ at large $k$. So ${\mathfrak p}_k(\alpha)-{\mathfrak p}_k(\beta)$ converges to $0$ almost surely, and the rate is (at least) arbitrarilyy close to that of $Z_{k-2}^{-1/2}$. 
}
\vspace{.3cm}

The situation in \autoref{i:ascp} is not artificial. If $\beta$ is any element of ${\mathfrak A}^1$ we may be interested in the fluctuations of the linear combinations of chain functions $F_k(\beta):=\sum_{m\in {\mathbb I}_k} \beta_{k,m}  F^{(m)}_k$ at large $k$. On a natural scale, the fluctuations are given by the normalization. We assume that, at least from some $k$ on, $F_k(\beta)$ has strictly positive variance. We observe that the normalization $\widehat{F}_k(\beta)$ of $F_k(\beta)$ is then given by
\[ \widehat{F}_k(\beta)=\frac{1}{\sqrt{\var{F_k(\beta)}}}\sum_{m\in {\mathbb I}_k} \beta_{k,m}\sqrt{\var{F^{(m)}_k}}\widehat{F}^{(m)}_k.\]
In particular, we have $\widehat{F}_k(\beta)={\mathfrak p}_k(\alpha)$ for an appropriate $\alpha \in {\mathfrak A}^1$. Moreover, the components of $\beta$ and $\alpha$ have the same sign, so $\alpha$ is positive if $\beta$ is, and by construction, ${\mathfrak p}_k(\alpha) $ has unit variance, at least from some $k$ on.

The simplest chain function is arguably $F^{(1)}_k=H^{(1)}_k$ which simply counts the number of elements of a sample in ${\mathcal S}_k$. Thus the sequence $\alpha^{(1)}:=(\delta_{m,1})_{m\in {\mathbb I}_\infty}$, with associated process ${\mathfrak p}_k(\alpha^{(1)})=\widehat{F}^{(1)}_k$, is a convenient supplement to the hyperplane of equation $s_{\infty}=0$, and it is convenient to use $(H^{(1)}_k)_{k\geq 2}$ as a standard to compare the other normalized chain functions with.  As a corollary of \autoref{i:ascp} we obtain

\myitem{Important special cases of almost sure convergence\label{i:ascpcor}}{For $n\in{\mathbb I}_\infty$ the $k$-sequences $\widehat{F}^{(n)}_k-\widehat{F}^{(1)}_k$ converge to $0$ in quadratic mean and almost surely, which is the content of \autoref{enumitem:part} in \autoref{i:lbcf}. The same holds for the $k$-sequences $\widehat{F}^{>0}_k-\widehat{F}^{(1)}_k$ and $\widehat{F}^{< -2}_k-\widehat{F}^{(1)}_k$ where  $F^{>0}_k:=\sum_{k>0} F^{(n)}_k=H_k-H^{(0)}$ counts the chains of length $>0$ in members of ${\mathcal S}_k$ and  $F^{< -2}_k:=\sum_{k<-2} F^{(n)}_k=G_k-G^{(1)}_k-G^{(2)}_k$ counts the number of maximal chains of length $>2$ in members of ${\mathcal S}_k$. These two examples are closely related to the total chain counting functions. 

\myproof Indeed, it is plain to express $F^{(n)}_k$, $n\in{\mathbb I}_\infty$, $F^{>0}_k$ or $F^{< -2}_k$ as $F_k(\beta):=\sum_{m\in {\mathbb I}_k} \beta_{k,m}  F^{(m)}_k$ for an appropriate $\beta \in {\mathfrak A}^1$.

Of course, as convergence to $0$, in quadratic mean or almost sure, is stable under finite linear combinations, we get for free many other cases of convergence, for example that for $n,n'\in{\mathbb I}_\infty$ the $k$-sequences $\widehat{F}^{(n)}_k-\widehat{F}^{(n')}_k$ converge to $0$ in quadratic mean and almost surely.   
}

\vspace{.3cm}
An obvious remark is that $H^{(0)}_k$ is a constant, so $H_k-H^{(0)}_k$ and $H_k$ have the same normalized versions, and the above implies that $\widehat{H}_k$ is tied up at large $k$ with $\widehat{H}^{(1)}_k$ (and then with the other normalized $F^{(n)}_k$s,  $n\in{\mathbb I}_\infty$). It is slightly less clear, but true, that $\widehat{G}_k$ is also tied up at large $k$ with $\widehat{H}^{(1)}_k$: an explicit computation shows that at large $k$ the fluctuations of $G^{(1)}_k+G^{(2)}_k$ become negligible with respect those of $G_k$. We do not provide details because \autoref{i:iee} contains a direct proof, with bounds that in fact ensure that $\widehat{G}_k-\widehat{H}^{(1)}_k$ goes to $0$ at large $k$ not only in mean square but almost surely.

\section{Large deviations}\label{sec:ld}

This final section is of more heuristic and conjectural nature than the previous ones.

Our main focus up to now has been on Gaussian fluctuations. We have established the alignment at large $k$ of the normalized fluctuations of generic linear combinations of chain counting functions on those of the simplest one, $H^{(1)}_k$. We would like to go beyond in two (related) directions.

\vspace{.3cm}
The main is large deviations. The \textit{conjectural} message is that large deviations of chain functions exhibit a similar alignment phenomenon.

At the moment, we can support this conjecture via concrete computations only on very simple special cases, when we have a closed form of the cumulant generating functions at our disposal. This means essentially \autoref{eq:cgf123}. For simplicity, we restrict concrete computations to the simpler \autoref{eq:cgf12}.

Even for this particular case, we content mostly with the easy part of large deviation bounds. What we mean by this is the following: if $H$ is a \rv\ (on some unspecified probability space) such that $\expec{e^{uH}}$ is well-defined for $u\in [0,+\infty[$ then
\[\expec{e^{uH}}\geq \expec{e^{uH}\ind{H\geq h}} \geq e^{uh} p(H\geq h).\footnote{The fist inequality hold whenever $u$ is real and the \rv\ $e^{uH}$ has an expectation, but the last one requires  $u\geq 0$.}\]
This leads to 
\begin{equation}\label{eq:eldb}
  \log p(H\geq h)\leq \inf \lb \varphi_H(u)-hu,\, u \geq 0 \rb, \end{equation}
where $\varphi_H(u):=\log \expec{e^{uH}}$. 

\vspace{.3cm}
The second direction we would like to explore concerns sub-leading fluctuations.
The general context is the following. One can view $H^{(n)}_k$ or $G^{(n)}_k$ as the size of a sample at the $n^{\text{th}}$ level of a sample in ${\mathcal S}_k$ and the statistics of $H^{(1)}_k$, which is more easily accessible, as a good proxy for the statistics of $H^{(n)}_k$ or $G^{(n)}_k$. This is justified by a special case of our main result: $\widehat{H}^{(n)}_k-\widehat{H}^{(1)}_k$ and $\widehat{G}^{(n)}_k-\widehat{H}^{(1)}_k$ go to $0$ in mean square at large $k$. However, this proxy is only an approximation because the \rv s $\widehat{H}^{(n)}_k-\widehat{H}^{(1)}_k$ or  $\widehat{G}^{(n)}_k-\widehat{H}^{(1)}_k$ still fluctuate, albeit with a smaller scale, at large $k$. We expect that this scale is controlled by the rate of convergence to $0$ in mean square and that the fluctuations are asymptotically Gaussian. The remarks in \autoref{i:evan} allow in fact to use simpler but essentially equivalent combinations, replacing for instance the study of $\widehat{H}^{(n)}_k-\widehat{H}^{(1)}_k$ by that of $\sqrt{Z_{k-1}}\left(\frac{H^{(n)}_k}{\expec{H^{(n)}_k}}- \frac{H^{(1)}_k}{\expec{H^{(1)}_k}}\right)$. As for large deviations, concrete computations are limited to cases when we have a closed form of the cumulant generating functions at our disposal. 

\vspace{.3cm}
To understand these severe limitations, note that we established a generic Gaussian asymptotic behavior of linear combinations of chain functions in a rather indirect way. The first chain function, $H^{(1)}_k$, has a binomial law whose Gaussian asymptotic behavior is a consequence of the usual central limit theorem,\footnote{In fact, the oldest occurrence of the CLT due to Abraham De Moivre, dating back to $1733$!} and the result for general chain functions is via the alignment with $H^{(1)}_k$. We have not been able to use directly the recursion relations satisfied by  cumulant generating functions\footnote{For example \autoref{eq:fic1}, remember $Y_k:=H^{(n)}_k$.} to establish simply the Gaussian asymptotics, though the result is suggested by scaling arguments.

\vspace{.3cm}
We can now turn to concrete computations. We rewrite \autoref{eq:cgf12} with some obvious changes of notation as
\[\expec{e^{u_1H^{(1)}_k+u_2H^{(2)}_k}}=\frac{1}{Z_k} \prod_l(1+e^{u_1+lu_2})^{\binom{Z_{k-2}}{l}}.\]
For later use, it is often convenient to remove expectations, i.e. to center, leading to
\begin{equation}  \label{eq:cgf12exp} \expec{e^{u_1(H^{(1)}_k-m^{(1)}_k)+u_2(H^{(2)}_k-m^{(2)}_k)}}=\prod_l(\cosh (u_1+lu_2)/2)^{\binom{Z_{k-2}}{l}},\end{equation}
where $m^{(1)}_k:=\expec{H^{(1)}_k}=Z_{k-1}/2$ and $m^{(2)}_k:=\expec{H^{(2)}_k}=Z_{k-1}Z_{k-2}/4$ (see e.g. \autoref{i:h1h2}).

\vspace{.3cm}
We start with a classical result to illustrate our naive approach.

\myitem{Large deviations of $H^{(1)}_k$\label{i:ldhun}}{The chain function $H^{(1)}_k$ is the sum of $Z_{k-1}$ symmetric Bernoulli random variables, so $\expec{e^{u(H^{(1)}_k-m^{(1)}_k)}}=(\cosh u/2)^{2m^{(1)}_k}$, as follows also from \autoref{eq:cgf12exp}. Using \autoref{eq:eldb} for $H:=H^{(1)}_k-m^{(1)}_k$ and $h=xm^{(1)}_k$ we infer
\begin{equation}\label{eq:ldhun}
\frac{1}{m^{(1)}_k} \log \mu(H^{(1)}_k-m^{(1)}_k\geq x m^{(1)}_k) \leq \inf \lb 2\log \cosh u/2-ux,\, u\geq 0\rb.   
\end{equation} 
For $x < 0$ the infimum is $0$ (reached for $u=0$, that is, on the boundary), a trivial bound. For $x > 1$ the infimum is $-\infty$ (reached for $u\to +\infty$), which fits with the obvious fact that $H^{(1)}_k\leq 2m^{(1)}_k$. For $x=1$, $\frac{1}{m^{(1)}_k} \log \mu(H^{(1)}_k-m^{(1)}_k\geq m^{(1)}_k) =\frac{1}{m^{(1)}_k} \log \mu(H^{(1)}_k=2m^{(1)}_k)=-2\log 2$. Finally, for $0\leq x < 1$ the infimum is reached for the only solution $u$ of the equation $\tanh u/2=x$, leading to
\[ \frac{1}{m^{(1)}_k} \log \mu(H^{(1)}_k-m^{(1)}_k\geq x m^{(1)}_k) \leq -(1+x)\log (1+x)+(1-x)\log (1-x).\]
When $x\to 1$ the right-hand side goes to $-2\log 2$ which equals the left-hand side  at $x=1$. The case of large deviations on the opposite side, i.e. of $\mu(H^{(1)}_k-Z_{k-1}/2\leq x Z_{k-1}/2)$ is dealt with similarly, and the final formul\ae\ are the same.  

Note that the right-hand side of \autoref{eq:ldhun} does not depend on $k$ and in particular
\[ \limsup_{k\to +\infty} \frac{1}{m^{(1)}_k} \log \mu(H^{(1)}_k-m^{(1)}_k\geq x m^{(1)}_k) \leq \inf \lb 2\log \cosh u/2-ux,\, u\geq 0\rb.\] 
Of course, in the case at hand, $H^{(1)}_k-m^{(1)}_k$ can be seen as a sum of independent identically distributed random variables with a finite cumulant generating function on the real line (here $\log \cosh u/2$ for $u\in {\mathbb R}$) so that  by Cram\'er's theorem the  $\limsup$ on the right-hand can be replaced by a $\lim$ and the inequality by an equality. In more complicated cases, we shall content with simple but widely applicable manipulations as above.
}

\myitem{Bounds on large deviations of $H^{(2)}_k$\label{i:ldeux}}{From \autoref{eq:cgf12exp} we infer
\[ \begin{aligned}
\expec{e^{um^{(1)}_k(H^{(2)}_k-m^{(2)}_k)/m^{(2)}_k}}& =\prod_l(\cosh ((lum^{(1)}_k)/(2m^{(2)}_k)))^{\binom{Z_{k-2}}{l}}\\ & =\prod_l(\cosh (lu/Z_{k-2}))^{\binom{Z_{k-2}}{l}}\\ &=e^{Z_{k-1} \varphi^{(2)}_k(u)}
 \end{aligned}\]
 where 
 \[ \varphi^{(2)}_k(u):= \frac{1}{Z_{k-1}} \sum_l \binom{Z_{k-2}}{l}\log \cosh (lu/Z_{k-2}),\]
 Using \autoref{eq:eldb} for $H:=m^{(1)}_k(H^{(2)}_k-m^{(2)}_k)/m^{(2)}_k$ and $h:=xm^{(1)}_k$ leads to
 \[ \frac{1}{m^{(1)}_k} \log \mu(H^{(2)}_k-m^{(2)}_k)\geq xm^{(2)}_k) \leq \inf \lb 2\varphi^{(2)}_k(u)-ux,\, u\geq 0\rb.\]
 where we have used that $\lb m^{(1)}_k(H^{(2)}_k-m^{(2)}_k)/m^{(2)}_k \geq xm^{(1)}_k\rb=\lb H^{(2)}_k-m^{(2)}_k)\geq xm^{(2)}_k\rb$.

 To take the $k\to \infty$ limit, observe that $\varphi^{(2)}_k(u)$ can be interpreted  as the expectation of a \rv\, $\log \cosh (L_ku/Z_{k-2})$, with $L_k$ a binomial \rv\ such that  $L_k=l$ with probability $\frac{1}{Z_{k-1}} \binom{Z_{k-2}}{l}$ (recall that $Z_{k-1}=2^{Z_{k-2}}$). As $L_k/Z_{k-2}$ converges in law (in fact not only in law!) to $1/2$ we infer that 
 \[ \lim_{k\to\infty} \varphi^{(2)}_k(u)=\log \cosh (u/2)=:\varphi^{(2)}(u).\]
 We conclude that
 \[ \limsup_{k\to +\infty} \frac{1}{m^{(1)}_k} \log \mu(H^{(2)}_k-m^{(2)}_k\geq x m^{(2)}_k) \leq \inf \lb 2\log \cosh u/2-ux,\, u\geq 0\rb.\]
 Thus the bound we obtain for large deviations of $H^{(2)}_k$ at large $k$ is the same as the one for $H^{(1)}_k$. This is no coincidence: we shall see in \autoref{i:jldh1h2} that the large deviations of $H^{(1)}_k$ and $H^{(2)}_k$ are aligned (in a sense to be made precise), so that the $\limsup$ on the right-hand can be replaced by a $\lim$ and the inequality by an equality, showing at the same time that large deviations of $H^{(1)}_k$ and $H^{(2)}_k$ are aligned. 
}

\myitem{Sub-leading fluctuations, the case of $2H^{(2)}_k-Z_{k-2}H^{(1)}_k$\label{i:slfh12}}{A consequence of \autoref{i:evan} is that the \rv s $\sqrt{Z_{k-1}}\left(\frac{H^{(2)}_k}{\expec{H^{(2)}_k}}-1\right)$ and $\sqrt{Z_{k-1}}\left(\frac{H^{(1)}_k}{\expec{H^{(1)}_k}}-1\right)$ are good substitutes for $\widehat{H}^{(2)}_k$ and $\widehat{H}^{(1)}_k$ at large $k$. In particular their variances goes to $1$ at large $k$ but in this limit their difference goes to $0$ in mean square. Forgetting about the overall normalizations, this means that $2H^{(2)}_k-Z_{k-2}H^{(1)}_k$ is of significantly smaller size than its two constituting terms. \autoref{eq:cgf12exp} yields
\begin{equation}  \label{eq:cgf12expslf} \expec{e^{u(2H^{(2)}_k-Z_{k-2}H^{(1)}_k)}}=\prod_l (\cosh (2l-Z_{k-2})u/2)^{\binom{Z_{k-2}}{l}}.
\end{equation}

We note that the fourth derivative of the even function $t \mapsto \log \cosh \frac{t}{2}$, namely $t\mapsto \frac{2 \cosh^2  \frac{t}{2} -3}{8\cosh^4 \frac{t}{2}}$,
is continuous and vanishes at infinity, so it is bounded. Its image is $[-1/8,1/24]$. Taking the logarithm of \autoref{eq:cgf12expslf}, we infer from Taylor's formula that
\[ -\frac{t^4}{192}\leq \log \cosh \frac{t}{2}-\frac{t^2}{8} \leq  \frac{t^4}{576} \text{ for } t\in \mathbb{R}.\]
Using the interpretation of the sum over $l$ as a binomial expectation, we infer easily that 
\[ \sum_l \binom{Z_{k-2}}{l}(2l-Z_{k-2})^2= Z_{k-1}Z_{k-2}\]
and
\[ \sum_l \binom{Z_{k-2}}{l}(2l-Z_{k-2})^4= Z_{k-1}Z_{k-2}(3Z_{k-2}-2).\]
Consequently,
\[ -\left(\frac{3Z_{k-2}-2}{Z_{k-1}Z_{k-2}}\right)\frac{u^4}{12}\leq \log \expec{e^{2u(2H^{(2)}_k-Z_{k-2}H^{(1)}_k)/\sqrt{Z_{k-1}Z_{k-2}}}}-\frac{u^2}{2} \leq \left(\frac{3Z_{k-2}-2}{Z_{k-1}Z_{k-2}}\right)\frac{u^4}{36} \text{ for } u\in \mathbb{R}.\]
In particular, the large $k$ limit in law of $2(2H^{(2)}_k-Z_{k-2}H^{(1)}_k)/\sqrt{Z_{k-1}Z_{k-2}}$ exists and is a normalized Gaussian $N(0,1)$. The scaling is indeed consistent with the formula
\[ \expec{\left(\frac{H^{(2)}_k}{\expec{H^{(2)}_k}}- \frac{H^{(1)}_k}{\expec{H^{(1)}_k}}\right)^2} =  R_{k-2,k}(0)-R_{k-1,k}(0)\]
established in \autoref{i:evan}: recall that
\[ \expec{H^{(2)}_k}=Z_{k-1}Z_{k-2}/4 \quad \expec{H^{(1)}_k}=Z_{k-1}/2 \quad R_{k-2,k}(0)-R_{k-1,k}(0)=1/(Z_{k-1}Z_{k-2}).\]
}

\myitem{Bounds on  $2H^{(2)}_k-Z_{k-2}H^{(1)}_k$\label{i:cbh12}}{We have seen in \autoref{i:slfh12} that in terms of variance $2H^{(2)}_k-Z_{k-2}H^{(1)}_k$ is of significantly smaller size than its two constituting terms.  For this item, we concentrate on the extreme values of this \rv.

We have established in \autoref{i:anc} that the \rv s $H^{(2)}_k$ and $H^{(1)}_k$ are maximal when evaluated on ${\mathcal S}_{k-1}$, with
\[H^{(2)}_k({\mathcal S}_{k-1}) = Z_{k-1}Z_{k-2}/2 \text{ for } k\geq 2 \qquad H^{(1)}_k({\mathcal S}_{k-1})=Z_{k-1} \text{ for } k\geq 1.\]
In particular, the integer-valued \rv\ $2H^{(2)}_k-Z_{k-1}H^{(1)}_k$ vanishes on ${\mathcal S}_{k-1}$.

But in fact, the compensations are much more drastic. From \autoref{eq:cgf12expslf} we infer that $\expec{e^{u(2H^{(2)}_k-Z_{k-2}H^{(1)}_k)}}$ is a symmetric Laurent polynomial in the variable $e^u$ of degree
\[\sum_{l}\binom{Z_{k-2}}{l}|l-Z_{k-2}/2|=\sum_{l> Z_{k-2}/2}\binom{Z_{k-2}}{l}(2l-Z_{k-2}).\]
The last sum is telescopic\footnote{Just as $xe^{-x^2/2}dx=d(-e^{-x^2/2})$, the continuous analog.} because $(2l-n)\binom{n}{l}=n\left(\binom{n-1}{l-1}-\binom{n-1}{l}\right)$ leading to
\[\sum_{l> Z_{k-2}/2}\binom{Z_{k-2}}{l}(2l-Z_{k-2})=Z_{k-2}\binom{Z_{k-2}-1}{\floor{(Z_{k-2}-1)/2}}. \]
The right-hand side equals $Z_{k-2}\binom{Z_{k-2}-1}{Z_{k-2}/2-1}=Z_{k-2}/2\binom{Z_{k-2}}{Z_{k-2}/2}$ as soon as $k\geq 3$.

Thus the possible values of $2H^{(2)}_k-Z_{k-2}H^{(1)}_k$ are strictly restricted to the interval $\intt{-Z_{k-2}\binom{Z_{k-2}-1}{\floor{(Z_{k-2}-1)/2}},Z_{k-2}\binom{Z_{k-2}-1}{\floor{(Z_{k-2}-1)/2}}}$. The large $k$ asymptotics for the upper bound of $2H^{(2)}_k-Z_{k-2}H^{(1)}_k$ turns out to be
\begin{equation}
\label{eq:mh12}
\max_{{\mathcal S}_k} 2H^{(2)}_k-Z_{k-2}H^{(1)}_k \sim 2H^{(2)}_k({\mathcal S}_{k-1})/\sqrt{2\pi Z_{k-2}}
\end{equation}
which means a reduction factor of $\sqrt{2\pi Z_{k-2}}$ compared to the maximum of $2H^{(2)}_k$ or $Z_{k-2}H^{(1)}_k$ taken separately. Massive compensations indeed! A direct combinatorial understanding of these compensations and their possible generalization to deeper chain functions would clearly be desirable.

Another inspection of \autoref{eq:cgf12expslf} shows that for $k\geq 2$ the probability of the largest (or the lowest) value of $2H^{(2)}_k-Z_{k-2}H^{(1)}_k$ is $2^{-Z_{k-1}}$ if $Z_{k-2}$ is odd, i.e. for $k=2$, and $2^{\binom{Z_{k-2}}{Z_{k-2}/2}-Z_{k-1}}$ if $Z_{k-2}$ is even, i.e. as soon as $k \geq 3$. In particular, for later reference,
\begin{equation}
  \label{eq:pmh12}
  \lim_{k\to \infty}\frac{1}{Z_{k-1}}\log \mu(2H^{(2)}_k-Z_{k-2}H^{(1)}_k \text{ is maximal})=-\log 2.
\end{equation}
}
\myitem{Bounds on large deviations of $2H^{(2)}_k-Z_{k-2}H^{(1)}_k$\label{i:bldh12}}{We now turn to large deviations proper. Using \autoref{eq:eldb} for $H:=(2H^{(2)}_k-Z_{k-2}H^{(1)}_k)\sqrt{\frac{4}{Z_{k-2}}}$ and $h=Z_{k-1}x$ leads to
\begin{eqnarray*}
\frac{1}{Z_{k-1}}\log \mu\left(2H^{(2)}_k-Z_{k-2}H^{(1)}_k\geq x Z_{k-1}\sqrt{\frac{Z_{k-2}}{4}}\right) & \leq & \\
& & \hspace{-7cm} \inf \lb -ux+\frac{1}{Z_{k-1}}\sum_l \binom{Z_{k-2}}{l} \log \cosh (l-Z_{k-2}/2)u\sqrt{\frac{4}{Z_{k-2}}},\, u\geq 0\rb.  
\end{eqnarray*}
By the central limit theorem for the binomial distribution (in the case at hand, the mean is $Z_{k-2}/2$ and the variance $Z_{k-2}/4$), we infer that
\[ \lim_{k\to \infty} \frac{1}{Z_{k-1}}\sum_l \binom{Z_{k-2}}{l} \log \cosh (l-Z_{k-2}/2)u\sqrt{\frac{4}{Z_{k-2}}} =\int_{-\infty}^{+\infty} \frac{d\lambda}{\sqrt{2\pi}} e^{-\lambda^2/2}\log \cosh \lambda u.\]
Finally
\begin{equation} \label{eq:ldbh12}
\begin{aligned}
\limsup_{k\to +\infty} \frac{1}{Z_{k-1}}\log \mu\left(2H^{(2)}_k-Z_{k-2}H^{(1)}_k\geq x Z_{k-1}\sqrt{\frac{Z_{k-2}}{4}}\right) & \leq & \\ & & \hspace{-6cm} \inf \lb \int_{-\infty}^{+\infty} \frac{d\lambda}{\sqrt{2\pi}} e^{-\lambda^2/2}\log \cosh \lambda u-ux,\, u\geq 0\rb.
\end{aligned}
\end{equation}
The variational equation $\int_{-\infty}^{+\infty} \frac{d\lambda}{\sqrt{2\pi}} e^{-\lambda^2/2}\lambda \tanh \lambda u=x$ has a single solution $u\in [0,+\infty[$ for $x\in [0,\sqrt{2/\pi}[$, with $u\to +\infty$ for $x\to\sqrt{2/\pi}$. But
\[ \inf \lb \int_{-\infty}^{+\infty} \frac{d\lambda}{\sqrt{2\pi}} e^{-\lambda^2/2}\log \cosh \lambda u-ux,\, u\geq 0 \rb =-\infty \text{ for } x >\sqrt{2/\pi}.\]
Finally, rewriting $\sqrt{2/\pi}=\int_{-\infty}^{+\infty} \frac{d\lambda}{\sqrt{2\pi}} e^{-\lambda^2/2}|\lambda|$ (a fact we already used silently above) we see by dominated convergence that
\[ \lim_{u\to +\infty} \int_{-\infty}^{+\infty} \frac{d\lambda}{\sqrt{2\pi}} e^{-\lambda^2/2}\log \cosh \lambda u-u\sqrt{2/\pi}=-\log 2. \]

Thus, the naive large deviation bound says that asymptotically\\
-- The \rv\ $2H^{(2)}_k-Z_{k-2}H^{(1)}_k$ cannot exceed
\[\sqrt{\frac{2}{\pi}}Z_{k-1}\sqrt{\frac{Z_{k-2}}{4}}=Z_{k-1}Z_{k-2}/\sqrt{2\pi Z_{k-2}}=2H^{(2)}_k({\mathcal S}_{k-1})/\sqrt{2\pi Z_{k-2}},\]
recovering the exact bound from \autoref{eq:mh12}\\
-- The corresponding probability satisfies
\[ \limsup_{k\to +\infty} \frac{1}{Z_{k-1}}\log \mu\left(2H^{(2)}_k-Z_{k-2}H^{(1)}_k\geq 2H^{(2)}_k({\mathcal S}_{k-1})/\sqrt{2\pi Z_{k-2}}\right) \leq -\log 2\]
and the bound $-\log 2$ is the exact value in \autoref{eq:pmh12}.

These two facts give some hope that in \autoref{eq:ldbh12} the $\limsup$ can be replaced by a $\lim$ and the inequality by an equality, or to say it differently some hope that  
\[\inf \lb \int_{-\infty}^{+\infty} \frac{d\lambda}{\sqrt{2\pi}} e^{-\lambda^2/2}\log \cosh \lambda u-ux,\, u\geq 0\rb\]
is the exact large deviation function for $2H^{(2)}_k-Z_{k-2}H^{(1)}_k$.
}

\myitem{Bounds on joint large deviations of $H^{(1)}_k$ and $H^{(2)}_k$\label{i:jldh1h2}}{The fact that the extremal values of $2H^{(2)}_k-Z_{k-2}H^{(1)}_k$ are much smaller than the extremal values of the individual terms was established in \autoref{i:cbh12}, in particular \autoref{eq:mh12}. This fact implies immediately that the joint large deviations of $H^{(1)}_k$ and $H^{(2)}_k$ exhibit an alignment phenomenon:\footnote{recall that $m^{(1)}_k:=\expec{H^{(1)}_k}=Z_{k-1}/2$ and $m^{(2)}_k:=\expec{H^{(2)}_k}=Z_{k-1}Z_{k-2}/4$.} for large enough $k$\\
-- The probability $\mu(H^{(1)}_k-m^{(1)}_k\geq x_1 m^{(1)}_k \text{ and } H^{(2)}_k-m^{(2)}_k\geq x_2 m^{(2)}_k)$ is exactly $\mu(H^{(1)}_k-m^{(1)}_k\geq x_1 m^{(1)}_k)$ for $x_1>x_2$ and $\mu(H^{(2)}_k-m^{(2)}_k\geq x_2 m^{(2)}_k)$ for $x_1<x_2$.\\
--  The probability $\mu(H^{(1)}_k-m^{(1)}_k\geq x_1 m^{(1)}_k \text{ and } H^{(2)}_k-m^{(2)}_k\leq x_2 m^{(2)}_k)$ is exactly $0$ for $x_1>x_2$.\\
--  The probability $\mu(H^{(1)}_k-m^{(1)}_k\leq x_1 m^{(1)}_k \text{ and } H^{(2)}_k-m^{(2)}_k\geq x_2 m^{(2)}_k)$ is exactly $0$ for $x_1<x_2$.

In particular $H^{(1)}_k$ and $H^{(2)}_k$ have the same large deviation function, as announced at the end of \autoref{i:ldeux}.
}

\section{Conclusions}

In this study of finite pure sets from the point of view of statistical mechanics, we have established a number of Gaussian limit theorems for chain functions. The main result is that for large depth, the joint fluctuations of all chain functions (except the simplest, $H^{(0)}_k$ which is constant, and  $G^{(1)}_k$ which is a symmetric Bernoulli \rv ) are described by just a pair of Gaussian random variables. We have argued on examples that this simplicity extends to large deviations. We have also been able to compute a number of explicit characteristics of finite pure sets, like the number of transitive pure sets at a given depth, and we have met examples of Liouville numbers. 

We have worked with the uniform probability measure all along. But the formalism of weights and generating functions suggests that we could use weights (with indeterminates replaced by numbers) as Boltzmann weights for pure sets. The simplest case, one that should lead to explicit results and must be considered first, is weighing by size. This means to assign a Boltzmann weight\footnote{Remember $H^{(1)}_k$ counts the number of chains of length $1$, i.e. the number of elements, of a pure set.} $e^{-\beta H^{(1)}}$ to pure sets in ${\mathcal S}_k$ and study the $\beta$ dependence, $\beta=0$ being the uniform measure. We leave this generalization for a future exploration.  

\vspace{.3cm}
We conclude with a more philosophical question. Along its history, physics has often fruitfully accessed to features of the infinite via the finite. The solution of (ordinary or partial) differential equations via discretization is a common example (not specific to physics). A more striking instance, though in the same spirit, is perhaps the use of finite portions of lattices, when the limits of larger and larger finite lattices and of smaller and smaller mesh, taken either simultaneously or one after the other in arbitrary order, have given access to deep properties of quantum field theories and phase transitions. The discretization of quantum gravity, and its spectacular successes in two dimensions via (planar) maps, is another deep illustration. In these notes, an analogous limiting procedure, the approximation of ${\mathcal S}_{\infty+1}=2^{{\mathcal S}_{\infty}}$ by the sequence $({\mathcal S}_k)_{\kN}$ has led to a number of (mostly Gaussian) limit theorems incorporating a perhaps unexpected rigidity. Whether we have, or could have, learned something about set theory from a more general perspective via this limiting procedure is another matter. It seems unlikely that the approach of these notes could give access to more than the first two naive levels in the vast hierarchy of infinities hosted by set theory.

We use the qualifier ``naive'' because they are the first two levels only if one takes the view that there is no kind of infinity strictly between the countable and the continuum, the view that $2^{\mathbb N}$ (or equivalently ${\mathbb R}$) comes right after ${\mathbb N}$ in terms of size. This is a famous undecidable question of standard set theory which dates back to Cantor. Understanding that there could be ``many'' in-between sizes is one of the great achievements of G\"odel and Cohen, see e.g. \cite{cohen-2008}. The so-called continuum hypothesis is that there is none. It can be added as a new axiom, but conflicting axioms that there are intermediates of various kinds could be added as well. These new axioms have sometimes implications for natural questions in everyday mathematics  (see e.g. \cite{godefroy-2022}) but what the ``right'' choice is, if any, remains unclear. It might be however that one day physics will give some motivations to prefer one alternative to the others. This possibility has been suggested, rather vaguely, in the context of quantum measurement \cite{bossche-grangier-2023}.

Baire's lemma allows to prove easily that some sets (with finite approximations for instance) are not countable, but it often happens that with some more work those sets can be proven to have the size of the continuum without the need to appeal to the continuum hypothesis (see e.g. \cite{godefroy-2022}). It is unknown to the author whether, under the negation of the continuum hypothesis, finite sets could be used in clever ways to approximate all or certain intermediate sizes between those of $\mathbb N$ and $\mathbb R$.

\section*{Acknowledgments} The author thanks the members of the quantum reading group at IPhT for the stimulating environment.

\newpage

\printbibliography

\end{document}